# integrable models in condensed matter physics

Natan Andrei

DEPARTMENT OF PHYSICS, RUTGERS UNIVERSITY
PISCATAWAY, N.J. 08854

**Lecture 1: INTRODUCTION**

The study of integrable models has a long and rich history in condensed matter physics, beginning with Bethe's solution of the one dimensional Heisenberg Model [1] and extending to our days when a variety of soluble models provide the paradigms that form much of our physical intuition.

Integrable models are typically defined on a line; they may thus describe the physics of actually linear systems such as organic conductors or the physics of higher dimensional systems where a particular rotational mode has been isolated, as is the case with impurity models. Higher dimensional models, though not integrable, often exhibit properties found in lower-dimensional relatives which can be studied non-perturbatively. This provides guidance to strongly coupled, often inaccessible, hamiltonians required to describe properties of the new materials such as heavy fermions systems and High-Temperature Superconductors.

Rather than offer an overview of the theory of integrability and a list of solved models, we have chosen to present here the detailed solutions of two models (with an occasional side glance at others) which play an important role in modern condensed matter physics. The first is the Hubbard hamiltonian, the prime example of a model incorporating strong correlations, and thus playing an important role in the efforts to understand Cuperate Superconductors. The second model is the Kondo hamiltonian, describing dilute magnetic alloys, whose properties are basic to the understanding of heavy Fermions. The models also provide examples of solutions on the lattice and in the continuum respectively.

In the first two lectures we discuss the steps leading from a given (integrable) hamiltonian to a set of algebraic equations that encapsulate the physics they contain. In the third lecture we solve the equations governing the Kondo model, and in the fourth those governing the Hubbard model. We discuss the structure and nature of the elementary excitation and their interactions, and derive the thermodynamic properties of the models by summing over all excitations.

We intended originally to include a fifth lecture to situate the exact solution in the framework of the Renormalization Group (RG) approach [2]. Time constraints, however, will not allow it. Instead, let me provide a brief summary. The RG approach provides a systematic way to explore the low energy physics of a hamiltonian by successively integrating out high energy modes and incorporating their effects in the resulting theory. This way one constructs a series of hamiltonians all of which have the same physics in the infra-red. This renormalization group "flow" may tend to a fixed point hamiltonian which by construction is scale invariant, and typically much simpler than the starting hamiltonian. In many cases it is also conformally invariant and then can be completely specified in terms of very few parameters [?]. When an exact solution is available these parameters can be computed quite easily, without resorting to the much more arduous task of constructing the RG flow. Once the fixed point has been identified, correlation functions can be written down which describe the asymptotic behavior of the model. The calculation of the full correlation function from the exact solution, valid on all energy scales, is still a major open problem.

**The Hubbard Model.**

The solution of the model was obtained by Lieb and Wu [4] in the form of a set of coupled integral equations applying a method due to Yang [5]. They also analyzed the equations for the repulsive interaction at half filling, and showed that a charge gap opens no matter how weak the interaction.

Investigating the spectrum of the model, one finds it contains spinless charge excitations (in modern terminology refered to as *holons* ), as well as spin-$\frac{1}{2}$ excitations carrying no charge (*spinons*). In the repulsive case away from half filling both types of excitations are gapless, while at half filling a charge gap opens, inducing a Mott transition at zero coupling. In the attractive case a spin gap is always present and the charge excitations are gapless.

In the low energy limit the excitation spectrum becomes linear. This fact allows a simple description of the low energy properties of the model: the fixed point hamiltonian to which the system flows at long distances is expected to be conformally invariant.

An important property of the model with, perhaps, analogues in higher dimensions is the charge-spin decoupling. One finds that charge and spin degrees of freedom propagate with different velocities and are characterized by different couplings. Thus an electron added to the system will not survive as a single particle and hence the fermion correlation function will not possess single particle poles (i.e. Z=0), signaling a breakdown of the Fermi-liquid assumptions.

A related, but not identical, phenomenon of complete decoupling occurs in the low energy limit. What one observes is that the charge and spin excitations do not interact at all and can be treated separately. (A simple way to understand the complete decoupling is to consider a continuum version of the model which would describe it at weak coupling and low energies. Away from half filling one obtains the g-ology model [6], another integrable model, solved under the name of the Chiral Gross-Neveu model [7] [8]. In the continuum model the decoupling occurs on all scales, and is due to simple kinematical considerations.) This decoupling may be an important clue in understanding how a Fermi liquid is destroyed; Already in the free field theory there is a complete charge-spin decoupling (however, both degrees of freedom are chracterized by the same velocities) in the Bethe-Ansatz basis of the Hilbert space, obtained by taking the $U \to 0$ limit in the equations below. The conventional Fermi liquid description, on the other hand, corresponds to the choice of the occupation number basis (Fock basis) in the Hilbert space. Both descriptions are valid. When $U$ is turned on, the charge and spin excitations still maintain their separate identities, but are modified. They acquire different velocities and couplings, they interact with each other and among themselves, and the Fermi liquid picture breaks down. Even in the low energy limit the spectrum can no longer be described in terms of fermionic quasi particles. Put differently, the model flows (in the sense of Renormalization Group) to a non Fermi liquid fixed point, the Luttinger liquid [9] (a particular c=1 conformal field theory.)

The Hamiltonian is well known

$$H = -t\sum_{i=1}^{L}(\psi_{ia}^*\psi_{i+1a} + h.c.) + U\sum_{i=1}^{L}n_{i\uparrow}n_{i\downarrow}. \qquad (1)$$

The fermionic field $\psi_{ia}$ annihilates a particle with spin component $a$ at site $i$ on a chain with $L$ sites. The first term in the Hamiltonian describes hopping from site $i$ to site $i+1$ and back, while the second term is a crude approximation to a Coulomb repulsion.

The model has an obvious $U(1) \times SU(2)$ symmetry,

$$\begin{aligned}\psi_{ia} &\longrightarrow e^{i\theta}\psi_{ia} \\ \psi_{ia} &\longrightarrow U_{ab}\psi_{ib}\end{aligned}$$

expressing the charge consevation and invariance under spin rotation. The associated generators are given by the spin operators,

$$S_z = \frac{1}{2}\sum_{i=1}^{L}(n_{i\uparrow} - n_{i\downarrow}), \quad S^+ = \sum_{i=1}^{L}\psi_{i\uparrow}^*\psi_{i\downarrow}, \quad S^- = (S^+)^* \qquad (2)$$

and the number operator,

$$\mathcal{N} = \sum_{i=1}^{L}(n_{i\uparrow} + n_{i\downarrow}). \qquad (3)$$

The conservation law associated with the U(1) symmetry $\mathcal{N}$ allows us to study the hamiltonian for a fixed number $N$ of electrons. We shall label the states with the quantum numbers $M$ and $M' = N - M$ of the down-and up spins, $|F(N-M,M)>$, and the corresponding energies $E(N-M,M;U)$. The $z$-component of the total spin is $S_z = \frac{1}{2}(M' - M) = \frac{1}{2}(N - 2M)$. By construction, this is also the value of the total spin $S$, $S = S_z$, since the $B$-operators we shall use in Lecture 2 to build the eigenstates commute among themselves and thus specify a Young-tableau with definite transformation properties. The states we construct are therefore $SU(2)$ highest weight states and the rest of the multiplet is obtained by repeated action of the lowering operators $S^-$. Beyond $N$ and $M$, each state is labeled by an infinite set of quantum numbers which we shall specify later.

There is another, less obvious, charge SU(2) invariance (of which the U(1) is a subgroup) present in a slightly modified version of the model [10],

$$H' = H - \frac{U}{2}\sum_{i=1}^{L}(n_{i,\downarrow} + n_{i,\uparrow}). \qquad (4)$$

We added a chemical potential term to the hamiltonian. In a grand canonical ensemble the model will be half filled. Equivalently, the symmetry will show up if we work in the canonical ensemble and choose the filling appropriately. The symmetry is realized by number density and pair creation and annihilation operators,

$$C_z = \frac{1}{2}\sum_{i=1}^{L}(n_{i\uparrow} + n_{i\downarrow}) - \frac{L}{2}, \quad C^+ = \sum_{i=1}^{L}(-1)^i\psi_{i\uparrow}\psi_{i\downarrow}, \quad C^- = (C^+)^*. \qquad (5)$$

As the number operator does not commute with $C^\pm$ and the symmetry manifests itself only upon comparing excitations in systems with different number of electrons. Still we shall find some consequences of the symmetry even though we mostly work with a fixed number of particles.

We shall discuss the repulsive as well as the attractive model. The following particle-hole $Z_2$ transformation

$$\psi_{i\downarrow}^* \to (-1)^i \psi_{i\downarrow}, \quad \psi_{i\uparrow}^* \to (-1)^i \psi_{i\uparrow}^* \tag{6}$$

$$\psi_{i\downarrow} \to (-1)^i \psi_{i\downarrow}^*, \quad \psi_{i\uparrow}^* \to (-1)^i \psi_{i\uparrow}^* \tag{7}$$

leads to the relation [4],

$$E(N-M, M; U) = (N-M)U + E(N-M, L-M; -U), \tag{8}$$

between the the energies of states in the two cases. The eigenstates, we shall see, are related in a more complicated way. The eigenvalues of the the modified hamiltonian are particle-hole symmetric at half filling and in particular one has

$$E'(N-M, M; U) = E'(N-M, N-M; -U). \tag{9}$$

We turn now to study the hamiltonian in a Hilbert space $\mathcal{H}_N$ of $N$ particles, defined with respect to the vacuum state $|0>$ containing none,

$$\psi_{ia}|0> = 0. \tag{10}$$

The states that span $\mathcal{H}_N$ are of the form

$$|F> = \sum_{a_1...a_N} \sum_{n_1...n_N} F_{a_1...a_N}(n_1...n_N) \prod_{i=1}^{N} \psi_{a_i n_i}^* |0>, \tag{11}$$

and the Fock eigenvalue problem

$$H|F> = E|F> \tag{12}$$

turns into its $N$-particle version

$$hF = EF, \tag{13}$$

with the first quantized hamiltonian,

$$h = -t \sum_{j=1}^{N} \Delta_j + U \sum_{j<l} \delta_{n_j n_l}, \tag{14}$$

acting on the wave function $F_{a_1..a_N}(n_1...n_N)$. The hopping operator $\Delta$ (the discrete version of the Laplacian ) is given by:

$$\Delta_j F_{a_1..a_N}(n_1...n_j...n_N) = F_{a_1..a_N}(n_1...n_j+1...n_N) + F_{a_1..a_N}(n_1...n_j-1...n_N) \tag{15}$$

Let us note that when a model is studied perturbatively the starting point is the hamiltonian

$$H_0 = -t\sum_i (\psi^*_{i+1\,a}\psi_{ia} + h.c.), \tag{16}$$

whose ground state is the Fermi-sphere. Subsequently, the interaction is turned on and the energy levels are corrected order by order. In some cases, when nothing dramatic (such as a phase transition) happens, this procedure leads to $|\Omega>$, the true ground state.

Our approach here is different. In the presence of a finite volume (infrared) cut-off $L$, and in the presence of an ultraviolet cut-off (the lattice spacing, in this case) the true ground state and the empty state are in the same Hilbert space. Therefore one can use the representation of $\psi$ and $\psi^*$ as creation and annihilation operators with respect to $|0>$ to construct a full set of eigenvectors and, in particular, determine $|\Omega>$.

We proceed, then, to diagonalize $h$ within $\mathcal{H}_N$. We shall do it for finite $N$, and eventually take the thermodynamic limit: $L, N \to \infty$, with $n = N/L$ fixed.

Begin by considering the case $N=1$. Now

$$h = -t\Delta \tag{17}$$

with the obvious solution:

$$F_a(n) = A_a e^{ikn}, \quad E = -2t\cos k$$

Proceed to consider the case $N = 2$. Then,

$$h = -t(\Delta_1 + \Delta_2) + U\delta_{n_1 n_2}. \tag{18}$$

The particles interact only when $n_1 = n_2 = n$. Away from this boundary the Hamiltonian is free, and the wave function is given as a product of single particle solutions:

$$\begin{aligned}F_{a_1 a_2}(n_1, n_2) &= \mathcal{A} e^{ik_1 n_1 + ik_2 n_2}[A_{a_1 a_2}\theta(n_1 - n_2) + B_{a_1 a_2}\theta(n_2 - n_1)]\\ &= e^{ik_1 n_1 + ik_2 n_2}[A_{a_1 a_2}\theta(n_1 - n_2) + B_{a_1 a_2}\theta(n_2 - n_1)]\\ &\quad - e^{ik_1 n_2 + ik_2 n_1}[A_{a_2 a_1}\theta(n_2 - n_1) + B_{a_2 a_1}\theta(n_1 - n_2)]\end{aligned}$$

where $\mathcal{A}$ is the antisymmetrizer, and $\theta(n)$ is a step function. The corresponding eigenvalue is
$$E = -2t(\cos k_1 + \cos k_2). \tag{19}$$

We introduce now the $S$-matrix relating the amplitudes in the two regions
$$B_{a_1 a_2} = S^{b_1 b_2}_{a_1 a_2} A_{b_1 b_2}. \tag{20}$$

This is the *bare* S-matrix, later we shall also discuss the dressed or physical S-matrix. To determine it we have to impose two conditions:

(1) *Uniqueness*, to ensure that the value of $F(n,n)$ is defined independently of the region. This leads to
$$A_{n_1 n_2} - B_{n_2 n_1} = B_{n_1 n_2} - A_{n_2 n_1} \tag{21}$$

in other words
$$I - PS = S - P \tag{22}$$

where $P$ is the spin exchange operator, and $I$ is the spin identity operator
$$(PA)_{a_1 a_2} = P^{b_1 b_2}_{a_1 a_2} A_{b_1 b_2} = A_{a_2 a_1}$$
$$(IA)_{a_1 a_2} = I^{b_1 b_2}_{a_1 a_2} A_{b_1 b_2} = A_{a_1 a_2}$$

Clearly $P^{b_1 b_2}_{a_1 a_2} = \delta^{b_2}_{a_1} \delta^{b_1}_{a_2}$ and $I^{b_1 b_2}_{a_1 a_2} = \delta^{b_1}_{a_1} \delta^{b_2}_{a_2}$. Expression (22) can be rearranged to
$$\frac{1}{2}(1+P) \, S \, \frac{1}{2}(1+P) = \frac{1}{2}(1+P), \tag{23}$$

indicating that $S = 1$ in the symmetric channel. This is not surprising since an on-site interaction operates only in the anti-symmetric spin channel. The S-matrix therefore is of the form
$$S = \frac{1}{2}(1+P) + \frac{1}{2}(1-P) \, s \tag{24}$$

where s is a scalar in spin space.

(2) *The Schrodinger equation on the boundary $n_1 = n_2$*. That is,

$$-t[F(n+1,n) + F(n-1,n) + F(n,n+1) + F(n,n-1)] + UF(n,n) = EF(n,n).$$

Explicitly

$$e^{i(k_1+k_2)n}[(-t)((e^{ik_1}+e^{-ik_2})s - e^{-ik_1} - e^{ik_2}) + U\frac{1+s}{2}](I-P)A$$
$$= -2t(\cos k_1 + \cos k_2)e^{i(k_1+k_2)n}\frac{1+s}{2}(I-P)A.$$

Solving for $s$ we find
$$s = \frac{i(\sin k_1 - \sin k_2) + \frac{u}{2}}{i(\sin k_1 - \sin k_2) - \frac{u}{2}}, \qquad (25)$$

where
$$u = \frac{U}{t}, \qquad (26)$$

and the $S$-matrix $S^{b_j b_l}_{a_j a_l}$ becomes (when particles $j$ and $l$ interact),

$$S^{jl} \equiv S^{b_j b_l}_{a_j a_l} = \frac{(\sin k_j - \sin k_l)I^{b_j b_l}_{a_j a_l} + i\frac{u}{2}P^{b_j b_l}_{a_j a_l}}{(\sin k_j - \sin k_l) + i\frac{u}{2}}. \qquad (27)$$

Let us generalize the construction to $N$ particles. Begin by dividing the configuration space into $N!$ regions according to the ordering of the particles on the line, and label them by elements of the permutation group. For example, the region ($n_3 < n_1 < n_5...$) will be labeled by the permutation $Q = (Q_1 = 3, Q_2 = 1, Q_3 = 5...)\epsilon S_N$. As there is no interaction in the interior of these regions the wave function will given as a sum over a product of single particle wave functions. In our case $F = \sum$ PLANE WAVES. When a boundary is crossed, two particles interact (note that mutliparticle interaction is forbidden by Fermi statistics ) and hence the amplitudes in the regions across the boundary will be related by the *two particle* S-matrix just determined. We thus consider wave functions of the Bethe-form (The Bethe Ansatz):

$$F_{a_1...a_N}(n_1...n_N) = \mathcal{A}e^{i\sum_j k_j n_j}\sum_Q A_{a_1...a_N}(Q)\theta(n_Q), \qquad (28)$$

with the energy and momentum given by

$$E = \sum_j -2t\cos k_j \qquad (29)$$
$$P = \sum_j k_j. \qquad (30)$$

In eq(28) the $Q$-sum runs over all the $N!$ regions, $\theta(x_Q)$ is equal to 1 if the particles are ordered according to $Q$ and vanishes otherwise, and $A_{a_1..a_N}(Q)$ is the spin amplitude in region $Q$. The amplitude in region $Q$ is related to an adjacent amplitude $Q'$, differing from it by the exchange of neighboring particles $i$ and $j$, via the S-matrix $S^{ij}$,

$$A_{a_1..a_i..a_j..a_N}(Q') = (S^{ij})^{b_1...b_N}_{a_1...a_N} A_{b_1...b_N}(Q) = (S^{ij})^{b_i b_j}_{a_i a_j} A_{a_1..b_i..b_j..a_N},(Q)$$

where for convenience we regard the 2-particle S-matrix $S^{ij}$ as carrying $N$ indices operating in an $N$ particle spin space but acting non trivially only on particles $i$ and $j$,

$$(S^{ij})^{b_1...b_N}_{a_1...a_N} = (S^{ij})^{b_i b_j}_{a_i a_j} \prod_{k \neq i,j} \delta^{b_k}_{a_k} \tag{31}$$

We labeled the S-matrix by the particles it acts on. Let me be more explicit: We are considering a region $Q = (Q1, Q2 \cdots Ql, Q(l+1) \cdots QN)$ and an adjacent region $Q' = (Q1, Q2 \cdots Q(l+1), Ql \cdots QN)$. In region $Q$ the particle $i = Ql$ is to the left of particle $j = Q(l+1)$ while in region $Q'$ the particle $j$ is to the left of particle $i$. In other words, $Q' = P^{ij}Q$. Hence to move from region $Q$ to region $Q'$ we apply $S^{ij}$, $A(Q') = S^{ij}A(Q)$ and similarly $A(Q) = S^{ji}A(Q')$. Note that $S^{ij} = (S^{ji})^{-1}$, and in the amplitude on which $S^{ij}$ acts particle $i$ is to the left of particle $j$.

How are regions $Q_1$ and $Q_2$ related when they are not adjacent, that is, not related by a single transposition? There is always a path (and usually more than one) in the permutation group, given as a product of transpositions, leading from $Q_1$ to $Q_2$. To relate the regions we take the corresponding product of the S-matrices. If, for example

$$Q_1 = P^{ij} P^{jk} P^{kl} Q_2 \tag{32}$$

then

$$A(Q_1) = S^{ij} S^{jk} S^{kl} A(Q_2). \tag{33}$$

If the path is not unique, consistency requires that the result be path independent. Consider the $N = 3$ case. Configuration space is divided into 3!=6 regions, not all of which are adjacent. We label the regions by $(ijk)$ for the ordering $(n_i < n_j < n_k)$, and draw a line between adjacent regions. Thus one obtains the diagram,

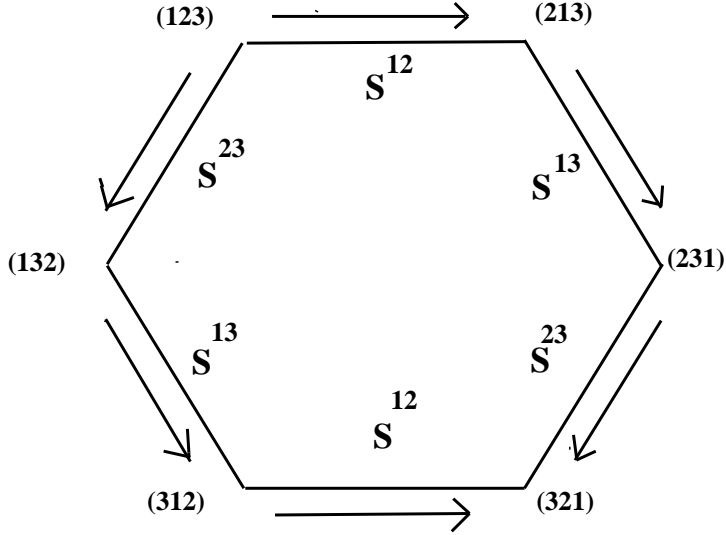

Figure 1: The regions in the Configuration Space of 3 Particles.

Here region (123) is related to region (213) by exchanging particles 1 and 2. Therefore $A(213) = S^{12} A(123)$. Continuing this process one may, starting from the wave function in region (123), construct the wavefunction in any other region. However, note that region (321) can be reached by two different paths and hence for the procedure to be consistent these paths must yield the same result. Thus consistency in the three particle case leads to the Yang-Baxter Equation (YBE) [5]

$$S^{23} S^{13} S^{12} = S^{12} S^{13} S^{23} \tag{34}$$

in the notation of eq (31), or explicitly

$$(S^{23})^{d_2 d_3}_{a_2 a_3} (S^{13})^{d_1 b_3}_{a_1 d_3} (S^{12})^{b_1 b_2}_{d_1 d_2} = (S^{12})^{d_1 d_2}_{a_1 a_2} (S^{13})^{b_1 d_3}_{d_1 a_3} (S^{23})^{b_2 b_3}_{d_2 d_3}, \tag{35}$$

where both sides act on $A_{b_1 b_2 b_3}(123)$ to produce $A_{a_1 a_2 a_3}(321)$.

Similar considerations in the case of $N=2$ and $N=4$ require that the matrices also satisfy

$$\begin{aligned} S^{ij} S^{ji} &= I \\ S^{ij} S^{kl} &= S^{kl} S^{ij} \end{aligned} \qquad (36)$$

Furthermore, it can be shown that these relations are sufficient to guarantee consistency, that is, path independence, for all $N$ [11].

We may therefore rewrite eq(28) as

$$F_{a_1...a_N}(n_1...n_N) = \mathcal{A} e^{i \sum_j k_j n_j} \sum_Q S(Q) A_{a_1...a_N} \theta(n_Q) \qquad (37)$$

with $A_{a_1...a_N}$ being the amplitude in some reference region labeled $I$, $Q$ is given as a product of successive transpositions, and $S(Q)$ is the corresponding product of S-matrices.

To Summarize: if the $S$-matrix derived from the Hamiltonian satisfies the YBE, then the Bethe-form of the wave functions is consistent and the model is integrable. A direct calculation shows that indeed the $S$-matrix we derived for the Hubbard model does satisfy the YBE.

We may stop at this point to ask how a model can fail to have eigenstates of the Bethe-form. The answer is that we insisted that the single particle solutions whose product forms the wavefunction in the interior of each region be labeled by the *same set* of momenta $k_j$. This goes beyond energy and momentum conservation which place the milder requirement that $\sum_j \cos k_j$ and $\sum_j k_j$ be the same in each region. The fact that the eigenstates can be built with momenta $k_j$, individually preserved under the interaction, is a special feature of an integrable model and reflects the fact that the model possesses additional dynamical symmetry, expressed by an infinite number of commuting conserved charges. A familiar example in quantum mechanics is the hydrogen atom where the conserved Lenz vector is responsible for its integrability. The existence of these conservation laws restricts the dynamics of the model and has many consequences: There is no particle production in on-shell collisions, multi-particle (physical) S-matrices factorize into prodcucts of two-particle matrices, the Thermodynamics can be consistently expressed in terms of these on-shell S-matrices. We shall not pursue this direction here but concentrate on the off-shell consequences of the conservation laws which are manifested by the YBE guaranteeing consistent wave function of the Bethe-form.

*Periodic Boundary Conditions:* To properly quantize the model one needs to introduce a volume cut off to regulate the infra-red behaviour. We shall study the model on a finite ring with length $L$ :

$$F_{a_1...a_N}(...n_j = 0...) = F_{a_1...a_N}(...n_j = L...). \tag{38}$$

This condition, which is imposed in different regions in configuration space, can be translated to conditions on the wave function in a single region,

$$(Z_j)_{a_1...a_N}^{b_1...b_N} A_{b_1...b_N}(Q) = e^{-ik_j L} A_{a_1...a_N}(Q), \tag{39}$$

where

$$(Z_j)_{a_1...a_N}^{b_1...b_N} = (S^{jj-1}...S^{j1}S^{jN}...S^{jj+1})_{a_1...a_N}^{b_1...b_N} \tag{40}$$

To derive this expression consider an amplitude $A_{a_1...a_N}(Q)$ and act on it first with $S^{jj+1}$ to exchange particle $j$ with particle $j+1$ to its right, then act with $S^{jj+2}$ to exchange it with the particle $j+2$ which is adjacent to its right after the previous exchange with $j+1$, and so on till it is brought to the right end of the system. Now repeat the operation, moving it to the left; act on $A_{a_1...a_N}(Q)$ with $S^{j-1j}$, then with $S^{j-2j}$ and so on, till the particle $j$ reaches the left end of the system. From periodicity

$$S^{1j}....S^{j-2j}S^{j-1j}A(Q) = S^{jN}....S^{jj+2}S^{jj+1}A(Q)e^{ik_j L} \tag{41}$$

leading to eq(39).

We need to diagonalize the spin operator $Z_j$ acting in the $N$ spin space, $V^N$ (in other words, $Z_j$ acts on spin functions $A_{b_1...b_N} \in V^N = \prod_j V_j$, where $V_j = \mathcal{C}^2$ is the spin space of particle $j$). From the eigenvalues of the new spin hamiltonian $Z_j$ we can find the energy and momentum of the Fock eigenstate $|F>$ constructed from the spin eigenfunction $A_{b_1...b_N}$. Note that the $Z_j$ commute, $[Z_j, Z_l] = 0$ as a result of YBE , and can be simultaneously diagonalized.

The diagonalization of the (spin-Hamiltonian) $Z_j$ was achieved by Yang [5] by means of another Bethe Ansatz built out of single "particle" spin functions. The role of the vacuum now is played by the ferromagnetic state (all spins aligned), a single particle state corresponds to a single spin flip and a general eigenstate is built out of products of single particle solutions. This approach is reviewed in [12][13]. A related problem arose in the study of the 6-vertex model where the spin operator $Z$ plays the role of the transfer matrix [14]. Algebraic methods for its diagonalization were developed by Baxter [15], and further extended by the St Petersburg school under the name of Inverse Scattering Method [16]. We shall discuss the diagonalization of $Z$ using the latter approach. But before doing so let us turn to the Kondo model and construct its eigenstates. We shall find that similar spin problem $Z$ arises.

**The Kondo Model.**

The Kondo model describes the interaction of a conduction band with a localized spin impurity. The litterature on the subject is immense [17]; here we concentrate on some of the theoretical aspects .

The conduction band is described by the Hamiltonian,

$$H_0 = \sum_{\vec{k}} \epsilon(k) c^*_{\vec{k},a} c_{\vec{k},a} \tag{42}$$

with $c_{\vec{k},a}$ annihilating an electron with momentum $\vec{k}$ and spin component $a$.

The conduction band is coupled via spin exchange interaction to a spin $\sigma_0$ localized at $\vec{r} = 0$,

$$H_I = J \Psi^*_a(\vec{r}=0) \vec{\sigma}_{ab} \Psi_b(\vec{r}=0) \cdot \vec{\sigma}_0 \tag{43}$$

The field $\Psi_a(\vec{r})$ is the Fourier transform of $c_{\vec{k},a}$.

Since the model we consider is rotationally but not translationally invariant, an appropriate basis for the electron annihilation operators is $c_{klm,a}$, expanded in angular modes around the impurity, rather than $c_{\vec{k},a}$. Of these modes we assume now that only the s-wave modes have non-zero coupling to the impurity. Later we shall discuss the case where higher orbital modes couple to an impurity leading to the Multi-Channel Kondo model.

We further restrict our attention to low energy phenomena, entitling us to retain only momenta $k$ close to the Fermi surface: $k = k_F + q$, $|q| \ll D$, where $D$ is a cut-off, of the order of $k_F$, which will be considered large compared to any other physical scale in the problem. Linearizing the energy, $\epsilon(k) = \epsilon(k_F) + v_F q$, and Fourier transforming with respect to $q$ we find that the $s$-component of $H_0$ becomes in the limit $D \to \infty$,

$$H_0 = -i \int \psi^*_a(x) \partial_x \psi_a(x) dx. \tag{44}$$

We have chosen our units so that $v_F = 1$. The field $\psi_a(x)$ is the Fourier transform of $c_{k_F+q,00,a}$, where $x$ is the variable conjugate to $q$, $-\infty < x < \infty$.

We obtain a 1-d field theory with only right moving electrons, a reflection of the fact that the Fermi surface is simply connected. One may also carry out an analogous derivation in real space, where the problem is defined on the half line with incoming and outgoing waves, then mapping the outgoing to incoming waves but defined for $x < 0$, one obtains a full line problem with only one kind of movers.

Adding the interaction term to $H_0$ we obtain the Kondo model which is the starting point of our investigation:

$$H = -i \int \psi_a^*(x) \partial_x \psi_a(x) dx + J \psi_a^*(0) \vec{\sigma}_{ab} \psi_b(0) \cdot \vec{\sigma}_0 + J' \psi_a^*(0) \psi_a(0). \quad (45)$$

We also included a term that couples to the electron charge density at $x = 0$ with strength $J'$. Since we have linearized the model charge and spin degrees of freedom completely decouple on all scales in $H_0$. This is not modified by the interaction terms which couple separately to the spin and to the charge of the conduction electrons.

The linearization procedure is valid only when all energy scales (such as temperature $T$, magnetic field $h$, excitation energy $\epsilon$) are small compared to the cutoff. Otherwise the linearization breaks down, and details of the band structure (reflected in the cutoff procedure) become relevant. We shall consider only quantities that characterize low-energy properties of the model, and are independent of the cutoff scheme. These quantities we shall call universal.

As long as we consider only universal quantities we do not have to insist on a particular cutoff scheme. Different schemes may be employed to give the same universal quantities, though outside their domain of applicability universality may break down and results may vary; if we want to analyze properties at $T \sim D$, much more care must be taken in the construction of the model, and there may be only one physically acceptable cutoff scheme.

The above observations allow us to apply the considerations and methods of quantum field theory to the problem. Thus, as the coupling constant $J$ is dimensionless, the Hamiltonian is renormalizable, and divergences are expected in the calculations. The divergences are absorbed within the cut-off scheme chosen[18], and any particular numerical value assigned to $J$ is defined only with respect to that scheme.

One might think that no scale remains in the problem, as the coupling constant is dimensionless and the cutoff is considered infinite if we restrict ourselves to the low energy regime.

One of the fundamental properties of the model, however, is the appearance of dynamically generated scale $T_o$ (to be defined later) which uniquely determines the low-energy physics. This scale depends on the cutoff $D$ and the coupling constant $J$ in the following generic way: $T_o = D \exp[-a/\lambda(J)]$, where $\lambda(J) \to J$ as $J \to 0$. The explicit form of $\lambda(J)$ depends on the scheme used. In the conventional momentum cutoff scheme, $p \leq D_M$, one finds,

$$T_o = D_M e^{-(\pi/J_M)+(1/2)\ln J_M + \cdots}. \qquad (46)$$

This scheme, however, spoils intergrability which is restored only when $D_M$ is taken to infinity. We impose therefore another cut-off respecting integrability while still finite [19]. In this scheme one finds

$$T_o = D e^{-\pi/J}. \qquad (47)$$

Still, both constructions are characterized by the choice of, say, $T_o = 0.0007$ eV. This value is the only relevant scale in the scaling (universal) regime which defines the low-temperature and low magnetic field properties of the model. In this region the free energy $F$ takes the form

$$\frac{F}{T}(T,h;D,J) \to f\left[\frac{T}{T_o}, \frac{h}{T}\right], \qquad h, T \ll D \qquad (48)$$

where the function $f$ is universal in the sense that it is independent of the particular scheme used to define the model. The cutoff and coupling constant enter only in the combination determining $T_o$. Also, any other scale must be related to $T_o$ by pure numbers that are directly calculable. These numbers are universal.

The part of the scaling region where $T \gg T_o$ will be called the high-temperature region (still $T \gg D$). As we shall see, this is the weak coupling regime, where the effective coupling constant is small and the physics can be captured by perturbing around the weak coupling fixed point hamiltonian, $H_0$. The low-temperature region ($T \ll T_o$), however, is a strong coupling regime governed another fixed point hamiltonian, $H^*$ describing a local Fermi liquid. The *crossover* in behavior from the strong coupling regime to the weak coupling, can be described as a renormalization group flow in the space of effective hamiltonians, and is the essence of the Kondo problem [20][21][22].

The crossover can be driven by any physical parameter. Let us discuss it as a function of the temperature. Consider the impurity susceptibility $\chi^i$, which is the term in the susceptibility left over after subtracting from the total susceptibility $\chi = \partial M/\partial H$ the contribution of the electrons. (We take electrons and impurity to have the same $g$ factor.)

As we shall see, the high-temperature region lies in the weak coupling regime allowing us to apply perturbation theory to find that the impurity susceptibility attains its free value $\chi^i = \mu^2/T$ (Curie law) up to corrections that vanish logarithmically at high temperatures:

$$\chi^i \to \frac{\mu^2}{T} \left\{ 1 - \frac{1}{\ln \frac{T}{T_k}} - \frac{1}{2} \frac{\ln \ln \frac{T}{T_k}}{\ln^2 \frac{T}{T_k}} + \left( \frac{1}{\ln \frac{T}{T_k}} \right)^3 \right\}, \quad T \ll T_o \qquad (49)$$

where a new scale $T_k$ has been defined by requiring that the $1/[\ln^2(T/T_k)]$ term be absent. This is a normalization condition on $T_k$, the high-temperature or perturbative scale, which is conventionally referred to as the Kondo temperature.

While the high-temperature region is thus accessible by perturbation theory, the system enters a strong coupling regime at low temperatures and its properties change drastically. Due to the strong coupling to the electrons the impurity spin will be screened, leading to a *finite* susceptibility, $\chi_0^i$, at zero temperature. Thus define the scale $T_0$,

$$\chi_o^i = \frac{\mu^2}{\pi T_o}, \qquad (50)$$

for the low-energy regime of the model. The ratio

$$W = \frac{T_k}{T_o} \qquad (51)$$

is a universal number characterizing the temperature crossover. The crossover occurs, of course, as a function of other quantities, be it the excitation energy, the magnetic field or any other energy scale, each crossover having its own universal number. These numbers relate different asymptotic regimes and as such cannot be calculated from an effective hamilonian probing only the neighborhood of one fixed point. Instead, a complete construction is required, valid over all scales. This was first carried out numerically by Wilson [21].

We mentioned already that the physics of the strong coupling regime of the model is Fermi liquid like; it is characterized by a local potential center (the remnant of the impurity that was screened), and by induced weak interactions among the electrons. We shall discuss now a genralization, the *multichannel Kondo model*, where other, non-Fermi liquid fixed points may be reached in the infra-red.

The model was introduced by Nozieres and Blandin to describe "real metals"[23]. Taking account of the the orbital structure of the impurity they derived the most general exchange hamiltonian to describe the Kondo effect. When the atomic shell (with orbital quantum number $l$) is half filled, Hund's rule indicates that the ground state is an orbital singlet with total spin $S = (2l+1)/2$. The electrons scattering off the impurity then also carry the orbital quantum number $m$, $-l \leq m \leq l$, and one ends up with a multichannel version of the Kondo model,

$$H = -i \int \psi^*_{a,m}(x)\partial_x \psi_{a,m}(x)dx + J\psi^*_{a,m}(0)\vec{\sigma}_{ab}\psi_{b,m}(0) \cdot \vec{S} \qquad (52)$$

Here $m = 1, ..., f = 2l+1$ is the orbital channel (or flavor) index and the spin operator $\vec{S}$ is in spin-$S$ representation of $SU(2)$. In the hamiltonian the values of $f$ and $S$ are unrestricted, though in a magnetic impurity hamiltonian $f = 2S$. Other non-magnetic applications of the model exist with other values of spin and flavor [24].

The nature of the infra-red fixed point depends on those values [23]: for $f \leq 2S$ the coupling $J$ flows to infinity leading to a screened impurity in the case $f = 2S$, and to a partially screened impurity $S' = S - \frac{1}{2}$ in the case $f < 2S$. The strong coupling fixed point becomes unstable when $f > 2S$ and the infra-red physics is controled by a new, finite coupling fixed point. This new fixed point is expected to describe non Fermi-liquid behavior.

We shall show next that the Kondo hamiltonian is integrable [25] [26], and construct a complete set of eigenstates. We shall find that the spectrum consists of spin-1/2 uncharged excitations, spinons, as well as spinless particles carrying the charge degrees of freedom, holons [25]. The spectrum bears similarities to the spectrum of the Hubbard model we discussed earlier. In fact, in all integrable models belonging to this class, in a sense to be discussed, the fundamental excitations carry the same quantum numbers. They may differ in their dynamics, though; unlike the spinons and holons in the Hubbard model, these excitations decouple on all scales as result of our choice of a linear spectrum. The spinons however interact with each other and we shall calculate the physical S-matrix. We shall also derive an expression for the phase shift a spinon undergoes as it passes the impurity. Then we shall proceed to calculate the free energy, and explore the crossover as a function of thermodynamic parameters such as temperature and magnetic field, or dynamic parameters as excitation energy and momentum. In particular, we shall find an analytic expression for $W$, as well as for other crossover numbers.

We turn now to the diagonalization of the model. In the Hilbert space $\mathcal{H}_{N^e}$ of $N^e$ electrons the hamiltonian takes the form

$$h = -i \sum_{j=1}^{N^e} \partial_{x_j} + J \sum_j \delta(x_j) \vec{\sigma}_j \cdot \vec{\sigma}_0 + J' \sum_j \delta(x_j) \tag{53}$$

A single electron interacting with the impurity will be described by a wave function

$$F_{a_j a_0}(x_j) = e^{ikx_j}[A_{a_j a_0}\theta(-x_j) + B_{a_j a_0}\theta(x_j)] \tag{54}$$

here $a_j$ and $a_0$ are the spin indices of the elecrton and the impurity respectively. This function obviously satisfies the Hamiltonian for $x_j \neq 0$ and has the eigenvalue:

$$E = k \tag{55}$$

Applying the Hamiltonian and evaluating it at $x_j = 0$ we have

$$(h - E)F(0) = -i(B - A)\delta(0) + (J\vec{\sigma}_j \cdot \vec{\sigma}_0 + J')(1/2)(A + B)\delta(0) = 0 \tag{56}$$

where we chose the convention (or rather the renormalization prescription) $\theta(x)\delta(x) = (1/2)\delta(x) = \theta(-x)\delta(x)$. Hence the S-matrix that relates the amplitude $B_{a_j a_0}$ to the amplitude $A_{a_j a_0}$ is,

$$S^{j0} = \frac{i + (1/2)J\sigma_j \cdot \sigma_0 + (1/2)J'}{i - (1/2)J\sigma_j \cdot \sigma_0 - (1/2)J'} = \frac{i + JP^{j0} + J''}{i - JP^{j0} - J''} \tag{57}$$

where we used $P^{j0} = (1/2)(I + \sigma_j \cdot \sigma_0)$, and set $J'' = (1/2)(J' - J)$. The S-matrix can be brought to the explicit form,

$$S^{j0} = e^{-i\phi}[a_{jo}I^{j0} + b_{j0}P^{j0}] = e^{-i\phi}\frac{I^{j0} - icP^{j0}}{1 - ic} \tag{58}$$

with

$$\begin{aligned} c &= \frac{2J}{1 + J''^2 - J^2} \\ e^{i\phi} &= \frac{1 + J''^2 - J^2 + 2iJ}{1 - J''^2 + J^2 - 2iJ''} \end{aligned}$$

When we wish to proceed and construct the $N$-electron wave function a problem arises: the hamiltonian does not contain any interaction terms among electrons. That might induce us to adopt $S^{ij} = I$ as the scattering matrix of electron $i$ and $j$, but this choice would not satisfy the YBE

$$S^{ij}S^{i0}S^{j0} = S^{j0}S^{i0}S^{ij}. \tag{59}$$

In fact the non commutativity of $S^{j0}$ and $S^{i0}$ captures some important aspects of the model: after electron $i$ crossed the impurity the latter is left ia a different state then before. Hence the state in which electron $j$ finds the impurity depends on whether it crosses the impurity before or after electron $i$. Herein lies the differnce between a system of electrons interacting with a fixed potential (a one-body problem since all electron see the same potential) and a Kondo system, where the impurity correlates the motion of all electrons.

This non-commutativity, however, does not ruin the integrability of the model. Considering more carefully the model for two electrons away from the impurity

$$h = -i(\partial_i + \partial_j) \tag{60}$$

we note that an arbitrary electron-electron S-matrix may be introduced, and one is allowed to consider a *basis* of free eigenstates of the form

$$F_{a_i a_j}(x_i x_j) = \mathcal{A}e^{i(k_i x_i + k_j x_j)}[A_{a_i a_j}\theta(x_i - x_j) + (SA)_{a_i a_j}\theta(x_j - x_i)]. \tag{61}$$

This function obviously satisfies the free hamiltonian with eigenvalue $E = k_i + k_j$ for any choice of $S$, but is not an eigenstate of the individual momentum operators unless $S = I$. Let me elaborate this simple point. One would tend to write a solution for $h$, in the form

$$F_{a_i a_j}(x_i x_j) = \mathcal{A}e^{i(k_i x_i + k_j x_j)}A_{a_i a_j} \tag{62}$$

which is indeed an eigenstate of individual momentum operators. As this state is degenerate with

$$F^p_{a_i a_j}(x_i x_j) = \mathcal{A} e^{i((k_i+p)x_i+(k_j-p)x_j)} A_{a_i a_j} \tag{63}$$

for any $p$, one may sum over $p$ with appropriate coefficients to form eq (61). What one is doing, in fact, is to find the zero order approximation in a degenerate perturbation theory. Physically, the eigenfunction can be constructed with arbitrary S-matrix because the particles move with the same velocity and never cross.

We conclude, then, that we are at liberty to choose as a basis of states one which is determined by an $S^{ij}$ satisfying the YBE. It is clear that the choice

$$S^{ij} = P^{ij} \tag{64}$$

does so, and thus leads to a consistent solution, with

$$F_{a_1...a_{N^e},a_0}(x_1...x_{N^e}) = \mathcal{A} e^{i \sum_{j=1}^{N^e} k_j x_j} \sum_Q A^Q_{a_1...a_{N^e},a_0} \theta(x_Q) \tag{65}$$

corresponding to the eigenvalue

$$E = \sum_{j=1}^{N^e} k_j. \tag{66}$$

In eq (65) $Q \in S_{N^e+1}$ describes the ordering of the $N^e$ electrons and of the impurity, localized at $x = 0$. The antisymmetrizer acts on the electron variables only.

We have thus constructed a consistent Bethe-Ansatz for the model, with the S-matrix given by

$$S^{j,\alpha} = \begin{cases} e^{-i\phi \frac{I^{j0} - icP^{j0}}{1-ic}} & \text{electron-impurity} \\ P^{jl} & \text{electron-electron} \end{cases} \tag{67}$$

indicating that it is integrable.

Again, imposing periodic boundary conditions one is led to to the problem of diagonalizing the operator $Z$ defined in eq(40) but now constructed with the (bare) S-matrices corresponding to the Kondo-model. It takes the form

$$(Z_j)^{b_1...b_N}_{a_1...a_N} = \left( P^{jj-1}...P^{j1}P^{jN}...e^{i\phi \frac{I^{j0} - icP^{j0}}{1-ic}}...P^{jj+1} \right)^{b_1...b_N}_{a_1...a_N} \tag{68}$$

We denoted $N = N^e + 1$, the number of spins in the problem.

Unlike the case of the Hubbard model, there is only one Z Hamiltonian to diagonalize, so that in a given state all electron momenta will be shifted by the same amount. In other words, the phase shift of the electrons due to their interactions is independent of their motion. This circumstance is due to the fact that the coupling constant $J$ is dimensionless, hence the S-matrix cannot depend on the momenta. In the Hubbard model, $U$ is dimensionful and the S-matrix is of the form $S = S(U/(k_j - k_l))$ leading to a coupling of all the modes. In the next lecture we turn to the problem of diagonalization of the Z-hamiltonian.

# References

[1] H. Bethe, Z. Physik **71**, 205 (1931).

[2] K. G. Wilson and J. Kogut, Phys. Rep. **C 12**, 75 (1974)

[3] see e.g. P. Ginsparg, in *Fields, Strings, and Critical Phenomena*, Eds. E Brezin and J. Zinn-Justin, North Holland, Amsterdam, 1990.

[4] E.L. Lieb and F.Y. Wu, Phys. Rev. Lett **20**, 1445 (1968).

[5] C.N. Yang, Phys. Rev. Lett. **19**, 1312 (1967).

[6] V. J. Emery, in *Highly Conducting One-Dimensional Solids*, J. T. Devreese et al. eds, Plenum, N. Y. 1979. J. Solyom, Adv. in Phys. **28**, 201 (1979).

[7] N. Andrei and J.H. Lowenstein, Phys. Rev. Lett. bf 43, 1693 (1979).

[8] A. A. Belavin, Phys. Lett. **B 87**, 117 (1979).

[9] F. D. M. Haldane, J. Phys. **C 14**, 2585 (1981).

[10] C. N. Yang, Phys. Rev. Lett.**63**, 2144 (1989). I. Affleck, talk given at Nato Advanced Study Institute on *Physics, Geometry and Topology*, Banff, (1989)

[11]  J. Zinn-Justin and E. Brezin, C.R. Acad. Sci. **263**, 670 (1966).

[12]  B. Sutherland, in *Exactly Solvable Problems in Condensed Matter and Relativistic Field Theory*, Eds. B. S. Shastry and V. Singh, Lecture Notes in Physics, vol. 42, Springer (1985).

[13]  D. Baeriswyl, in *Le Model de Hubbard*, Troisieme Cycle de la Physique en Suisse Romande (1991).

[14]  E. Lieb and F. Y. Wu, in *Phase Transitions and Critical Phenomena*, Vol. 1, eds. C. Domb and M. S. Green.

[15]  R. Baxter, Ann. Phys. **70**, 193 (1972); **70**, 323 (1972); **76**, 1 (1972).

[16]  L. Faddeev and L. Takhtajan, Usp. Mat. Nauk **34**, 15 (1979).

[17]  See e.g. G. Gruner and A. Zawadowski, Rep. Prog. Phys. **37**, 1497 (1974). A.C. Hewson, *The Kondo Problem to Heavy Fermions*, Cambridge University Press.

[18]  See any book on Quantum Field Theory.

[19]  N. Andrei, K. Furuya and J.H. Lowenstein, Rev. Mod. Phys. **55**, 331 (1983).

[20]  P. W. Anderson, J. Phys. C 3, 2346 (1970).

[21]  K. G. Wilson, Rev. Mod. Phy. **47**, 773 (1975).

[22]  P. Nozieres, in Proceedings of LT 14, edited by M. Krusius and M. Vuorio, p.339 (1975).

[23]  P. Nozieres and A. Blandin, J. Phys. (Paris) **41**, 193 (1980).

[24]  A. Zawadowski and N. Vladar, Sol. Stat. Com. **35**. D. L. Cox, Phys. Rev. Lett. **59**, 1240 (1987).

[25]  N. Andrei, Phys. Rev. Lett. **45**, 379 (1979).

[26]  P. B. Wiegmann, JETP Lett. **31**, 392 (1980).

## Lecture 2: The Quantum Inverse Method

We wish to find the eigenvalues (and eigenunctions) of the operator

$$(Z_j)_{a_1...a_N}^{b_1...b_N} = (S^{jj-1}...S^{j1}S^{jN}...S^{jj+1})_{a_1...a_N}^{b_1...b_N} \qquad (1)$$

where

$$S^{ij} = \frac{(\alpha_i - \alpha_j)I^{ij} + icP^{ij}}{\alpha_i - \alpha_j + ic} \qquad (2)$$

This operator occurs in the solution of the Hubbard model where $\alpha_j = \sin k_j$ and of the Kondo model where $\alpha_j = 1$ or $0$ depending on whether $j$ refers to an electron or to the impurity, respectively. The solution of other integrable models possessing SU(2) symmetry leads again to the same operator, differing only as to the values taken by the variables $\alpha_j$. For example in the Heisenberg model $\alpha_j = 0$ [1], and in the Backscattering model Model $\alpha_j = \pm 1$ depending whether $j$ refers to a left mover or a right mover [2].

The natural question to ask is whether the Hamiltonians $Z_j$ are integrable. That this is the case was shown by Yang [4] who solved it by means of another Bethe Ansatz. Let us follow a different path, and ask whether the hamiltonians $Z_j$ possess an infinite set of conserved charges, whose presence would guarantee integrability.

We have already seen that $[Z_i, Z_j]=0$. It will prove very useful to obtain a continuous version of this statement, since then we may expand in the continuous parameter (the one continuing the discrete index $j$) and obtain a set of charges commuting with $Z_i$. One is led thus, following Baxter [5], to introduce a continuous parameter, usually refered to as the spectral parameter, into the definition of the S-matrix

$$S(\alpha) = \frac{\alpha I + icP}{\alpha + ic} \equiv a(\alpha)I + b(\alpha)P, \qquad (3)$$

in a way that a continuous version of YBE is also satisfied

$$S^{kj}(\alpha - \beta)S^{ki}(\alpha)S^{ji}(\beta) = S^{ji}(\beta)S^{ki}(\alpha)S^{kj}(\alpha - \beta). \qquad (4)$$

We proceed now to introduce the *monodromy matrix* [6]. Its construction is natural when starting from the 6-vertex model, although its introduction here may seem ad-hoc. To begin with, introduce an auxiliary spin space $V_A$ which will help us keep track of the proliferating spin indices, and define an S-matrix acting in $V_j \times V_A$, where $V_j$ is the spin space of particle $j$.

$$(S^{jA}(\alpha))_{a,u}^{b,v} = \frac{\alpha (I^{jA})_{a,u}^{b,v} + ic(P^{jA})_{a,u}^{b,v}}{\alpha + ic} \tag{5}$$

the variables $u, v$ live in the auxiliary spin space, and $a, b$ live in the physical spin space. Now, the monodromy matrix $\Xi(\alpha)$ is defined by:

$$\Xi(\alpha) = S^{1A}(\alpha - \alpha_1) S^{2A}(\alpha - \alpha_2) \ldots\ldots S^{NA}(\alpha - \alpha_N) \tag{6}$$

Where the $\alpha_j$ are the physical values appropriate to the model, and the product is carried out only in the auxiliary space. Explicitly

$$(\Xi)_{a_1 \ldots a_N, u}^{b_1 \ldots b_N, v} = \sum_{s_1 \ldots s_{N-1}} (S^{1A})_{a_1, u}^{b_1, s_1} (S^{2A})_{a_2, s_1}^{b_2, s_2} \ldots\ldots (S^{NA})_{a_N, s_{N-1}}^{b_N, v} \tag{7}$$

where we suppressed the spectral parameters. It is convenient to represent the monodromy matrix graphically as:

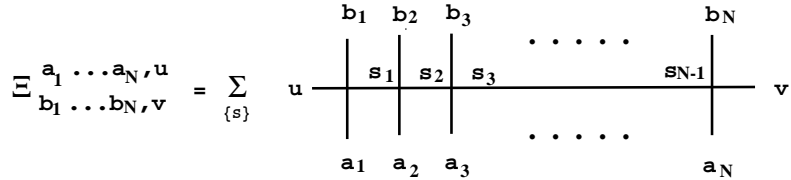

Figure 2: The Monodromy Matrix.

where the $j^{th}$ vertex represents the $j^{th}$ S-matrix with the horizontal lines carrying the auxiliary variables and the vertical lines at the $j$-site carring the variables of $V^j$.

Define also the *transfer matrix* $Z(\alpha) = tr_A \Xi(\alpha)$, by taking a trace over the auxiliary variables. Then using: $S^{jA}(o) = P^{jA}$, it follows [3],

$$Z(\alpha = \alpha_j) = Z_j. \tag{8}$$

We can now reformulate our question about the integrability of the spin problem. If we can show that for $\beta \neq \alpha$, $[Z(\alpha), Z(\beta)] = 0$, then expanding in powers of $\beta$ we produce an infinite set of charges commuting with the hamiltonian $Z(\alpha)$ guaranteeing integrability.

A sufficient condition to assure the commutativity of the operators $Z(\alpha)$ and $Z(\beta)$ is the existence of a matrix $R$ acting in $V_A \times V_A$ satisfying

$$R \ (\Xi(\alpha) \times \Xi(\beta)) = (\Xi(\beta) \times \Xi(\alpha)) \ R \tag{9}$$

namely,

$$R^{s,t}_{u,w} \ \Xi(\alpha)^v_s \ \Xi(\beta)^z_t = \Xi(\beta)^{s'}_u \ \Xi(\alpha)^{t'}_w \ R^{v,z}_{s',t'} \tag{10}$$

since taking the trace over the auxiliary variables leads to the desired expression. In eq (10) the physical indices are contracted in the usual quantum mechanical way. Thus rewriting it in its full index glory we have

$$R^{s,t}_{u,w} \ \Xi(\alpha)^{d_1...d_N,v}_{a_1...a_N,s} \ \Xi(\beta)^{b_1...b_N,z}_{d_1...d_N,t} = \Xi(\beta)^{c_1...c_N,s'}_{a_1...a_N,u} \ \Xi(\alpha)^{c_1...c_N,t'}_{c_1...c_N,w} \ R^{v,z}_{s',t'} \tag{11}$$

To show that eq(11) holds it is sufficient to examine it locally since the monodromy-matrix is given as an ordered product of the S-matrices. In other words, consider the problem where $\Xi$ is built of one vertex only,

$$R \ S \times S = S \times S \ R \tag{12}$$

namely

$$R^{s,t}_{u,w} \ S(\alpha)^{d,v}_{a,s} \ S(\beta)^{b,z}_{d,t} = S(\beta)^{c,s'}_{a,u} \ S(\alpha)^{b,t'}_{c,w} \ R^{v,z}_{s',t'} \tag{13}$$

from which eq(9) can be deduced by applying eq(12) repeatedly. Comparing eq(13) with eq(4) we see that it is again the ubiquitous YBE if we identify

$$R^{s,t}_{u,w} = S(\alpha - \beta)^{t,s}_{u,w}. \tag{14}$$

In other words, we have shown that a martix $R$, implementing eq(9), does exist and is given by (all action is in the auxiliary spaces $V_A \times V_A$)

$$R = S(\alpha - \beta) \ P = \frac{(\alpha - \beta)P + icI}{(\alpha - \beta) + ic} \tag{15}$$

As a consequence, we have shown the commutativity of the Z-matrices and the integrability of the spin problem.

Let us consider the construction more explicitly in the auxiliary space. The 4x4 matrix $R$ takes the form

$$R(\alpha) = \begin{pmatrix} a+b & 0 & 0 & 0 \\ 0 & b & a & 0 \\ 0 & a & b & 0 \\ 0 & 0 & 0 & a+b \end{pmatrix} = \begin{pmatrix} 1 & 0 & 0 & 0 \\ 0 & \frac{ic}{\alpha+ic} & \frac{\alpha}{\alpha+ic} & 0 \\ 0 & \frac{\alpha}{\alpha+ic} & \frac{ic}{\alpha+ic} & 0 \\ 0 & 0 & 0 & 1 \end{pmatrix}, \quad (16)$$

and the monodromy matrix,

$$\Xi(\alpha)_{a_1...a_N}^{b_1...b_N} = \begin{pmatrix} A_{a_1...a_N}^{b_1...b_N}(\alpha) & B_{a_1...a_N}^{b_1...b_N}(\alpha) \\ C_{a_1...a_N}^{b_1...b_N}(\alpha) & D_{a_1...a_N}^{b_1...b_N}(\alpha) \end{pmatrix} \quad (17)$$

with the operators $A, B, C$ and $D$ acting in the physical space $V^N$. Carrying out the products as indicated the fundamental relation eq(10), and equating term by term left and right hand sides we obtain the following algebraic relations (we display only those necessary to our purposes):

$$A(\alpha)B(\beta) = u(\beta-\alpha)B(\beta)A(\alpha) + v(\beta-\alpha)B(\alpha)A(\beta) \quad (18)$$
$$D(\alpha)B(\beta) = u(\alpha-\beta)B(\beta)D(\alpha) + v(\alpha-\beta)B(\alpha)D(\beta) \quad (19)$$
$$\quad (20)$$

where

$$u(\alpha) = \frac{1}{a(\alpha)} = \frac{\alpha+ic}{\alpha}$$
$$v(\alpha) = -\frac{b(\alpha)}{a(\alpha)} = -\frac{ic}{\alpha}$$

Furthermore,

$$B(\alpha)B(\beta) = B(\beta)B(\alpha) \quad (21)$$
$$A(\alpha)A(\beta) = A(\beta)A(\alpha) \quad (22)$$
$$D(\alpha)D(\beta) = D(\beta)D(\alpha) \quad (23)$$

We shall use these relations to diagonalize the transfer matrix $Z(\alpha)$, explicitly given by

$$Z_{a_1...a_N}^{b_1...b_N}(\alpha) = A_{a_1...a_N}^{b_1...b_N}(\alpha) + D_{a_1...a_N}^{b_1...b_N}(\alpha). \tag{24}$$

The key point in the diagonalization procedure is the observation that the operator $B(\beta)$ plays a role of a creation operator with respect to the Hamiltonian $A(\alpha) + D(\beta)$, up to "unwanted terms" (generated by the second term on the right hand side of eqs(18,19), when acting on a ferromagnetic "up" vacuum $|\omega>$,

$$|\omega> = \prod_{j=1}^{N} \begin{pmatrix} 1 \\ 0 \end{pmatrix}_j. \tag{25}$$

We may show that this ferromagnetic vacuum is an eigenstate of the $Z$ hamiltonian using the fact that the latter is given as a product of local vertices $S^{jA}$ which, when written out in auxiliary space, take the form

$$S^{jA}(\alpha) = (a+b/2)(\alpha)1_j 1_A + b/2(\alpha)\sigma_j \cdot \sigma_A$$
$$= \begin{pmatrix} (a+b/2)(\alpha)1_j + (b/2)(\alpha)\sigma_j^z & b(\alpha)\sigma_j^- \\ b(\alpha)\sigma_j^+ & (a+b/2)(\alpha)1_j - (b/2)(\alpha)\sigma_j^z \end{pmatrix} \tag{26}$$

We used the identity $(P^{jA})_{a,u}^{b,v} = (1/2)(\delta_a^b \delta_u^v + (\sigma_j)_a^b \cdot (\sigma_A)_u^v)$ to separate the physical and auxiliary spaces. $1_j$ denotes the two dimensional unit matrix acting in $V_j$. Acting on $\begin{pmatrix} 1 \\ 0 \end{pmatrix}_j$ the vertex $S^{jA}$ becomes triangular

$$S_{jA}(\alpha - \alpha_j)\begin{pmatrix} 1 \\ 0 \end{pmatrix}_j = \begin{pmatrix} 1\begin{pmatrix} 1 \\ 0 \end{pmatrix}_j & b(\alpha - \alpha_j)\begin{pmatrix} 0 \\ 1 \end{pmatrix}_j \\ 0 & \frac{\alpha - \alpha_j}{\alpha - \alpha_j + ic}\begin{pmatrix} 1 \\ 0 \end{pmatrix}_j \end{pmatrix}, \tag{27}$$

so that

$$Z(\alpha)|\omega> = tr_A(S^{1A}.....S^{NA})|\omega> = tr_A\left(S^{1A}\begin{pmatrix} 1 \\ 0 \end{pmatrix}_1 .....S^{NA}\begin{pmatrix} 1 \\ 0 \end{pmatrix}_N\right) \tag{28}$$

and we see that $|\omega>$ is an eigenstate of of $A(\alpha)$ and of $D(\alpha)$

$$A(\alpha)|\omega> = |\omega> \tag{29}$$
$$D(\alpha)|\omega> = \Delta(\alpha)|\omega> \tag{30}$$

where
$$\Delta(\alpha) = \prod_{j=1}^{N} \frac{\alpha - \alpha_j}{\alpha - \alpha_j + ic}. \tag{31}$$

To proceed and find all eigenstates of $Z(\alpha)$ we consider states formed by repeated application of the flipping operator $B$ on the ferromagnetic eigensate

$$|A(\beta_1...\beta_M)> = B(\beta_1)...B(\beta_M)|\omega> = \sum_{j_1...j_M} A_{j_1...j_M} \sigma_{j_1}^-...\sigma_{j_M}^- |\omega> \tag{32}$$

where the usual spin amplitude notation $A_{a_1...a_N}$ is written as $A_{j_1...j_M}$ by specifying the position of the $M$ down spins. Acting on the state with the Hamiltonian, and applying the algebraic relations eqs(18-23) we find that in addition to terms of the form of the original state, we also obtain unwanted terms preventing the state from being an eigenstate. However these unwanted terms can be removed by a proper choice ot the parameters $\beta_1....\beta_M$. To illustrate the procedure consider $M=2$. Then, moving $A(\alpha)$ past $B(\beta_1)$ and $B(\beta_2)$ we have

$$\begin{aligned}
(A(\alpha) + D(\alpha))&B(\beta_1)B(\beta_2)|\omega> = \\
&u(\beta_1 - \alpha)u(\beta_2 - \alpha)B(\beta_1)B(\beta_2)A(\alpha)|\omega> \\
&+ u(\alpha - \beta_1)u(\alpha - \beta_2)B(\beta_1)B(\beta_2)D(\alpha)|\omega> \\
&+ [u(\beta_1 - \alpha)v(\beta_2 - \alpha) + v(\beta_1 - \alpha)v(\beta_2 - \beta_1)]B(\alpha)B(\beta_1)A(\beta_2)|\omega> \\
&+ [u(\alpha - \beta_1)v(\alpha - \beta_2) + v(\alpha - \beta_1)v(\beta_1 - \beta_2)]B(\alpha)B(\beta_1)D(\beta_2)|\omega> \\
&+ v(\beta_1 - \alpha)u(\beta_2 - \beta_1)B(\alpha)B(\beta_2)A(\beta_1)|\omega> \\
&+ v(\alpha - \beta_1)u(\beta_1 - \beta_2)B(\alpha)B(\beta_2)D(\beta_1)|\omega> \\
= \lambda(\alpha, \beta_1\beta_2)&B(\beta_1)B(\beta_2)|\omega> + \lambda_1(\alpha, \beta_1\beta_2)B(\alpha)B(\beta_2)|\omega> + \lambda_2(\alpha, \beta_1\beta_2)B(\alpha)B(\beta_1)|\omega>
\end{aligned}$$

where

$$\begin{aligned}
\lambda(\alpha, \beta_1\beta_2) &= u(\beta_1 - \alpha)u(\beta_2 - \alpha) + \Delta(\alpha)u(\alpha - \beta_1)u(\alpha - \beta_2) & (33) \\
\lambda_1(\alpha, \beta_1\beta_2) &= v(\beta_1 - \alpha)[u(\beta_2 - \beta_1) - u(\beta_1 - \beta_2)\Delta(\beta_1)] & (34) \\
\lambda_2(\alpha, \beta_1\beta_2) &= v(\beta_2 - \alpha)[u(\beta_1 - \beta_2) - u(\beta_2 - \beta_1)\Delta(\beta_2)] & (35)
\end{aligned}$$

The condition for $B(\beta_1)B(\beta_2)|\omega>$ to be an eigenstate is that $\beta_1$ and $\beta_2$ be chosen so that
$$\lambda_\gamma(\alpha, \beta_1\beta_2) = 0 \qquad \gamma = 1, 2 \, . \tag{36}$$

The generalization to arbitrary $M$ is straightforward [6]: $B(\beta_1)....B(\beta_M)|\omega>$ is an eigenstate of $Z(\alpha) = A(\alpha) + D(\alpha)$, with eigenvalue

$$\lambda(\alpha, \beta_1...\beta_M) = \prod_{\gamma=1}^{M} u(\beta_\gamma - \alpha) + \Delta(\alpha) \prod_{\gamma=1}^{M} u(\alpha - \beta_\gamma) \qquad (37)$$

if the paramaters $\beta_1....\beta_M$ are chosen so as to eliminate the "unwanted terms", namely:

$$\lambda_\gamma(\alpha, \beta_1....\beta_M) = 0 \qquad \gamma = 1....M, \qquad (38)$$

where,

$$\lambda_\gamma(\alpha, \beta_1...\beta_M) = v(\beta_\gamma - \alpha)\left[\prod_{\delta=1,\delta\neq\gamma}^{M} u(\beta_\delta - \beta_\gamma) - \Delta(\beta_\gamma) \prod_{\delta=1,\delta\neq\gamma}^{M} u(\beta_\gamma - \beta_\delta)\right] (39)$$

Recall now that our original goal was to find the eigenalues of the oerator $Z_j = Z(\alpha = \alpha_j)$. Denoting these by $z_j$, we have (see eq(37))

$$z_j = \lambda(\alpha_j, \beta_1....\beta_M) = \prod_{\gamma=1}^{M} \frac{\beta_\gamma - \alpha_j + ic}{\beta_\gamma - \alpha_j} \qquad (40)$$

with the parameters $\beta_1...\beta_M$ satisfying

$$\prod_{\delta=1,\delta\neq\gamma}^{M} \frac{\beta_\delta - \beta_\gamma + ic}{\beta_\delta - \beta_\gamma - ic} = \prod_{i=1}^{N} \frac{\beta_\gamma - \alpha_i}{\beta_\gamma - \alpha_i + ic}. \qquad (41)$$

We may cast these equations in a more appealing form by changing variables: $\beta_\gamma = \Lambda_\gamma - ic/2$, and recalling that periodic boundary conditiond impose $z_j = e^{-ik_j L}$, we finally obtain:

$$e^{ik_j L} = \prod_{\gamma=1}^{M} \frac{\Lambda_\gamma - \alpha_j - ic/2}{\Lambda_\gamma - \alpha_j + ic/2} \qquad (42)$$

and

$$\prod_{\delta=1,\delta\neq\gamma}^{M} \frac{\Lambda_\delta - \Lambda_\gamma + ic}{\Lambda_\delta - \Lambda_\gamma - ic} = \prod_{i=1}^{N} \frac{\Lambda_\gamma - \alpha_i - ic/2}{\Lambda_\gamma - \alpha_i + ic/2}. \qquad (43)$$

We have now completed the diagonalization of the $Z$-operator, and have solved at the same time the underlying spin problem for the whole class of integrable SU(2)-invariant models built with the $R$-matrix in eq(16).

For the Hubbard model $\alpha_j = \sin k_j$ and the Bethe-Ansatz equations take the form [7]:

$$e^{ik_j L} = \prod_{\gamma=1}^{M} \frac{\Lambda_\gamma - \sin k_j - ic/2}{\Lambda_\gamma - \sin k_j + ic/2} \tag{44}$$

and

$$\prod_{\delta=1,\delta\neq\gamma}^{M} \frac{\Lambda_\delta - \Lambda_\gamma + ic}{\Lambda_\delta - \Lambda_\gamma - ic} = \prod_{i=1}^{N} \frac{\Lambda_\gamma - \sin k_i - ic/2}{\Lambda_\gamma - \sin k_i + ic/2} \tag{45}$$

with $c = \frac{u}{2} = \frac{U}{2t}$.

For the Kondo model $\alpha_j = 1, 0$ and the equations become [8] [9]:

$$e^{ik_j L} = \prod_{\gamma=1}^{M} \frac{\Lambda_\gamma - 1 + ic/2}{\Lambda_\gamma - 1 - ic/2} \tag{46}$$

and

$$\prod_{\delta=1,\delta\neq\gamma}^{M} \frac{\Lambda_\delta - \Lambda_\gamma + ic}{\Lambda_\delta - \Lambda_\gamma - ic} = \left(\frac{\Lambda_\gamma - 1 - ic/2}{\Lambda_\gamma - 1 + ic/2}\right)^{N^e} \left(\frac{\Lambda_\gamma - ic/2}{\Lambda_\gamma + ic/2}\right) \tag{47}$$

with the coupling $c$ given by

$$c = \frac{2J}{1 + J''^2 - J^2}. \tag{48}$$

We note that the equations have decoupled; the spin variables $\Lambda_\gamma$ are determined independently of the momenta $k_j$, reflecting the charge-spin decoupling discussed earlier.

We proceed now to solve the equations, and discuss the physics of each model.

# References

[1]     H. Bethe, Z. Physik **71**, 205 (1931).

[2]     N. Andrei and J.H. Lowenstein, Phys. Rev. Lett. **43**, 1693 (1979).

[3]     A. A. Belavin, Phys. Lett. **B 87**, 117 (1979).

[4]     C.N. Yang, Phys. Rev. Lett. **19**, 1312 (1967).

[5]     R. Baxter, Ann. Phys. **70**, 193 (1972); **70**, 323 (1972); **76**, 1 (1972).

[6]     L. Faddeev and L. Takhtajan, Usp. Mat. Nauk **34**, 15 (1979.

[7]     E.L. Lieb and F.Y. Wu, Phys. Rev. Lett **20**, 1445 (1968).

[8]     N. Andrei, Phys. Rev. Lett. **45**, 379 (1979).

[9]     P. B. Wiegmann, JETP Lett. **31**, 392 (1980).

## Lecture 3: The Kondo Model

We now turn to derive the physics of the Kondo model from the Bethe Ansatz equations. We shall mainly discuss the excitation spectrum and the thermodynamics, with only a casual discussion of transport properties.

We show here that the fundamental spin excitations are interacting spin-1/2 particles, *spinons*, carrying no charge; while the charge excitations, the *holons*, carry no spin, and are non-interacting [1]. More complex excitations are superpositions of the fundamental ones. The spin and charge sectors in the model decouple completely. A physical electron, however, is a coherent superposition of states in both sectors. Hence the electron-electron correlation function has no single particle poles.

We then proceed to obtain the free energy of the model in terms of coupled integral equations [2],[3]. The equations are studied in the utraviolet (high temperature, or large magnetic fields) as well as in the infra-red regime, and a remarkable change in the impurity properties is observed. In the ultraviolet limit the impurity behaves as an almost free-spin, given by an effective weakly coupled theory, while in the infra-red it is completely screened, becoming a non-magnetic scattering center. The properties of the impurity in this regime are given by another effective theory, strongly coupled, describing a local Fermi liquid. This *crossover* from one regime to another, the Kondo Effect, can be driven by any energy parameter - be it the temperature, the magnetic field or an excitation energy. It represents, in the language of Renormalization Group, a flow from from one fixed fixed point to another. Understanding the crossover requires the construction of the model over all energy scales, which is not feasible by methods valid only in the vicinity of one fixed point, such as perturbation theory, strong coupling expansion or conformal field theory. The crossover was first carried out numerically by explicitly constructing the flow in the space of effective hamiltonians [5]. Here we study it in the framework of the Bethe ansatz [6],[7].

We shall end this lecture with a brief discussion of the Multichannel Kondo model. Again a crossover takes place from a weak coupling regime in the ulraviolet to a new regime in the infrared. The latter is typically non Fermi liquid.

Let us now determine the energy eigenvalues. Consider a system of $N^e$ electrons on a ring of length $L$ interacting with an impurity localized at $x = 0$. There are therefore $N = N^e + 1$ spins in the problem, with $N$ to be taken even. An eigenstate will be also labeled by the quantum number $M$, the conserved number of down spins, as well as by an infinite set of local quantum numbers $\{n_j, I_\gamma\}$ defined below. The $z$ component of the spin of the state is given by $S_z = \frac{1}{2}(N - 2M)$. This is also the total spin of the state since by construction the $B(\Lambda_\gamma)$ operators (from Lecture 2) commute among themselves and thus specify a unique Young Tableau. The states we construct are therefore highest weight states, $|S, S_z = S>$, and the rest of the the states in the multiplet are obtained by the action of lowering operators $S^-$.

In previous lectures we found that the energy eigenvalues are given by

$$E = \sum_{j=1}^{N^e} k_j \tag{1}$$

with the momenta $k_j$ obtained from the eigenvalue $z$

$$z = e^{ik_j L} = \prod_{\gamma=1}^{M} \frac{\Lambda_\gamma - 1 + ic/2}{\Lambda_\gamma - 1 - ic/2}. \tag{2}$$

Hence,

$$k_j = \frac{2\pi}{L} n_j + \frac{1}{L} \sum_{\gamma=1}^{M} [\Theta(2\Lambda_\gamma - 2) - \pi], \tag{3}$$

with $n_j$ an integer arising from taking the logarithm, and

$$\Theta(x) = -2 \tan^{-1}(x/c). \tag{4}$$

The expression for the energy becomes (dropping inessential terms)

$$E = \sum_{j=1}^{N^e} \frac{2\pi}{L} n_j + D \sum_{\gamma=1}^{M} [\Theta(2\Lambda_\gamma - 2) - \pi], \tag{5}$$

where $D = N^e/L$ is the electron density.

The spin momenta $\Lambda_1....\Lambda_M$ are found from the condition guaranteeing the cancellation of the "unwanted terms" (see Lecture 2),

$$-\prod_{\delta=1}^{M} \frac{\Lambda_\delta - \Lambda_\gamma + ic}{\Lambda_\delta - \Lambda_\gamma - ic} = \left(\frac{\Lambda_\gamma - 1 - ic/2}{\Lambda_\gamma - 1 + ic/2}\right)^{N^e} \left(\frac{\Lambda_\gamma - ic/2}{\Lambda_\gamma + ic/2}\right). \tag{6}$$

Note that the equations determining the $\{\Lambda_\gamma\}$ have decoupled from those determining the momenta $\{k_j\}$, reflecting the complete charge-spin decoupling discussed earlier.

Upon taking the logarithm of eqs(6) we find that the variables $\{\Lambda_\gamma\}$ satisfy the following set of coupled equations.

$$N^e \Theta(2\Lambda_\gamma - 2) + \Theta(2\Lambda_\gamma) = -2\pi I_\gamma + \sum_{\delta=1}^{M} \Theta(\Lambda_\gamma - \Lambda_\delta), \qquad \gamma = 1....M \quad (7)$$

The numbers $I_\gamma$ are even or odd half integers depending on $N - M - 1$ being even or odd, and together with the integers $n_j$ they specify a solution of eqs(3,7).

Each *allowed* choice (see below) of $\{n_j, I_\gamma\}$ uniquely determines an eigenstate of the Hamiltonian. We shall refer to the $\{n_j, I_\gamma\}$ configuration as the quantum numbers of the state they determine.

These quantum numbers replace, for example, the $\{n_j^+, n_j^-\}$ quantum numbers of the free electron gas, where in the conventional Fock-basis each level $n_j$ could be populated by a spin up and a spin down particle. When the impurity is removed the Bethe ansatz equations describe a free electron gas. Indeed, in the absence of the term $\Theta(2\Lambda)$ in eq(7) we have,

$$\begin{aligned}
E &= \sum_{j=1}^{N^e} \frac{2\pi}{L} n_j + D \sum_{\gamma=1}^{M} [\Theta(2\Lambda_\gamma - 2) - \pi] \\
&= \sum_{j=1}^{N^e} \frac{2\pi}{L} n_j + \frac{1}{L} \sum_{\gamma=1}^{M} [-2\pi I_\gamma + \sum_{\delta=1}^{M} \Theta(\Lambda_\gamma - \Lambda_\delta)] \\
&= \sum_{j=1}^{N^e} \frac{2\pi}{L} n_j + \sum_{\gamma=1}^{M} -\frac{2\pi}{L} I_\gamma,
\end{aligned}$$

namely, a noninteracting gas given in a basis which is already charge-spin decoupled, and therefore adapted to turning on the spin exchange interaction which couples to the spin sector only.

What restrictions are there on the choice of configurations $\{n_j, I_\gamma\}$ ? Obviously, the spectrum is unbound from below, as the integers $n_j$ can take arbitrarily large and negative values. To define the model we introduce a "bottom to the sea", a cutoff $K$, taken to be very large compared with any physical parameter in the theory. Then, since we are interested in the low energy properties of the model, we may study it in the limit where the cutoff is taken to infinity as long as the physical quantities of the model have a well defined limit. We impose the cutoff as follows:

$$|\frac{2\pi}{L} n_j| < K. \tag{8}$$

The cutoff $K$ is imposed on the eigenstates of the fully interacting Hamiltonian and thus differs from the conventional cutoffs that are imposed on the eigenfunctions of the free Hamiltonian. The choice of cutoff is irrelevant in models, such as the Kondo model, that are renormalizable. It affects the way physical scales (such as the Kondo temperature) depend on *bare* parameters, but not the way physical functions depend on physical scales [1].

From the action of the spin flip operators $B(\Lambda_\gamma)$ on $|\omega>$ it is obvious that the state $B(\Lambda_1)...B(\Lambda_M)|\omega>$ vanishes if two of the $\Lambda$'s coincide. Further, as $|\Theta(x)| \leq \pi$, it is clear that the $I_\gamma$ must satisfy the restriction

$$I^-(N,M) = -(N-M-1)/2 \leq I_\gamma \leq (N-M-1)/2 = I^+(N,M). \tag{9}$$

We shall call a configuration $\{I_\gamma\}$, for which a solution exists with all $\Lambda_\gamma$ distinct, *allowed*. Counting all allowed configurations (using the string hypothesis, see below) one finds that there are, indeed, $2^N$ configurations, as required by the dimensionality of the spin space.

We turn now to the determination of the eigenstates begininnig with the ground state configuration $\{n_j^o, I_\gamma^0\}$.

**The ground state.**

The state with the lowest energy is a spin singlet, $M_0 = N/2$, induced by a configuration of consecutive $\{I_\gamma^0\}$,

$$I_{\gamma+1}^0 = I_\gamma^0 + 1 \tag{10}$$

with the $I_\gamma$ filling all the slots from $I^+$ to $I^-$

$$I^- \leq I_\gamma \leq I^+, \qquad I^\pm = \pm(N/2 - 1)/2. \tag{11}$$

There are $N/2$ slots from $I^+$ to $I^-$ and all are occupied by the $M_0 = N/2$ spin quantum numbers $I_\gamma$.

The charge quantum numbers $\{n_j^0\}$ are taken the minimum allowed by the cutoff. They are all distinct and run from $-KL/2\pi$ upwards. Setting $E_F = 0$ we have $K = 2\pi D$, $D = N^e/L$.

We shall be interested in solving the equations in the thermodynamic limit: $N, L \to \infty$, with $D = N^e/L$ held fixed (later we shall also take the scaling limit: $K = \pi D \to \infty$ to achieve universality). Therefore, rather than finding the actual solutions $\{\Lambda_\gamma\}$ we consider their density $\sigma(\Lambda)$, describing the number of solutions in the interval $(\Lambda, \Lambda + d\Lambda)$. In other words,

$$\sigma(\Lambda_\gamma) = 1/(\Lambda_{\gamma+1} - \Lambda_\gamma). \tag{12}$$

When all $\Lambda$-solutions are real (which is the case of the ground state) eqs. (5) and (7) can be rewritten in terms of the $\Lambda$-density

$$E = \sum_{J=1}^{N^e} \frac{2\pi}{L} n_j + D \int d\Lambda \sigma(\Lambda)[\Theta(2\Lambda - 2) - \pi] \tag{13}$$

and

$$N^e \Theta(2\Lambda_\gamma - 2) + \Theta(2\Lambda_\gamma) = \int d\Lambda' \sigma(\Lambda') \Theta(\Lambda_\gamma - \Lambda') - 2\pi I_\gamma. \tag{14}$$

An equation for the $\Lambda$-density in the ground state, $\sigma_o(\Lambda)$, is obtained by subtracting eq(14) written for $\Lambda_\gamma$ from that written for $\Lambda_{\gamma+1}$ and expanding in the difference $\Delta\Lambda = \Lambda_{\gamma+1} - \Lambda_\gamma$, assumed to be of order $1/N$. One then finds

$$\sigma_o(\Lambda) = f(\Lambda) - \int K(\Lambda - \Lambda')\sigma_o(\Lambda')d\Lambda', \tag{15}$$

where

$$\begin{aligned} f(\Lambda) &= \frac{2c}{\pi}\left[\frac{N^e}{c^2 + 4(\Lambda-1)^2} + \frac{1}{c^2 + 4\Lambda^2}\right], \\ K(\Lambda) &= \frac{1}{\pi}\frac{c}{c^2 + \Lambda^2} \equiv K_2(\Lambda). \end{aligned} \tag{16}$$

Here we used $\Theta'(\Lambda) = -2c/(c^2 + \Lambda^2)$, and the fact that, for the ground state, $I_{\gamma+1} - I_\gamma = 1$ for all $\gamma$. We also define, for future use,

$$\begin{aligned} K_n(x) &= \frac{1}{\pi}\frac{n\frac{c}{2}}{(n\frac{c}{2})^2 + x^2} = -\frac{1}{2\pi}\Theta'_n(x) \\ \Theta_n(x) &= \Theta\left(\frac{2x}{n}\right) \end{aligned} \tag{17}$$

The solution of eq(15) by means of Fourier transform is immediate,

$$\sigma_o(\Lambda) = \frac{1}{2c}\left[\frac{N^e}{\cosh\frac{\pi}{c}(\Lambda-1)} + \frac{1}{\cosh\frac{\pi}{c}\Lambda}\right]. \qquad (18)$$

The transformation properties of the ground state are found by calculating $M_o = \int \sigma_o(\Lambda)d\Lambda = \frac{1}{2}N$ in accordance with the consideration in eqs(9,11) for finite $N$ and $M$. The state has a Young tableau of two equal-length rows and is a SU(2) singlet.

The ground-state energy is given by

$$\begin{aligned} E_o &= \sum_j \frac{2\pi}{L} n_j + D\int d\Lambda \sigma_o(\Lambda)\left[\Theta(2\Lambda-2) - \pi\right] \\ &= -\frac{\pi}{2L}(N^e)^2 - iD\ln\frac{\Gamma(1+ic)\Gamma(\frac{1}{2}-ic)}{\Gamma(1-ic)\Gamma(\frac{1}{2}+ic)}, \end{aligned} \qquad (19)$$

To show that, indeed, this is the lowest energy state we shall study variations from the ground-state configuration $\{n_j^0, I_\gamma^0\}$. These correspond to excited states.

### Elementary Excitations

*Charge excitations (particle-hole)* — obtained by exciting the charge degrees of freedom. Thus excite a given $n_j^0$, where $-K \leq (2\pi/L)n_j^0 < 0$, to $n_j' = n_j^0 + \Delta n \geq 0$. The change in energy involved is

$$\Delta E = \frac{2\pi}{L}\Delta n > 0. \qquad (20)$$

Obviously $M$, which depends only on the $\{I_\gamma\}$ quantum numbers, does not change and neither does the spin.

We see that the charge spectrum of the theory is that of a decoupled free gas, a result of the interaction acting only on the spin degrees of freedom.

*Spin excitations* — obtained by varying the $\{I_\gamma^0\}$ sequence from its ground-state configuration, leaving the charge quantum numbers, $n_j$, unchanged. One way to modify the sequence is to put "holes" into it, namely to have unfilled slots, and correpondingly omit $\Lambda$'s. Another way is to add complex conjugate pairs of $\Lambda$'s.

*The triplet*: The simplest excitation (keeping the number of electrons fixed) is obtained by considering the state with $M = M_o - 1$. This is a spin triplet since $S = N/2 - M = 1$. Equivalently, $\Delta M = -1$ means that one box is moved from the lower to the upper row in the Young tableau.

This choice induces two holes since now $I^\pm = \pm N/4$ yielding $N/2 + 1$ slots for the $M = N/2 - 1$ $I_\gamma$'s leaving two slots unfilled. To find the effect of a hole suppose we choose a sequence $\{I_\gamma\}$ such that $I_{\gamma_0+1} = I_{\gamma_0} + 2$, and $I_{\gamma+1} = I_\gamma + 1, \gamma \neq \gamma_0$, omitting the integer $I^h = I_{\gamma_0+1}$. The spin momentum corresponding to it, $\Lambda^h$, constitutes a "hole". This means that we have to solve eq(14) in the presence of a (bare) hole density $\sigma^h(\Lambda) = \delta(\Lambda - \Lambda^h)$. To be more precise consider an $\{I_\gamma\}$ sequence with holes in it, denoting the omitted integers by $\{I_j^h, j = 1, ..., m\}$. The BAE with the prescribed quantum numbers $\{I_\gamma\}$ determine the corresponding set $\{\Lambda_\gamma\}$. Now define the function

$$\nu(\Lambda) = -\frac{1}{2\pi}[N^e \Theta(2\Lambda - 2) + \Theta(2\Lambda) - \sum_{\gamma=1}^{M} \Theta(\Lambda - \Lambda_\gamma)], \qquad (21)$$

constructed with the determined values $\Lambda_\gamma$. Those values of the variable $\Lambda$ that satisfy

$$\nu(\Lambda_\gamma) = I_\gamma, \qquad (22)$$

where $I_\gamma$ is an integer belonging to the sequence, are the solutions we began with, while those values of $\Lambda$ satisfying

$$\nu(\Lambda_j^h) = I_j^h, \qquad (23)$$

where $I_j^h$, the integers omitted from the $I_\gamma$ sequence, are the holes. Introducing the distribution functions $\sigma(\Lambda)$ and $\sigma^h(\Lambda)$ of the $\Lambda$-solutions and $\Lambda$-holes, respectively, we have

$$\frac{d\nu}{d\Lambda} = \sigma(\Lambda) + \sigma^h(\Lambda), \qquad (24)$$

since the number of holes and $\Lambda$'s in the interval $d\Lambda$ is given, on the one hand, by $[\sigma(\Lambda) + \sigma^h(\Lambda)]d\Lambda$ and, on the other hand, by the number of values of $I_j^h$ and $I_\gamma$ which $\nu(\Lambda)$ takes as it ranges over the interval $d\Lambda$.

The equation for the density $\sigma(\Lambda)$ in the presence of holes is obtained from eq(21) by taking the derivative with respect to $\Lambda$,

$$\sigma(\Lambda) + \sigma^h(\Lambda) = f(\Lambda) - \int K(\Lambda - \Lambda')\sigma(\Lambda')d\Lambda', \qquad (25)$$

where
$$\sigma^h(\Lambda) = \sum_{j=1}^{m} \delta(\Lambda - \Lambda_j^h). \tag{26}$$

The solution (in Fourier space) to the equation is given by
$$\tilde{\sigma}(p) = \tilde{\sigma}_o(p) + \Delta\tilde{\sigma}(p) \tag{27}$$
with
$$\Delta\tilde{\sigma}(p) = -\sum_{j=1}^{m} e^{-i\Lambda^h p} \frac{\exp\frac{c}{2}|p|}{2\cosh\frac{c}{2}p} \tag{28}$$
being the change induced in the density.

Thus the "bare" hole density $e^{-i\Lambda^h p}$ (in Fourier space) is "dressed" to $e^{-i\Lambda^h p}\frac{\exp\frac{c}{2}|p|}{2\cosh\frac{c}{2}p}$ by the back flow of the sea of spin momenta $\Lambda$: since all $\Lambda$ momenta are coupled through eq(14), removing one of them affects all and leads to the redistribution given by $\Delta\sigma(\Lambda)$.

Given the density we can calculate the properties of the state; $M$, the number of "down spins", or the length of the lower row in the Young tableau, is
$$M = \int \sigma(\Lambda)d\Lambda = \tilde{\sigma}(p=0) = \frac{1}{2}N - \frac{1}{2}m \tag{29}$$
so that each hole contributes $(\Delta M)_h = -\frac{1}{2}$ and corresponds to a spin-1/2 object, a *spinon* in modern parlance, since it obviously carries no charge.

The triplet excitation, being characterized by $\Delta M = -1$, is made of two holes, again in accord with the finite $N$ considerations. In the language of spin representations the state consists of a symmetrized product of two spin-1/2 objects yielding spin-one.

The triplet excitation energy $\Delta E^t$ for holes at $\Lambda_1^h$ and $\Lambda_2^h$ is given by
$$\begin{aligned}\Delta E^t &= D\int \Delta\sigma(\Lambda)[\Theta(2\Lambda - 2) - \pi]d\Lambda \\ &= 2D\tan^{-1} e^{(\pi/c)(\Lambda_1^h - 1)} + 2D\tan^{-1} e^{(\pi/c)(\Lambda_2^h - 1)}.\end{aligned} \tag{30}$$

It is a sum of two terms, each term being the energy carried by the spin-half spinon.

We claim, then, that the triplet is composed of two spin-1/2 uncharged objects whose spins are coupled symmetrically to form a spin-1 state. To confirm this interpretation we need to show that another state exists where the spins are coupled antisymmetrically to form a spin singlet. We are lead to consider configurations of $I_\gamma$ inducing complex $\Lambda_\gamma$ solutions. Since the energy is real these complex solutions occur in conjugate pairs. We shall show that a singlet excited state is composed of a "sea" of real $\Lambda$-solutions with two holes, at $\Lambda_1^h$ and $\Lambda_2^h$, and a *two-string*, namely, a pair of complex $\Lambda$'s located at $\Lambda^\pm = \bar{\Lambda} \pm ic/2$, where $\bar{\Lambda} = \frac{1}{2}(\Lambda_1^h + \Lambda_2^h)$.

*The singlet*: The equations governing the state are obtained from eq(7) written first for real $\Lambda$'s (1-strings) and then for the $\Lambda$'s in the 2-string. The first equation determines the density of real $\Lambda$-solutions $\sigma(\Lambda)$

$$\sigma(\Lambda) + \sigma^h(\Lambda) = f(\Lambda) - \int K(\Lambda - \Lambda')\sigma(\Lambda')d\Lambda' - \sigma^{st}(\Lambda), \qquad (31)$$

where

$$\sigma^h(\Lambda) = \delta(\Lambda - \Lambda_1^h) + \delta(\Lambda - \Lambda_2^h),$$
$$\sigma^{st}(\Lambda) = K_3(\Lambda - \bar{\Lambda}) + K_1(\Lambda - \bar{\Lambda}).$$

The second equation

$$N^e \Theta(\bar{\Lambda} - 1) + \Theta(\bar{\Lambda}) = -2\pi I^{(2)} + \sum_{\delta=1}^{M} \Theta_1(\bar{\Lambda} - \Lambda_\delta) + \sum_{\delta=1}^{M} \Theta_3(\bar{\Lambda} - \Lambda_\delta) \qquad (32)$$

fixes the position of the 2-string $\bar{\Lambda}$.

As before, $\sigma^h(\Lambda)$ arises by placing holes in the ground-state (consecutive) sequence. The string term $\sigma^{st}(\Lambda)$ is the contribution $\Theta(\Lambda_\gamma - \Lambda^+) + \Theta(\Lambda_\gamma - \Lambda^-)$ of the two-string at $\Lambda^\pm = \bar{\Lambda} \pm ic/2$ to the sum in eq (7). This contribution can be rewritten as $\Theta_1(\Lambda - \bar{\Lambda}) + \Theta_3(\Lambda - \bar{\Lambda})$ with $\bar{\Lambda}$ real, recall $\Theta_n(x) = \Theta(\frac{2}{n}x)$. When we convert the set of algebraic equations to an integral equation, by the method described before, we obtain eqs(31).

Its solution in Fourier space is

$$\tilde{\sigma}(p) = \tilde{\sigma}_o(p) + \Delta\tilde{\sigma}^{st}(p) + \Delta\tilde{\sigma}^h(p), \qquad (33)$$

with

$$\Delta\tilde{\sigma}^h(p) = -\frac{\exp\frac{c}{2}|p|}{2\cosh\frac{c}{2}p}(e^{-i\Lambda_1^h p} + e^{-i\Lambda_2^h p}),$$
$$\Delta\tilde{\sigma}^{st}(p) = -e^{-(c/2)|p|}e^{-i\bar{\Lambda}p}.$$

When $\sigma(\Lambda)$ is fed into eq(32) we find $\bar{\Lambda} = \frac{1}{2}(\Lambda_1^h + \Lambda_2^h)$, as stated.

This state is indeed a singlet, as can be deduced from calculating

$$M = 2 + \int d\Lambda \sigma(\Lambda) = 2 + \tilde{\sigma}(0)$$
$$= 2 + \tilde{\sigma}_o(0) + \Delta\tilde{\sigma}^h(0) + \Delta\tilde{\sigma}^{st}(0) = \frac{1}{2}N,$$

leading to a Young tableau with two rows of equal length.

The singlet excitation energy, $\Delta E^s$, is also easy to calculate, and is found to be degenerate (in the thermodynamic limit) with the triplet excitation energy,

$$\begin{aligned}\Delta E^s &= D \int \left(\Delta\sigma^h(\Lambda) + \Delta\sigma^{st}(\Lambda)\right) (\Theta(2\Lambda - 2) - \pi) \, d\Lambda \\ &\quad + D \left(\Theta(2\Lambda^+ - 2) + \Theta(2\Lambda^- - 2) - 2\pi\right) \\ &= 2D(\tan^{-1} e^{(\pi/c)(\Lambda_1^h - 1)} + \tan^{-1} e^{(\pi/c)(\Lambda_2^h - 1)}), \end{aligned} \quad (34)$$

since $\int d\Lambda \Delta\sigma^{st}(\Lambda)[\Theta(2\Lambda - 2) - \pi] + \Theta(2\Lambda^+ - 2) + \Theta(2\Lambda^- - 2) = 0$. We find that the string induces a change in the sea of real $\Lambda$ momenta which exactly cancels its direct contribution.

We thus confirmed the picture of spin-1/2 objects, each corresponding to a "hole" in the $I_\gamma$ sequence, being the underlying fundamental excitations in the spin sector [1]. But though the energy of a two spinon state does not depend on the total spin, it does not follow that the interaction between the spinons is spin-independent. Indeed, we shall calculate the phase shift when these excitations cross each other and find that they differ in the singlet and the triplet state.

*The doublet*: To have a single spinon excitation it is necessary to add an electron to the system whose total spin will then change to be one-half. This change induces a hole in the sequence since $M$ is unchanged and $N$ is increased by one so that there is another slot available which is unfilled. Denoting by $\Lambda^h$ the corresponding spin momentum we find that the energy associated with the adding an electron to form a doublet is composed of two terms

$$\Delta E^d = 2D \tan^{-1} e^{(\pi/c)(\Lambda^h - 1)} + \frac{2\pi}{L} n, \quad (35)$$

the first term being the *spinon* energy and the second the *holon* energy ($n$ is the level into which the electron was inserted). In other words, the added electron decomposes into two independent excitations, one carrying the spin content and one carrying the charge content of the bare electron.

Another way of creating a one-hole state is by removing a particle from the system. Thus consider $N \to N-1$ and $M \to M-1$. The counting argument indicates that a hole opens in the $I_\gamma$ sequence leading to a spinon excitation, but since an $n$ level is removed ($n$ is filled in the ground state, $n < 0$ ) we have an *antiholon*, a massless anticharged particle, rather than a *holon*

$$\Delta E^d = 2D \tan^{-1} e^{(\pi/c)(\Lambda^h - 1)} + \frac{2\pi}{L}|n|. \tag{36}$$

Thus the states obtained by adding an electron have the identical spin-content as the states obtained by removing an electron. The decoupled charge sector is, of course, different.

We proceed to show that this basic excitation, the spin-$\frac{1}{2}$ spinon is a massless relativistic right-mover, and calculate its scattering phase shift off the impurity as well as off another spinon. From eq(35), we learn that the Kondo system has no spin-gap, since there are arbitrarily low-lying spin-flip excitations for $\Lambda^h$ arbitrarily large and negative. The spinon excitation energy takes the form $\epsilon = 2T_0 e^{(\pi/c)\Lambda^h}$ with $T_0$, given by $T_0 = De^{-\pi/c}$. This is a dimensionful scale, dynamically generated by the model, and in terms of which all other dimensionful scales are measured.

Having identified a physical scale, $T_0$, we may take the scaling limit, $D \to \infty$ with the scale held fixed. The coupling acquires a dependence on the running cut-off $c = c(D) = \pi/\ln \frac{T_0}{D} \to 0$ as $D \to \infty$. In this limit, having removed the cut off dependence, all surviving quantities are universal. In particular, the excitation energy is

$$\epsilon = 2T_0 e^\theta \tag{37}$$

with $\theta = (\pi/c)\Lambda^h$ held finite in the limit. The form of the energy is typical of a relativistically right moving particle with $\theta$ being its *rapidity*. As such it is also its momentum, $p = \epsilon$.

Within the scaling regime we still distinguish between (universal) ultra-violet behaviour as $T, h, \epsilon, \ldots \gg T_o$ and (universal) infrared behaviour as $T, h, \epsilon, \ldots \ll T_o$. Of course, all energy scales are small compared with the cut off $D$ which has been taken to infinity. We shall find soon that the limiting procedure assures a universal limit to all excitations and thermodynamic quantities.

We now wish to compute the various spinon scattering phase shifts. To deduce them we return to the cut-off theory on a large but finite ring $L$, and note that $\Lambda^h$ is not a continuous, independent variable. It is determined by the $\{I_\gamma\}$ configuration, and takes values which are determined by the choice of the omitted $I^h$. This allows the calculation of the S-matrix, or the scattering phase shift $\delta(\theta)$, of the spinon off the inpurity, as well as the scattering of one spinon off another, $\phi(\theta_1 - \theta_2)$. The method we use [8] is based on the well known observation that when a particle is quantised on a finite ring of length $L$, the shift of its momentum from the free value $\frac{2\pi}{L}n$ is interpreted as a phase shift,

$$p = 2T_0 e^\theta = \frac{2\pi}{L}n + \frac{\delta(\theta)}{L}, \qquad \theta = \frac{\pi}{c}\Lambda^h \qquad (38)$$

We shall refer to this method of calculation as the momentum shift method and will use it also in the next lecture to calculate the S-matrix of the Hubbard model. Another method of calculation is due to Korepin [9]. More recently a bootstrap Kondo S-matrix approach was developed by Fendley [10].

We determine the phase-shift as follows: consider again eq(21)

$$\nu(\Lambda) = -\frac{1}{2\pi}[N^e \Theta(2\Lambda - 2) + \Theta(2\Lambda) - \int d\Lambda' \Theta(\Lambda - \Lambda')\sigma(\Lambda')] \qquad (39)$$

and evaluate $\nu(\Lambda)$ for a density $\sigma(\Lambda)$ in the presence of holes $\{\Lambda_j^h\}$, eq(27). We shall add 2-strings later. One finds

$$\begin{aligned}
2\pi\nu(\Lambda) = & -2N^e(\tan^{-1} e^{\frac{\pi}{c}(\Lambda-1)} - \frac{\pi}{2}) - 2(\tan^{-1} e^{\frac{\pi}{c}\Lambda} - \frac{\pi}{2}) \\
& + \frac{1}{i}\sum_j \ln \frac{\Gamma(1 - i\frac{\Lambda-\Lambda_j^h}{2c})\Gamma(\frac{1}{2} + i\frac{\Lambda-\Lambda_j^h}{2c})}{\Gamma(1 + i\frac{\Lambda-\Lambda_j^h}{2c})\Gamma(\frac{1}{2} - i\frac{\Lambda-\Lambda_j^h}{2c})}
\end{aligned} \qquad (40)$$

Consider first the case with one "hole" in the $\{I_\gamma\}$ sequence, a one spinon excitation (induced, for example, by adding an electron to the system). The corresponding $\Lambda^h$ is given by the value satisfying $2\pi\nu(\Lambda^h) = I^h$, where $I^h$ is the "hole", so that eq(40) takes the form (upon dividing by $L$ ):

$$2\frac{N^e}{L}(\tan^{-1} e^{\frac{\pi}{c}(\Lambda^h-1)} - \frac{\pi}{2}) = \frac{2\pi}{L}I^h + 2\frac{1}{L}(\frac{\pi}{2} - \tan^{-1} e^{\frac{\pi}{c}\Lambda^h}) \qquad (41)$$

In the scaling limit the left hand side contains the spinon momentum, and we find that its shift from the free value $\frac{2\pi}{L}I^h$, namely the the scattering phase shift, is

$$\delta(\theta) = 2(\frac{\pi}{2} - \tan^{-1} e^\theta) \tag{42}$$

and the spinon-inpurity S-matrix becomes

$$S = e^{i\delta(\theta)} = -\frac{1+ie^\theta}{1-ie^\theta} = -\tanh(\frac{\theta}{2} - \frac{i\pi}{4}). \tag{43}$$

The phase shift we found describes the one-dimensional scattering process. If we want to interpret it as an s-wave scattering in three dimensions, we have

$$S = e^{2i\delta_0} \tag{44}$$

with $\delta_0$ being the s-wave phase shift so that

$$\delta_0(p) = \frac{1}{2}\delta(p) = \frac{\pi}{2} - \tan^{-1}(\frac{p}{2T_0}). \tag{45}$$

As a function of the momentum the phase shift varies from $\delta_0 = \frac{\pi}{2}$ at low momenta to $\delta_0 = 0$ at large momenta. We shall interpret the the large momentum physics as being described by a weak coupling hamiltonian, $H_0$, while the physics in the infra-red is given by another hamiltonian $H^*$ which is strongly coupled producing a maximal phase-shift. This is our first example of a crossover in the properties of the model.

The phase shift describes the scattering of the spin content of the incoming electron. The charge does not scatter. Thus, after passing the impurity the emerging state is not an electron anymore, but a superposition of unscattered holons and a time delayed spinon.

The spinon-spinon S-matrix can be read off eq(40) by considering the configuration with two "holes" at $\Lambda_1^h$ and $\Lambda_2^h$ corresponding to the omitted integers $I_1^h$ and $I_2^h$. In this state the two spinon scatter in a triplet state. We have

$$\frac{2\pi}{L}I_1^h = p_1 + \delta(\theta_1) + \frac{1}{iL}\ln\frac{\Gamma(1-i\frac{\Lambda_1^h-\Lambda_2^h}{2c})\Gamma(\frac{1}{2}+i\frac{\Lambda_1^h-\Lambda_2^h}{2c})}{\Gamma(1+i\frac{\Lambda_1^h-\Lambda_2^h}{2c})\Gamma(\frac{1}{2}-i\frac{\Lambda_1^h-\Lambda_2^h}{2c})} \tag{46}$$

with a similar equation determining $p_2$. Hence, the spinon momentum, $p_1 = T_0 e^{\theta_1}$, is shifted by $\delta(\theta_1)$ in the presence of the impurity, and by $\phi^t(\theta_1 - \theta_2)$, the triplet spinon-spinon phase shift, in the presence of another spinon. The triplet scattering is

$$S^t(\theta_1 - \theta_2) = e^{i\phi^t(\theta_1 - \theta_2)} = \frac{\Gamma(1 - i\frac{\theta_1 - \theta_2}{2\pi})\Gamma(\frac{1}{2} + i\frac{\theta_1 - \theta_2}{2\pi})}{\Gamma(1 + i\frac{\theta_1 - \theta_2}{2\pi})\Gamma(\frac{1}{2} - i\frac{\theta_1 - \theta_2}{2\pi})} \qquad (47)$$

In a simialr way, to determine the singlet scattering phase-shift we have to evaluate eq(21) in the presence of two holes and a 2-string, using the density in eq(33). One finds

$$S^s(\theta_1 - \theta_2) = \frac{1 - i\frac{\theta_1 - \theta_2}{\pi}}{1 + i\frac{\theta_1 - \theta_2}{\pi}} S^t(\theta_1 - \theta_2), \qquad (48)$$

and hence the total spinon S-matrix in the Kondo model is

$$S(\theta_1 - \theta_2) = \left(\frac{\Gamma(1 - i\frac{\theta_1 - \theta_2}{2\pi})\Gamma(\frac{1}{2} + i\frac{\theta_1 - \theta_2}{2\pi})}{\Gamma(1 + i\frac{\theta_1 - \theta_2}{2\pi})\Gamma(\frac{1}{2} - i\frac{\theta_1 - \theta_2}{2\pi})}\right) \frac{\frac{\theta_1 - \theta_2}{2\pi} I^{12} + ic P^{12}}{\frac{\theta_1 - \theta_2}{2\pi} + ic}. \qquad (49)$$

The S-matrix we obtained again satisfies the YBE as does its bare counterpart. This guarantees that the spinon quasi-particles, although interacting, do not decay. In fact they are protected by the conservation laws. The same S-matrix was previously calculated for the spinons in the g-ology model [8], and will appear again in Lecture 4 to describe the scattering of the Hubbard spinons. The same S-matrix will describe the scattering of spinons in all models solved with the R-matrix of Lecture 2. It can be also deduced from general bootstrap considerations [11].

The spinon-holon S-matrix, $S^{s,h} = 1$, linearizing the spectrum has completely decoupled the charge from the spinon sector.

The full spin spectrum of the model can be constructed this way; excitations are built of holes and complex pairs. If we have only real $\Lambda$ solutions with $m$ holes in the $\Lambda$-sea, then the state will have spin $S = \frac{1}{2}m$. (With $N$ fixed $m$ must be even.) These are maximum-spin excitations. The energy associated with holes at $\Lambda_1^h, \cdots, \Lambda_m^h$ is

$$(\Delta E)^{m \ holes} = 2\sum_{j=1}^{m} D \tan^{-1}(e^{(\pi/c)(\Lambda_j^h - 1)}). \qquad (50)$$

When complex pairs are added they lower the total spin by coupling spins antisymmetrically. The various complex structures allowed will be discussed below, but they have the feature that their contribution to the energy cancels. The latter therefore is determined only by the holes in the sea of real $\Lambda$ momenta. The contribution of the complexes shows up, however, not only in the counting of states and their total spin but also in the $S$ matrix determining the interaction of the various excitations. We conclude that the spin excitations form a Fermi liquid.

The nature of the complex solutions of the equations is captured by the *string hypothesis*: the solutions of eq(7), in the limit $N \to \infty$, always occur in the form of $n$-strings, where an $n$-string is a complex of $n$ $\Lambda$-solutions given by

$$\Lambda_j^{(n)} = \Lambda^{(n)} + i\frac{c}{2}(n+1-2j), \quad j = 1, 2, \ldots, n. \tag{51}$$

This hypothesis has a long history[15],[12], but is not always valid; the hypothesis holds in the presence of a macroscopic number of holes, and is not necessarily true in case we consider excitations containing only a small number of them [13],[14]. Two-string solutions always exist (when two holes or more are present), but the conjugate pairs organize themselves into $n$-strings, $n > 2$, only if driven macroscopically. Thus more care has to be taken when analyzing scattering events of elementary excitations than in thermodynamic applications, where, in studying the response of the system to external probes, we excite a sufficiently large number of holes and the string hypothesis is valid.

Let us develop the form that eq(7) takes in this case. Consider the case with $M_n$ $n$-strings $\Lambda_{\gamma,j}^{(n)} = \Lambda_\gamma^{(n)} + i(c/2)(n+1-2j)$, $\gamma = 1, \ldots, M_n$, with $\Lambda_\gamma^{(n)}$ the real part of the $\gamma$th $n$-string. Equation (7) then becomes:

$$N^e \Theta_n(\Lambda_\gamma^{(n)} - 1) + \Theta_n(\Lambda_\gamma^{(n)}) = {\sum_{m,\delta}}' \Theta_{n,m}(\Lambda_\gamma^{(n)} - \Lambda_\delta^{(m)}) - 2\pi I_\gamma^{(n)}, \tag{52}$$

where the summation is over all strings different from the particular $\Lambda_\gamma^{(n)}$ string. The function $\Theta_{mn}(x)$ is defined as

$$\Theta_{mn}(x) = \begin{cases} \Theta_{|n-m|}(x) + 2\Theta_{|n-m|+2}(x) + \cdots + 2\Theta_{n+m-2}(x) + \Theta_{n+m}(x), & n \neq m \\ 2\Theta_2(x) + \cdots + 2\Theta_{2n-2}(x) + \Theta_{2n}(x), & n = m \end{cases} \tag{53}$$

and as a reminder, $\Theta(x) = -2\tan^{-1}(x/c) = \Theta_{11}(x)$. Equation (52) is obtained from eq(7) by summing over all members of a string. For example,

$$\sum_{j=1}^{n} \Theta(\Lambda - \Lambda_j^{(n)}) = \frac{1}{i}\sum_{j=1}^{n} \ln\frac{c - i(\Lambda - \Lambda_j^{(n)})}{c + i(\Lambda - \Lambda_j^{(n)})} = \Theta_{n+1}(\Lambda - \Lambda^n) + \Theta_{n-1}(\Lambda - \Lambda^n),$$

$$\sum_{j=1}^{n} \Theta(2\Lambda_j^{(n)} - 2) = \Theta_n(\Lambda^{(n)} - 1).$$

The integers $I_\gamma^{(n)}$ determine the allowed string solutions $\Lambda_{\gamma,j}^{(n)}$, and are the spin quantum numbers of the system.

The rest is just as before. For a chosen set of configurations $I_\gamma^{(n)}$ one determines the corresponding $\Lambda_{\gamma,j}^{(n)}$ complex spin solutions, and then forms the function

$$\nu_n(\Lambda) = -\frac{1}{2\pi}\left[N^e\Theta_n(\Lambda - 1) + \Theta_n(\Lambda) - \sum_{m,\delta}\Theta_{nm}(\Lambda - \Lambda_\delta^{(m)})\right], \quad (54)$$

with the $\Lambda_\delta^{(m)}$ determined earlier, so that solutions of $\nu_n(\Lambda_\gamma^{(n)}) = I_\gamma^{(n)}$ are the allowed strings, while solutions, $\Lambda_j^{h(n)}$, of

$$\nu_n(\Lambda_j^{h(n)}) = I_j^{h(n)}, \quad (55)$$

where $I_\gamma^{h(n)}$ are the integers omitted in the sequence, correspond to $n$-string holes. In the limit of $N \to \infty$ we may introduce $n$-string density $\sigma_n(\Lambda)$ and $n$-string hole density $\sigma_n^h(\Lambda)$, which obviously satisfy

$$\frac{d\nu_n(\Lambda)}{d\Lambda} = \sigma_n^h(\Lambda) + \sigma_n(\Lambda). \quad (56)$$

Combining the last expression with the derivative of eq(54) we find that the string densities obey the following set of equations

$$f_n(\Lambda) = \sigma_n^h(\Lambda) + \sum_{m=1}^{\infty} A_{nm}\sigma_m(\Lambda),$$

where

$$f_n(\Lambda) = N^e K_n(\Lambda - 1) + K_n(\Lambda),$$
$$A_{nm} = [|n-m|] + 2[|n-m|+2] + \cdots + 2[n+m-2] + [n+m], (57)$$

and $[n]$ is the operator given by

$$[n]f(\Lambda) = \int K_n(\Lambda - \Lambda')f(\Lambda')d\Lambda'. \tag{58}$$

In terms of the string variables the energy can be expressed as

$$E = D\sum_n \int d\Lambda \sigma_n(\Lambda)\left[\Theta_n(\Lambda - 1) - \pi\right] + \sum_j \frac{2\pi}{L}n_j. \tag{59}$$

Here we have performed the sum over the individual members of a string and are left with integration over the string locations only.

With this complete classification of states, we proceed to compute the partition function of the system at nonzero temperature $T$ and external magnetic field $h$. We shall deduce a set of coupled integral equations determining the free energy $F$, using a method originated by Yand and Yang [16], and generalized by Gaudin [17] and Takahashi [12]. We shall analyse the equations and demonstrate scaling behaviour and crossover properties in the full $h - T$ plane.

The Thermodynamics of the Kondo Model.

The formal expression for the partition function is

$$\begin{aligned} Z &= Tr \exp\left[-\frac{1}{T}(H - 2\mu h S_z)\right], \\ &= \sum_{S=0}^{N/2} \sum_{S_z=-S}^{S} Tr_{ss_z} \exp\left[-\frac{1}{T}(H - 2\mu h S_z)\right], \end{aligned} \tag{60}$$

where $H$ is the zero-field Hamiltonian, $\mu$ is the magnetic moment and we set $g = 2$ for convenience. $Tr_{ss_z}$ is the trace over all basis states with values S and $S_z$ of total spin and z component of the spin. Since $H$ is invariant under simultaneous rotations of all spins, we may split off the sum over $S_z$ to obtain

$$\begin{aligned} Z &= \sum_{S=0}^{N/2} \frac{\sinh\left[(2S+1)\frac{\mu h}{T}\right]}{\sinh\left[\frac{\mu h}{T}\right]} Tr_{ss} \exp\left[-\frac{H}{T}\right] \\ &\approx \sum_{S=0}^{N/2} Tr_{ss} \exp\left[-\frac{1}{T}(H - 2\mu h S)\right]. \end{aligned} \tag{61}$$

In the last approximation we have dropped terms proportional to $\exp(-Sh/T)$, as well as an overall factor $[2\sinh(h/T)]^{-1}$, since these terms contribute negligibly to the calculation of thermodynamic quantities in the limit $L \to \infty$ (note that S~L).

Now let us exploit the specific form of the energy for our basis states. Recall that each such state is labeled by a set of quantum numbers $\{n_j, I_\gamma\}$ with $n_j \geq -N^e$. The corresponding energy is

$$E = E^{(c)}(\{n\}) + E^{(s)}(\{\Lambda\}) \qquad (62)$$

with the charge energy

$$E^{(c)}(\{n\}) = \frac{2\pi}{L} \sum_{j=1}^{N^e} n_j \qquad (63)$$

and the spin energy

$$E^{(s)}(\{\Lambda\}) = D \sum_{\gamma=1}^{M} \left[\Theta(2\Lambda_\gamma - 2) - \pi\right] \qquad (64)$$

The partition function factorizes acordingly,

$$Z = Z^{(c)} Z^{(s)}, \qquad (65)$$

where the charge partition function

$$Z^{(c)} = \sum_{\{n_j\}, n_j \geq -N^e} \exp\left[-\frac{1}{T} \sum_{j=1}^{N^e} \frac{2\pi}{L} n_j\right] \qquad (66)$$

describes the thermodynamics of $N^e$ noninteracting *spinless* fermions with linear kinetic energy. In the limit $D \to \infty$ it leads to the free energy

$$F^{(c)} = -\frac{LT}{2\pi} \int_{-\infty}^{\infty} dk \ln\left(1 + e^{-\frac{k}{T}}\right) = -\frac{\pi}{12} LT^2 + \{\text{infinite constant}\} \qquad (67)$$

and is half the the free energy of a noninteracting electron gas at zero magnetic field. Note that the effects of a magnetic field are all included in $Z^{(s)}$.

The spin partition function is

$$Z^{(s)} = \exp\left[\frac{N\mu h}{T}\right] \sum_M \sum_{\{\Lambda_1, \ldots, \Lambda_M\}} \exp\left[-\frac{1}{T}[E^{(s)}(\{\Lambda\}) + 2M\mu h]\right]. \qquad (68)$$

It can be rewritten in terms of the $n$-string and $n$-string hole density $\sigma_n(\Lambda)$ and $\sigma_n^h(\Lambda)$ as

$$Z^{(s)} = \exp\left[\frac{N\mu h}{T}\right] \int \prod D\sigma_n D\sigma_n^h \exp \mathcal{S}(\{\sigma_n, \sigma_n^h\}) \, \exp\left[-\frac{1}{T}[E^{(s)}(\{\Lambda\}) + 2\mu h M]\right], \tag{69}$$

with

$$E^{(s)}(\{\Lambda\}) + 2\mu h M = \sum_n \int d\Lambda \sigma_n(\Lambda) g_n(\Lambda), \tag{70}$$

where

$$g_n(\Lambda) = D[\Theta_n(\Lambda - 1) - \pi] + 2\mu h n \tag{71}$$

and where $\mathcal{S}(\{\sigma_n, \sigma_n^h\})$ is the entropy associated with the densities $\{\sigma_n, \sigma_n^h\}$. In other words, $\exp \mathcal{S}(\{\sigma_n, \sigma_n^h\})$ is the functional, counting the number of configurations $\{I_\gamma\}$ leading to solutions $\{\Lambda_\gamma\}$ that are consistent with a given set of densities $\{\sigma_n, \sigma_n^h\}$. To determine $\mathcal{S}$ divide the $\Lambda$ axis into intervals $d\Lambda$, chosen sufficiently small so that the densities are approximately constant over each interval, yet sufficiently large that $(\sigma_n + \sigma_n^h)d\Lambda \ll 1$. The number of slots for $\Lambda$-strings in the interval $d\Lambda$ is $d\nu_n = (\sigma_n + \sigma_n^h)d\Lambda$, $\sigma_n d\Lambda$ of which are occupied, while $\sigma_n^h d\Lambda$ are empty; thus the number of ways of distributing the $n$-strings among the slots is

$$\frac{[(\sigma(\Lambda) + \sigma_n^h(\Lambda))d\Lambda]!}{[\sigma_n(\Lambda)d\Lambda]![\sigma_n^h(\Lambda)d\Lambda]!}. \tag{72}$$

Using Stirling's formula, we can simplify this to give

$$d\mathcal{S}_n = \ln \frac{[(\sigma_n + \sigma_n^h)d\Lambda]!}{[\sigma_n d\Lambda]![\sigma_n^h d\Lambda]!} = [(\sigma_n + \sigma_n^h)\ln(\sigma_n + \sigma_n^h) - \sigma_n \ln \sigma_n - \sigma_n^h \ln \sigma_n^h] d\Lambda$$

so that entropy, $\mathcal{S}$, becomes

$$\mathcal{S}(\{\sigma_n, \sigma_n^h\}) = \sum_n \int d\Lambda[(\sigma_n + \sigma_n^h)\ln(\sigma_n + \sigma_n^h) - \sigma_n^h \ln \sigma_n^h - \sigma_n \ln \sigma_n]. \tag{73}$$

In thermodynamic limit, $N \to \infty$, we may evaluate $Z^{(s)}$ by the method of stationary phase approximation. Varying the functional

$$\begin{aligned} F^{(s)}\{\sigma_n, \sigma_n^h\} &= E^{(s)} + 2\mu h M - T\mathcal{S} - Nh \tag{74}\\ &= \sum_n \int d\Lambda\left[\sigma_n g_n - T\sigma_n \ln\left[1 + \frac{\sigma_n^h}{\sigma_n}\right] - T\sigma_n^h \ln\left[1 + \frac{\sigma_n}{\sigma_n^h}\right]\right] - N\mu h \end{aligned}$$

subject to the constraint, $\delta\sigma_n^h = -\sum_m A_{nm}\delta\sigma_m$, one finds,

$$\ln[1+\eta_n(\Lambda)] = \frac{g_n(\Lambda)}{T} + \sum_{m=1}^{\infty} A_{mn}\ln[1+\eta_n^{-1}(\Lambda)], \tag{75}$$

where

$$\eta_n(\Lambda) = \frac{\sigma_n^h(\Lambda)}{\sigma_n(\Lambda)}. \tag{76}$$

This set of equations may be rewritten as

$$\begin{aligned}\ln\eta_n &= G[\ln(1+\eta_{n+1}) + \ln(1+\eta_{n-1})], \\ \ln\eta_1 &= -\frac{2D}{T}\tan^{-1}e^{(\pi/c)(\Lambda-1)} + G\ln(1+\eta_2),\end{aligned} \tag{77}$$

with the driving term, $2D\tan^{-1}e^{(\pi/c)(\Lambda-1)}$, familiar as the one-hole excitation energy $\Delta E^d$. The operator $G$ is defined by

$$Gf(\Lambda) \equiv \frac{[1]}{[0]+[2]}f(\Lambda) = \frac{1}{2c}\int d\Lambda' \frac{1}{\cosh\frac{\pi}{c}(\Lambda-\Lambda')}f(\Lambda'). \tag{78}$$

To close the set of eqs(77) one has to supply a boundary condition for $n \to \infty$. It turns out to be

$$\lim_{n\to\infty}([n+1]\ln(1+\eta_n) - [n]\ln(1+\eta_{n+1})) = -\frac{2\mu h}{T}. \tag{79}$$

Once a set of solutions $\eta_n$ satisfying the equations has been found, the spin free energy may be obtained from eq(74), and considerably simplified (we shall skip some of the steps),

$$\begin{aligned}F^{(s)} &= \sum_n \int d\Lambda\left[\sigma_n g_n - T\sigma_n\ln(1+\eta_n) - T\left[f_n - \sum_m A_{nm}\sigma_m\right]\ln(1+\eta_n^{-1})\right] - N\mu h \\ &= -\sum_n T\int d\Lambda f_n(\Lambda)\ln(1+\eta_n^{-1}) - N\mu h \\ &= -\sum_n T\int d\Lambda[\,N^e\delta(1-\Lambda) + \delta(\Lambda)][n]\ln(1+\eta_n^{-1}) - N\mu h \\ &= -T[N^e\delta(1-\Lambda) + \delta(\Lambda)]G\left[\ln(1+\eta_1) - \frac{g_1}{T}\right] - N\mu h \\ &= \int d\Lambda\sigma_0(\Lambda)\{g_1(\Lambda) - T\ln[1+\eta_1(\Lambda)]\} - N\mu h \\ &= E_o - T\int d\Lambda\sigma_o(\Lambda)\ln[1+\eta_1(\Lambda)],\end{aligned} \tag{80}$$

where

$$\sigma_o(\Lambda) = G[N^e\delta(1-\Lambda) + \delta(\Lambda)] = \frac{1}{2c}\left[\frac{N^e}{\cosh\frac{\pi}{c}(\Lambda-1)} + \frac{1}{\cosh\frac{\pi}{c}\Lambda}\right] \quad (81)$$

is the ground-state $\Lambda$ density, and $E_o$ is the is the ground state energy.

Adding the charge free energy $F^{(c)}$, eq(67), we have

$$F = F^{(c)} + F^{(s)} = E_0 - \frac{\pi L T^2}{12} - T\int d\Lambda \sigma_0(\Lambda)\ln[1+\eta_1(\Lambda)], \quad (82)$$

$E_o$ now containing also the temperature-independent contribution of the charge fluctuation.

We shall now demonstrate the scaling properties of the thermodynamic equations, describing the behaviour of the system in the regime where $T, h \ll D$. In this case the function $\eta_1$ has a very sharp decrease, proportional to $\exp[-(2D/T)\tan^{-1} z]$, where $z = \exp[(\pi/c)(\Lambda-1)]$, and will contribute of order $\exp(-2D/T)$ to the partition function except for small $z$. Hence in the scaling limit we may replace $\tan^{-1} z$ by $z$ in these integrals and compute $\eta_n$ from a modified version of the thermodynamic equations

$$\ln\eta_n = -2\delta_{n,1}e^\zeta + G[\ln(1+\eta_{n+1}) + \ln(1+\eta_{n-1})] \quad n = 1, 2, \ldots \quad (83)$$

where we now regard $\eta_n$ as a function of the new variable $\zeta$,

$$\zeta = \frac{\pi}{c}\Lambda - \ln\frac{T_0}{T}, \quad (84)$$

with

$$G(\zeta - \zeta') = \frac{1}{2\pi}\frac{1}{\cosh(\zeta-\zeta')} \quad (85)$$

and by convention we set $\eta_0 = 0$.

The free energy becomes,

$$F = E_0 - \frac{\pi L T^2}{12} - \frac{T}{2\pi}\int d\zeta \left\{\frac{N^e}{\cosh\left[\zeta - \ln\frac{T_0}{T} - \frac{\pi}{c}\right]} + \frac{1}{\cosh\left[\zeta - \ln\frac{T_0}{T}\right]}\right\}\ln[1+\eta_1(\zeta, \frac{h}{T})], \quad (86)$$

which we rewrite now in terms of impurity and electron contributions rather than in terms of spin and charge sectors,

$$F = E_0 + F^e + F^i. \quad (87)$$

The impurity free energy

$$F^i = -\frac{T}{2\pi} \int d\zeta \frac{\ln[1 + \eta_1(\zeta, \frac{h}{T})]}{\cosh[\zeta - \ln \frac{T_0}{T}]} \equiv Tf\left(\frac{T}{T_0}, \frac{h}{T}\right) \qquad (88)$$

depends on the coupling constant and the cutoff only through the combination $T_0 = De^{-\pi/c}$, the only scale in the problem. The free energy is universal to all materials or constructions with the same $T_0$.

The electron contribution

$$F^e = -\frac{\pi L T^2}{12} - TN^e f\left(\frac{T}{D}, \frac{h}{T}\right)\Big|_{D \to \infty} \qquad (89)$$

is the free energy of a non interacting gas of electrons, here expressed in the charge-spin decoupled basis. This is obviously so since it is the part of the free energy that survives when the impurity is removed.

Our task, then, is to solve the equations eq(83) and from the obtained $\eta_1(\zeta)$ to compute the impurity free energy $F^i$, eq(88). Some properties of the solutions are easy to establish, even though no analytic solution is yet available:

1. $\eta_n(\zeta)$ is monotonically decreasing in $\zeta$ ( fixed $n$).
2. $\eta_n(\zeta)$ is monotonically increasing in $n$ (fixed $\zeta$).
3. $\eta_n(\zeta)$ has finite asymptotic limits:

$$\eta_n \to \begin{cases} \eta_n^+ = \frac{\sinh^2(n+1)\frac{\mu h}{T}}{\sinh^2 \frac{\mu h}{T}} - 1, & \text{as } \zeta \to +\infty \\ \eta_n^- = \frac{\sinh^2 n\frac{\mu h}{T}}{\sinh^2 \frac{\mu h}{T}} - 1, & \text{as } \zeta \to -\infty \end{cases} \qquad (90)$$

These properties will be of use as we turn now to study the impurity physics in the high temperature and low temperature limits for a given magnetic field $h$, and then at zero temperature as a function of $h$.

Let us check that the electron free energy $F^e$ coincides in the limit $D \to \infty$ with the conventional expression derived in the Fock basis[2]. Begin from the expression for $F^{(s)} = -\sum_n T \int d\Lambda f_n(\Lambda) \ln(1 + \eta_n^{-1})$, and notice that now, as we are considering a free electron gas $f_n$ does not contain the impurity part, $f_n(\Lambda) = N^e K_n(\Lambda - 1)$, which is proportional to the derivative of $g_n$, $f_n(\Lambda) = -\frac{L}{2\pi} g'_n(\Lambda)$ which in turn via eq(75) allows us to write

$$\begin{aligned}
F^{e(s)} &= -\sum_n T^2 \frac{L}{2\pi} \int d\Lambda \frac{d}{d\Lambda}\left[\ln(1+\eta_n) - \sum_m A_{nm}\ln(1+\eta_m^{-1})\right]\ln(1+\eta_n^{-1}) \\
&= -\sum_n T^2 \frac{L}{2\pi} \int d\Lambda \frac{d}{d\Lambda}\left[\ln(1+\eta_n) - \ln(1+\eta_n^{-1})\right]\ln(1+\eta_n^{-1}) \\
&= -\sum_n T^2 \frac{L}{2\pi} \int d\Lambda \frac{\eta'_n}{\eta_n}\ln(1+\eta_n^{-1}) = -\sum_n T^2 \frac{L}{2\pi} \int_{\eta_n^-}^{\eta_n^+} d\eta_n \frac{1}{\eta_n}\ln(1+\eta_n^{-1}) \\
&= -T^2 \frac{L}{2\pi}\int_0^\infty d\eta \frac{1}{\eta}\ln(1+\eta^{-1}) = -T\frac{L}{2\pi}\int_{-\infty}^\infty dk \ln(1+e^{-\frac{k}{T}}). \quad (91)
\end{aligned}$$

We used the fact that $\eta_1^- = 0$, $\eta_n^+ = \eta_{n+1}^-$, and in the last step changed variables $\eta = e^{\frac{k}{T}}$, to obtain an expression recognizable as the free energy of a gas of non interacting spinless electron, and identical to $F^{(c)}$, both being half the conventional free energy. We turn now to investigate the impurity physics.

*Impurity behavior at high temperature.*

The impurity free energy at high temperature is determined by the function $\eta_1(\zeta)$ for large negative values of $\zeta$. In the limit it is found from $\eta_1^-$ to be

$$F^i = -\frac{T}{2\pi}\int d\zeta \frac{1}{\cosh[\zeta + \ln\frac{T}{T_0}]}\ln[1+\eta_1(\zeta,\frac{h}{T})] \to -T\ln\left[2\cosh\frac{\mu h}{T}\right], \quad (92)$$

This is the free energy of an isolated spin in the presence of a magnetic field $h$, and we are therefore in the neighborhood of the weak coupling fixed point. How rapidly is this point approached? Assuming an expansion of $\eta_n$ in inverse powers of $\zeta$ and $\ln \zeta$, inserting the expansion in the thermodynamic equations we can determine the asymptotic behaviour of $\eta_1$ and hence the impurity free energy

$$F^i = T\left[\ln(2\cosh\frac{\mu h}{T}) - \frac{1}{2}\frac{\mu h}{T}\tanh\frac{\mu h}{T}\left(\frac{1}{\ln T/T_k} + \frac{1}{2}\frac{\ln\ln(T/T_k)}{\ln^2 T/T_k}\right)\right] \quad (93)$$

leading to the susceptibility,

$$\chi^i = \frac{\mu^2}{T}\left[1 - \left(\frac{1}{\ln T/T_k} + \frac{1}{2}\frac{\ln\ln(T/T_k)}{\ln^2 T/T_k}\right) + O\left(\frac{\ln^2 \ln T/T_k}{\ln^3 T/T_k}\right)\right]. \qquad (94)$$

We have introduced a new scale the Kondo temperature $T_k$, defined by the requirement that no term of the form $1/\ln^2(T/T_k)$ appear in the expansion. In other words, the term $a/\ln^2(T/T_0)$ that actually does appear is absorbed by expressing the expansion in terms of a scale $T_k = e^a T_0$. This expression for the susceptibility can also be obtained by conventional perturbation theory, by studying the neighborhood of the weak coupling hamiltonian, $H_0$, the ultra-violet fixed point.

*Impurity behavior at low temperature.*

We shall see now that as $T$ tends to zero the system flows to a new fixed point that is Fermi liquid in character. Indeed, note that the function $f(t,x)$ which in eq(89) describes for small $t$ the physics of a free electron gas on all scales, also appears in eq(88), where it captures for small $t$ the low temperature physics of the impurity. Since the former is the prime example of a Fermi-liquid it follows that so is the latter. More precisely, the impurity physics at low temperatures is the same as the spin sector of the free electron gas, and as such it carries all the spin degrees of freedom but only half of the entropy. This is the origin of the value of the low-temperature Wilson ratio $R$ to which we turn.

Let us then carry out the limiting procedure $t = \frac{T}{T_0} \to 0$. For sufficiently small $t$ we may expand the kernel $1/\cosh(\zeta + \ln t) = 2t \exp\zeta (1 - t^2 \exp 2\zeta + t^4 \exp 4\zeta + \cdots)$, to obtain (we may perform the expansion within the integral as $\eta_1$ vanishes for large $\zeta$ as $e^{-2e^\zeta}$)

$$f(t,x) = \frac{x}{\pi}\int d\zeta\, e^\zeta \ln(1 + \eta_1(\zeta, x)) + O(t^2), \qquad (95)$$

so that

$$\frac{F^e}{L} = -\frac{\pi T^2}{12} - \frac{T^2}{\pi}\int d\zeta\, e^\zeta \ln\left(1 + \eta_1(\zeta, \frac{h}{T})\right) + O\left(\frac{T^4}{D^2}\right). \qquad (96)$$

As argued earlier, $F^e$ is the free energy, in the spin-charge decoupled Bethe basis, of a system of $N^e = DL$ noninteracting spin-$\frac{1}{2}$ electrons in the presence of a field $h$. Comparing it with $F^e/L$ calculated in the conventional manner in the the Fock basis (the identification is valid as $D \to \infty$),

$$\begin{aligned} \frac{F^e}{L} &= -\frac{T}{2\pi}\left[\int_{-(\pi D-\mu h)}^{\infty} dk \ln(1+e^{-k/T}) + \int_{-(\pi D-\mu h)}^{\infty} dk \ln(1+e^{-k/T})\right] \\ &= -\frac{\pi T^2}{6} - \frac{(\mu h)^2}{2\pi}, \end{aligned}$$

we obtain

$$\int d\zeta\, e^\zeta \ln\left(1+\eta_1(\zeta,\frac{h}{T})\right) = \frac{\pi^2}{12} + \frac{(\mu h)^2}{2T^2}. \tag{97}$$

The same integral appears in the impurity low-temperature expression, therefore we have

$$F^i = -\frac{T^2}{\pi T_0}\int d\zeta\, e^\zeta \ln\left(1+\eta_1(\zeta,\frac{h}{T})\right) = -\frac{1}{\pi T_0}\left[\frac{\pi^2}{12}T^2 + \frac{1}{2}(\mu h)^2\right]. \tag{98}$$

We find that the impurity contribution to specific heat at low temperature is Fermi-liquid like

$$C^i = \frac{\pi}{6T_0}T \tag{99}$$

as is the magnetic susceptibility

$$\chi^i = \frac{\mu^2}{\pi T_0}. \tag{100}$$

The zero temperature susceptibility is finite, indicating that the impurity spin manifest in the high temperature regime by Curie's law, $\chi^i = \frac{\mu^2}{T}$, is now completely screened. We shall interpret the effect as due to strong effective coupling between the impurity and the conduction electrons leading to the formation of a local singlet, and the infra-red physics is dominated by a strong coupling fixed point.

Although the strong coupling impurity physics is Fermi-liquid like, it is different from the electrons' Fermi-liquid. This can be brought out by comparing

$$U^i = \frac{T\chi^i}{C^i} = 6\frac{\mu^2}{\pi^2} \tag{101}$$

with the corresponding electronic value $U^e = 3\frac{\mu^2}{\pi^2}$. Hence Wilson's ratio $R = \frac{\chi^i/\chi^e}{C^i/C^e}$ takes the value,

$$R = 2. \qquad (102)$$

The main element that allowed the identification of the fixed point as a Fermi liquid was that the same function $\eta_1$ occurs in the description of both the electron gas and the impurity. At the same time we have $R = 2$ due to the decoupling of the charge degrees of freedom from the impurity. More generally, the natural basis in Hilbert space for the description of the infra-red is the Bethe-basis with charge and spin decoupled, and the impurity modifying only the spin sector, while in the ultra-violet the Fock basis is the natural one.

Let us study the crossover as a function of the magnetic field at zero temperature. In this case the infinite set of themodynamic equations collapses into a single one,

$$\sigma_B(\Lambda) + \int_B^\infty K(\Lambda - \Lambda')\sigma_B(\Lambda')d\Lambda' = f(\Lambda), \qquad (103)$$

where the lower bound $B$ is related to the magnetic field $h$. The magnetization equation can be derived directly by looking for the lowest energy state for a given spin. In this state the spinons are excited at the lower end up to $\Lambda = B$.

The equation can be solved by means of Wiener-Hopf technique and yields the magnetization, $\mathcal{M} = \mu S = \mu\frac{1}{2}(N - 2M)$, as a function of the magnetic field [4][3]. The solution is of form,

$$\mathcal{M} = \mathcal{M}^e + \mathcal{M}^i, \qquad (104)$$

where $\mathcal{M}^e = \mu\left[\frac{2}{\pi e}\right]^{1/2} LT_0 e^{\pi B/c}$ in the scaling limit. Identifying it as the magnetization of the free electron gas, $\mathcal{M}_{Pauli} = \mu h L/\pi$,

$$\mathcal{M}^e = \mu\left[\frac{2}{\pi e}\right]^{1/2} LT_0 e^{\pi B/c} = \mu\frac{hL}{\pi}, \qquad \text{where } B << 1 \text{ (i.e., } \mu h << D) \quad (105)$$

allows us to relate the parameter $B$ to the magnetic field $h$,

$$e^{\pi B/c} = \left[\frac{2}{\pi e}\right]^{1/2}\frac{h}{T_0} \equiv \frac{h}{T_1} \qquad (106)$$

The second terms is the impurity magnetization which upon the previous identification becomes,

$$\mathcal{M}^i = \begin{cases} \frac{\mu}{\sqrt{\pi}} \sum_{k=0}^{\infty} (-1)^k \frac{1}{k!}(k+\frac{1}{2})^{k-(1/2)} e^{-[k+(1/2)]} \left(\frac{\mu h}{T_1}\right)^{2k+1}, & \mu h \leq T_1 \\ \mu \left[1 - \pi^{-3/2} \int_0^\infty \frac{dt}{t} \sin(\pi t) e^{-t(\ln t - 1)} \left(\frac{T_1}{\mu h}\right)^{2t} \Gamma(t+\frac{1}{2})\right], & T_1 \leq \mu h \ll D. \end{cases} \quad (107)$$

This expression is valid over the entire range of the energies and we may read off the asymptotics:

In the ultra-violet,

$$\mathcal{M}^i \to \mu \left[1 - \frac{1}{2\ln\frac{\mu h}{T_1}} + \frac{\ln 2}{\ln^2 \frac{\mu h}{T_1}} - \frac{\ln\frac{\mu h}{T_1}}{2\ln^2 \frac{\mu h}{T_1}}\right] + \cdots \quad (108)$$

which is the magnetization of a free spin weakly interacting with the conduction band. This is in accord with our previous conclusions.

In the infra-red[1],

$$\mathcal{M}^i \to \frac{\mu^2}{\pi T_0} h \quad (109)$$

indicating a screened impurity.

This magnetic crossover also figures in another interesting quantity, the scattering phase shift of an electron on the Fermi surface off the impurity as a function of the magnetic field, $\delta_0(h)$ [18]. This quantity determines the transport relaxation time $\tau$

$$\frac{1}{2\tau} = -c\sin^2\delta_0(h) \quad (110)$$

and hence the magnetoresistance

$$\rho(h) = \frac{3c}{e^2 \mathcal{D}_0} \sin^2 \delta_0(h). \quad (111)$$

Here $c$ is the impurity concentration, and $\mathcal{D}_0$ the density of states at the Fermi level.

To calculate $\delta_0(h)$ we add an electron to the system which is at its lowest energy state in the presence of a magnetic field, $|\Omega_h>$, creating a doublet, a one-"hole" state $|k,a;\Omega_h>$. Denoting by $E_o(N^e, S)$ the lowest energy for a system with $N^e$ electrons and a total spin $S$, we have that the energy of the state $|k, a; \Omega_h>$ with $k$ on the Fermi-level is

$$E(k, \pm 1/2; \Omega_h)|_{k=k_F(h)} = E(N^e + 1, S \pm 1/2) \tag{112}$$

where $S$ is the spin corresponding to the field $h$. On the other hand we may add the electron far from the impurity, hence $E(k, \pm; \Omega_h) = k + E(\Omega_h)$ hence only those values of $k$ are allowed (on a finite ring $L$) which satisfy

$$k = E(k, \pm 1/2; \Omega_h) - E(\Omega_h) = E(N^e + 1, S \pm 1/2) - E(N^e, S) \tag{113}$$

By studying the deviation of the allowed incident momentum $k$ from free values we can determine, as before, the scattering phase shift,

$$\delta_{0,\pm}(h) = \delta_{0,\pm}(k_F(h)) = \frac{\pi}{2}[1 \pm M^i(h)/\mu]. \tag{114}$$

where the sign indicates whether the electron spin projection is parallel or antiparallel to $S$.

The phase shift reaches its unitarity limit, $\delta_0 = \pi/2$, at low magnetic fields and falls to zero logarithmically as the fields are increased. The corresponding behavior in the resistivity (albeit as a function of the temperature) was the experimental measurement that ignited the interest in the model.

We have explored the crossover as a function of temperature, magnetic field, and momentum concluding consistently that all infrared physics can be described by a strong coupling fixed point that produces a Fermi liquid. Let us turn now to discuss it further.

*The strong coupling fixed point.*

In the infra-red regime the system can be viewed as a gas of quasi particles scattering off a non-magnetic potetial center characterized by the phase shift $\delta(p)$, and weakly interacting among themselves. These are the spinons interacting via scattering matrix elements $S^s$ and $S^t$ and undergoing a phase shift $\delta_0(p)$ upon passing the screened impurity potential. The localized potential is due to a singlet formed by the impurity *strongly* interacting with the electrons in the infra-red and effectively capturing an electron to form a local singlet. In other words, from Friedel's sum rule we can argue that $\delta_0(p = 0) = \pi/2$ indicates an enhanced density of states around the impurity, tantamount to an electron (or its spin content) captured there, forming a singlet. Virtual transitions are then responsible for inducing the scattering of the spinon off the screened-impurity.

We are immediately led to the following expession (valid at low temperatures) for the impurity free energy

$$F^i = -2T \int_0^\infty \frac{dp}{2\pi} \delta'(p) \ln(1 + e^{-\frac{p}{T}}) \to -\frac{\pi}{12} \frac{T^2}{T_0} \qquad (115)$$

as sum over spinons (hence the lower limit of the integral is zero). Here $\frac{1}{\pi}\delta'(\epsilon) = \mathcal{D}^i = \frac{1}{\pi} \frac{2T_o}{(2T_0)^2 + \epsilon^2}$ is the impurity part of the one-particle density of states. (An equivalent way to obtain it is from $\mathcal{D} = \frac{\partial \nu}{\partial \epsilon} = \sigma_0(\Lambda)/\left(\frac{d\epsilon}{d\Lambda}\right)$)

More generally, Nozieres[19] has argued that the physics around this fixed point can be captured by a Landau expansion of the phase shift

$$\delta_0(\epsilon) = \delta_0 - \alpha\epsilon \pm \phi^a 2M^e \qquad (116)$$

with $\delta_0 = \frac{\pi}{2}$ from strong coupling arguments, and $\alpha$ and $\phi^a$ phenomenological parameters. These parameters determine all low temperature proprties of the model. In particular, resistivity specific heat and magnetic susceptibility are given by

$$\begin{aligned} \rho(T) &= \rho_0 \left(1 - (\alpha\pi T)^2\right) \\ \frac{C^i}{C^e} &= \frac{\alpha}{\pi \mathcal{D}_0} \\ \frac{\chi^i}{\chi^e} &= \frac{\alpha}{\pi \mathcal{D}_0} + \frac{2\phi^a}{\pi} \end{aligned}$$

where $\mathcal{D}_0$ denotes the density of states, here $\mathcal{D}_0 = \frac{L}{\pi}$. Wilson's Ratio is given by $R = 1 + 2\mathcal{D}_0 \phi^a/\alpha$. Assuming the Kondo singularity is tied to the Fermi level and only antiparallel spin interactions Nozieres concluded $2\mathcal{D}_0 \phi^a/\alpha = 1$ and hence $R = 2$ in accord with Wilson's result.

The explicit values of these parameters can be read off eq(45) and eq(114),

$$\alpha = \frac{1}{2T_0}, \quad \phi^a = \frac{\pi}{4} \frac{M^i}{M^e} = \frac{\pi}{4} \frac{\chi^i}{\chi^e} = \frac{\pi}{4T_0 L}.$$

yielding a complete characterization of the fixed point.

*The Kondo Problem.*

We now return to consider the main question: the characterization of the crossover by universal numbers [5][20]. Begin by introducing the high magnetic field scale $T_h$, parametrizing the weak coupling regime in the $h - T$ plane, by requiring that no term of $O[\ln^{-2}(h/T_h)]$ should appear in the expansion. One finds from eq(108),

$$\mathcal{M}^i \to \mu \left[ 1 - \frac{1}{2 \ln \frac{\mu h}{T_h}} + \frac{\ln \ln \frac{h}{T_h}}{4 \ln^2 \frac{\mu h}{T_h}} + 0(\frac{1}{\ln \frac{\mu h}{T_h}})^3 + \cdots \right],$$

with

$$T_h = \left( \frac{\pi}{e} \right)^{\frac{1}{2}} T_0. \tag{117}$$

The number $W_h = \frac{T_h}{T_0} = \sqrt{\frac{\pi}{e}}$ [3], defined as the ratio of the ultraviolet scale $T_h$ to the infrared scale $T_0$, relates to both fixed points and its computation requires the construction of the model on all energy scales.

Can one similarly understand the crossover as a function of the temperature? One wishes now to calculate $W = \frac{T_k}{T_0}$, where $T_k$ was defined at the weak coupling regime and characterizes the corrections to Curie's law, while $T_0$ defined at strong coupling sets the screening scale. We have encountered the number previously in the form $W = e^a$ in the discussion after eq(94), but were unable to determine it from the infinite set of coupled thermodynamic equations. We proceed to calculate it now, appealing to the idea of universality. Expressing $W$ as

$$W = \frac{T_k}{T_0} = \frac{T_k}{T_h} \frac{T_h}{T_0}, \tag{118}$$

we note that the universal number $T_k/T_h$ is completely defined in the weak coupling regime, and therefore can be calculated *exactly* by means of perturbation theory. We find,

$$\frac{T_k}{T_h} = 2\beta\gamma\sqrt{\pi}e^{-\frac{9}{4}} \tag{119}$$

where

$$\ln \beta = \int_0^1 dx (1-x^2) x \left( \frac{\pi^2}{\sin^2 \pi x} - \frac{1}{x^2} \right) = \ln \frac{e^{5/2}}{2\pi} \quad [21]$$

$$\ln \gamma = \mathcal{C} = \text{Euler's constant}.$$

Hence
$$\frac{W}{4\pi} = \frac{e^{(\mathcal{C}-1/4)}}{4\pi} = 0.102676... \quad [3]. \qquad (120)$$

This number was first calculated by RG techniques to be

$$\frac{W}{4\pi} = 0.1032 \pm 0.0005 \quad [5], \qquad (121)$$

in good agreement with the analytic result. Note that the three numbers $T_K/T_h$, $T_h/T_0$, $T_K/T_0$ were computed within three different constructions of the model. Their accord is due to the renormalizability of the model and the universality resulting from it.

We have seen that the Kondo model flows toward an infra-red behaviour that is Fermi-liquid in character. We wish to describe now some generalizations where this behaviour is modified.

*The multichannel Kondo model.*

The model was introduced by Nozieres and Blandin [22] to include the orbital structure of the impurity,

$$H = -i \int \psi^*_{a,m}(x)\partial_x \psi_{a,m}(x)dx + J\psi^*_{a,m}(0)\vec{\sigma}_{ab}\psi_{b,m}(0) \cdot \vec{S} \qquad (122)$$

here $m = 1, ..., f = 2l+1$ is the orbital channel (or flavor) index and the spin operator $S^i$ is in spin-$S$ representation of SU(2). In the hamiltonian the values of $f$ and $S$ are unrestricted. To describe a magnetic impurity $f = 2S$, but other non-magnetic applications of the model exist with other values of spin and flavor [23].

The nature of the infra-red fixed point depends on those values [22]: for $f \leq 2S$ the coupling $J$ flow to infinity leading to a screened impurity in the case $f = 2S$, and to a partially screened impurity $S' = S - \frac{1}{2}$ in the case $f < 2S$. The strong coupling fixed point becomes unstable when $f > 2S$ and the infra-red physics is then controled by a new, finite coupling fixed point. This new fixed point is expected to describe non Fermi-liquid behavior.

The model is integrable but presents some challenge as to the handling of the cut off scheme. The reason is that the effect of flavor enters through the Fermi statistics; it allows $f$ spin-up electrons to interact at a time with a spin down impurity. The cut off procedure, if improperly handled, may smear out the local interaction and lead to wrong results. At the same time it must respect integrability.

One procedure [25], valid only for $f = 2S$, uses a version of the Anderson model as a cut-off. A more gereral approach is not to linearize the spectrum ab initio, but to maintain some curvature which is removed at the same time as the cut-off. This way one obtains a set of Bethe- Ansatz equations coupling spin and flavor degrees of freedom [24], and a remarkable effect takes place when the cut-off is taken to infinity: the flavor singlet sector separates from the rest of the Hilbert space leading to a new "fused" set of BAE describing the interaction of effective spin-$f/2$ electrons with the spin-$S$ impurity. This dynamical fusion to form the spin-$f/2$ electron complexes, captures the full spin content of the model and underlies the appearance of the non-Fermi-liquid behavior.

We shall skip further details of the Bethe-Ansatz construction and write down directly the BAE for the model,

$$-\prod_{\delta=1}^{M} \frac{\Lambda_\delta - \Lambda_\gamma + ic}{\Lambda_\delta - \Lambda_\gamma - ic} = \left(\frac{\Lambda_\gamma - 1 - ifc/2}{\Lambda_\gamma - 1 + ifc/2}\right)^{N^e} \left(\frac{\Lambda_\gamma - icS}{\Lambda_\gamma + icS}\right), \qquad (123)$$

and the resulting thermodynamic equations in the scaling limit,

$$\ln \eta_n = -2\delta_{n,f} + G[\ln(1 + \eta_{n+1}) + \ln(1 + \eta_{n-1})], \qquad (124)$$

noting that the effect of flavor is to move the driving term to the $f$th row of the coupled set of equations. The asymptotic conditions on the equations are,

$$[n+1]\ln(1+\eta_n) - [n]\ln(1+\eta_{n+1}) \to 2\mu h/T, \quad n \to \infty. \qquad (125)$$

Having solved the equations one computes the spin-$S$ impurity free energy from $\eta_{n=2S}$,

$$F_{S,f}^i = -\frac{T}{2\pi} \int d\zeta \frac{\ln[1 + \eta_{2S}(\zeta, \frac{h}{T})]}{\cosh[\zeta - \ln \frac{T_0}{T}]}. \qquad (126)$$

The same set of equations determines the free energy for all values of the spin $S$. We shall see below that the overscreened solutions $\eta_n$, $n < f$ differ in character from those with $n \geq f$. The electronic properties are still calculated from $\eta_f$. Hence we may expect Fermi-liquid behaviour only in the case $f = 2S$.

The functions $\eta_n$ are analytic monotonically decreasing in $\zeta$ for all $n$ and tending to finite limlts $\eta^\pm$ as $\zeta \to \pm\infty$. The limits are given by

$$\eta_n^- = \frac{\sinh^2[(n+1)\mu h/T]}{\sinh^2(\mu h/T)}, \quad n = 1, 2, ..., \tag{127}$$

and

$$\eta_n^+ = \begin{cases} \frac{\sin^2[(n+1)\pi/(f+2)]}{\sin^2[\pi/(f+2)]} - 1, & \text{for } n < f \\ \frac{\sinh^2[(n+1-f)\mu h/T]}{\sinh^2(\mu h/T)} - 1, & \text{for } n \geq f \end{cases} \tag{128}$$

Consider now the high-temperature properties of the model. They are determined, for a spin-S impurity, by behavior of $\eta_{2S}$ in the limit $\zeta \to -\infty$. As in the one-flavor case we have discussed in detail thus far this limit is approached with power corrections leading to

$$F^i \to -T \ln \frac{\sinh(2S+1)\mu h/T}{\sinh \mu h/T} + \frac{B_1}{\ln T/T_0}. \tag{129}$$

This is the weak-coupling regime, governed by the fixed point at $J = 0$. The free energy is that of an isolated spin $S$ up to the usual logarithmic corections. The nature of this fixed point is unaffected by the introduction of flavor degrees of freedom.

On the other hand flavor affects significantly the low temperature properties of the model. These are determined by the behavior of of $\eta_{2S}$ in the limit $\zeta \to +\infty$. The nature of the limit and the approach to it depend on the flavor degrees of freedom.

The *underscreened* functions, $\eta_n$, $n = 2S > f$, attain their asymptotic limit $\eta_n^+$ with power corrections and we have,

$$F^i \to -T \ln \frac{\sinh(2S+1-f)\mu h/T}{\sinh \mu h/T} + \frac{C_1}{\ln T/T_0} + \ldots \text{ as } T \to 0. \tag{130}$$

This is the free energy of a spin $S' = S - \frac{1}{2}f$. In other words, the impurity spin is partially screened. The approach to the limiting value is logarithmic in the temperature indicating a fixed point at $J = \infty$.

The function $\eta_n$, $n = f$, describes complete screening and, as $T \to 0$, $F^i \sim (D_1 T^2 + Sh^2)/T_0$. $A_1, B_1, C_1$, and $D_1$ are numerical constants.

A new situation arises when we consider the *overscreened* functions, $\eta_n$, $n < f$. These functions approach their limit values exponentially, $\eta_n(\zeta) \to \eta_n^+ + c_n e^{-\tau\zeta}$, $\zeta \to \infty$, $n < f$, with $0 < \tau < 1$. Hence we may apply the operator $G$ directly to the asymptotic form $G\, e^{-\tau\zeta} = e^{-\tau\zeta}/2\cos(\pi\tau/2)$ and the thermodynamic equations reduce to an algebraic recursion relation for the coefficients $b_n = c_n/(1 + \eta_n^+)$,

$$\frac{\eta_n^+}{1+\eta_n^+}(b_{n+1} + b_{n-1}) = \lambda b_n, \quad n = 1, ..., f-1 \tag{131}$$

with the boundary conditions $b_0 = b_f = 0$. The eigenvalue $\lambda$ is related to $\tau$ via $\lambda = 2\cos\pi\tau/2$.

The solution to the resursion relation is

$$b_n = \frac{\sin[(n+2)\pi/(f+2)]}{\sin[n\pi/(f+2)]} \tag{132}$$

with the smallest eigenvalue

$$\tau = \frac{4}{f+2} \tag{133}$$

(In the two channel case more care needs to be taken since $\tau = 1$ and applying $G$ to the asymptotic form generates logarithmic corrections.)

The impurity free energy at low temperature takes the form,

$$F_{f,S}^i = -\frac{1}{2}T\ln(1+\eta_{2S}^+) - \frac{b_{2S}}{2\cos(\pi\tau/2)}\,T(T/T_0)^\tau + \ldots \tag{134}$$

from which we conclude that the infra-red physics is dominated by a new non-trivial fixed point generating power laws behavior for the specific heat and susceptibility

$$C^i \sim \left(\frac{T}{T_0}\right)^{\frac{4}{f+2}} \tag{135}$$

$$\chi^i \sim \frac{1}{T}\left(\frac{T}{T_0}\right)^{\frac{4}{f+2}} \tag{136}$$

A similar power law arises when one considers low-energy scattering of the spinon off the impurity.

These power laws allow the identification of the infra-red fixed point hamiltonian as the $SU(2)$ level-$f$ WZW conformal field theory where the combination $\frac{1}{f+2}$ is the coefficient of the hamiltonian in the Sugawara-form[27]. This coefficient determines essentially the conformal dimensions, and hence the the asymptotic form of the dynamic correlation functions [26][28]. In particular, the primary field $\Phi_1(z)$ has the dimension $\Delta = 2/(f+2)$ and can be identified as the spin operator from eq(136).

Actually, the appearance of some WZW model as the fixed point hamiltonian can be expected from symmetry arguments since it is the simplest model to incorporate scale invariance and $SU(2)$ symmetry. Detailed calculations are required to determine to which level (here level $f$) the full model will flow to in the infra-red.

The zero temperature entropy can immediately be read off eq(134),

$$\mathcal{S} = -\frac{1}{2}\ln(1 + \eta_{2S}^+) = -\ln\frac{\sin[(2S+1)\pi/(f+2)]}{\sin[\pi/(f+2)]} \qquad (137)$$

and is the logarithm of a fractional number! The appearance of fractional entropy is due to the solitonic nature of the excitations[10], and is general in conformal field theory [29].

The fixed point can also be approached as a function of the magnetic field at zero temperature. The thermodynamic equations collapse into a single equation which describing the maximum spin excitations above the ground state, consisting of an $f$-string configuration in the presence of a magnetic field. For small magnetic fields, however, we can deduce the asymptotic behavior from the ground state solution. Denoting by $\sigma_0^i$ the impurity contribution to the ground-state density of $f$ strings, we have for the impurity magnetization fields $h$

$$M^i(h \sim 0) = (\mu/2) \int_{-\infty}^{\ln h/T_0} dx\ \sigma_0^i(x). \qquad (138)$$

The density $\sigma_0^i$ is given in Fourier space by

$$\tilde{\sigma}^i(p) = \begin{cases} \sinh(SJP)/[2\cosh(Jp/2)\sinh(fJp/2)], & f > 2S \\ \exp[(f/2 - S)J|p|]/[2\cosh(Jp/2)], & f \leq 2S. \end{cases}$$

In the limit $h \to 0$, the magnetization is dominated by the properties of $\tilde{\sigma}^i(p)$ at $p = 0$. While for $f < 2S$, $\tilde{\sigma}^i_0(p)$ is discontinuous at $p = 0$ leading to $M^i_{(h\sim 0)} = 2\mu(S - \frac{1}{2}f) + 0(\ln h/T_0)$, for $f \geq 2S$ the transform is analytic in $p$ so that $M^i(h)$ is controled by the pole at $p = -2i/f$. Hence

$$M^i(h) \sim \mu(h/T_0)^{2/f}, \; h \to 0 \tag{139}$$

leading to the critical exponent $\delta = f/2$ (and logarithmic corrections for $f = 2$).

We have dealt thus far with the theoretical aspects of the Kondo model. It is, however, a model that describes an experimentally realizable system, and the mathematical structure we discussed can be confronted with reality. It is found that the theory provides a remarkably good fit for a large body of experimental data [30] with no adjustable parameter except the scale $T_k$. This is more remarkable still in view of the simplicity of the model, which nevertheless captures the essential physics of a rather complex system.

# References

[1]     N. Andrei, Phys. Rev. Lett. **45**, 379 (1980).

[2]     V. M. Filyov, A. M. Tsvelick and P. B. Wiegman, Phys. Lett. A **81** (1981).

[3]     N. Andrei and J.H. Lowenstein, Phys. Rev. Lett. bf 46,356 (1981).

[4]     P. B. Wiegmann, JETP Lett. **31**, 392 (1980).

[5]     K. G. Wilson, Rev. Mod. Phy. **47**, 773 (1975).

[6]     N. Andrei, K. Furuya and J.H. Lowenstein, Rev. Mod. Phys. **55**, 331 (1983).

[7]     A. M. Tsvelik and P. B. Wiegmann, Adv. in Phys.**32**, 453 (1983).

[8]     N. Andrei, J. H. Lowenstein, Phys. Lett. **B 91**, 401 (1980).

[9]     V. E. Korepin, Theo. Math. Phys.**41**,953 (1979).

[10] P. Fendley, Phys. Rev. Lett. **71**, 2485 (1993).

[11] V. Kurak and J. A. Swieca, Phys. Lett. **B 82**, 289 (1979)

[12] M. Tahahashi, Prog. Theo. Phys. **47**, 69 (1972).

[13] C. Destri and J. H. Lowenstein, Nuc. Phys. **B 205**, 369 (1982).

[14] F. Woynarovich, J. Phys **A 15**, 2985 (1982).

[15] H. Bethe, Z. Physik **71**, 205 (1931).

[16] C. N. Yang and C. P. Yang, J. Math. Phys. **10**, 1115 (1969)

[17] M. Gaudin, Phys. Rev. Lett.**26**, 1301 (1971).

[18] N. Andrei, Phys. Lett. **A 87**, 299 (1981).

[19] P. Nozieres, J Low Temp. **17**, 31 (1974)

[20] P. Nozieres, in Proceedings of LT 14, edited by M. Krusius and M. Vuorio, p.339 (1975)

[21] A. Hewson, private communication

[22] P. Nozieres and A. Blandin, J. Phys. (Paris) **41**, 193 (1980).

[23] A. Zawadowski and N. Vladar, Sol. Stat. Com. **35** (1980). D. L. Cox, Phys. Rev. Lett. **59**, 1240 (1987).

[24] N. Andrei and C. Destri. Phys. Rev. Lett. **52**, 364 (1984).

[25] A. M. Tsvelick and P. W. Wiegmann, Z. Phys. **54** , 201 (1984).

[26] I. Affleck and A. Ludwig, Nuc. Phys. **B352**, 849 (1991)

[27] P. Ginsparg, in *Fields, Strings, and Critical Phenomena*, Eds. E Brezin and J. Zinn-Justin, North Holland, Amsterdam, 1990.

[28] I. Affleck and A. Ludwig, Nuc. Phys. **B360**, 641 (1991)

[29] I. Affleck and A. Ludwig, Phys. Rev Lett. **67**, 161 (1991).

[30] T. Rajan, J. H. Lowenstein and N. Andrei, Phys. Rev. Lett.**49**, 497 (1982).

Lecture 4: written with Andres Jerez
# The Hubbard Model

In this lecture we analyse the spectrum of the Hubbard model. Many of the results have been derived in the past by Lieb and Wu [1], Ovchinikov [2], Coll [3], Shiba [4], Takahashi [5], Woynarovich [6], and Kluemper et al.[7]. We systematize the results, occasionally correct them, and derive new ones [8]. The latter category includes results concerning the spectrum and scattering of the excitation away from half filling, for the repulsive and attractive interaction. While these lecture notes were written up Korepin and Esseler also derived some of the results presented here [9]. Their work concentrates on the half filled case and emphasizes the charge $SU(2)$ symmetry that appears in this case [10]. This symmerty is explicitly broken away from half filling but we shall find some remnants of it.

The analysis in this lecture is similar to the one carried out in Lecture 3, but is much more involved due to the coupling of the Bethe-Ansatz equations. We find that the fundamental excitations are, as in Lecture 3, spin-$\frac{1}{2}$ uncharged *spinons* and spinless charged *holons* and *antiholons*. Indeed, any $SU(2)$ invariant integrable model characterized by the R-matrix discussed in Lecture 2 will have excitations carrying the same spin quantum numbers, since these are determined by counting. Also the spinon S-matrix, computed from a counting argument, has essentially the same form as we encountered earlier. The dynamics, on the other hand, captured by the function $\alpha_j = \alpha(k_j)$ determines the dispersion of the excitations and depends on the particular form of the interaction. In the repulsive Hubbard model we shall find that spin excitations are always gapless and have a spin Fermi momentum $k_F^s = \frac{\pi}{2}n$, where $n = N/L$ is the particle density. The charge sector is composed of gapless (holon-antiholon) as well as of gapful excitations (holon-holon). The gapless charge excitations are present only as long as the band is less than half filled, and their charge Fermi momentum is $k_F^c = \pi n$. When $n = 1$ no gapless charge excitations survive and the system ungergoes a metal-insulator transition.

In the attractive model the situation is somewhat reversed. Gapless charge excitations always exist: the holon-antiholon excitation is gapless but disappears from the spectrum at half filling, while the holon-holon excitation becomes gapless at this point. On the other hand the spin excitations are always gapful.

The nature of the spin gap can be easily understood by comparing eq(44) and eq(46) in Lecture 2. The couplings of the (antiferromagnetic) Kondo model and the attractive Hubbard are of the same sign. But as we saw in Lecture 3, the effective spin coupling then flows to the strong coupling fixed point generating a spin scale for the Kondo model, and a spin gap for the attractive Hubbard model. In the repulsive model the effective spin coupling flows to zero and the spin excitations are gapless.

The same argument applies to the charge gap at half filling due to the appearance of the charge $SU(2)$ symmetry, but the effective couplings are reversed; thus a charge gap opens in the repulsive case, but not in the attractive case. Away from half filling the charge symmetry is broken to $U(1)$ and gapless charge excitations are present (holons-antiholons).

The $Z_2$ particle-hole transformation [1] maps, at half filling, spinons making up the singlet excitation in the repulsive case (always gapless) to the objects that make up the the holon-holon excitation in the attractive case (gapless at half filling). This is quite remarkable in view of the very different structure of the wave functions of the respective states. Not only are the energy and momentum the same but also the scattering matrices. To complete the corespondance, the triplet of the repulsive model would be matched not with the holon-antiholon of the attractive model (which disappears from the spectrum at half filling) but with states obtained by removing/adding two electron to the system applying the operator $C^\pm$ mentioned in Lecure 1. As one moves away from half filling one can follow explicitly the breaking of the symmetry.

Spin and charge sectors are coupled; spin excitations involve rearrangement of charge degrees of freedom and vice versa. In the low energy limit, however, we shall see that complete decoupling takes place, similar to the one occuring on all energy scales for the Kondo model.

Let us proceed now to the construction of the eigenstates. We consider a system of $N$ electrons moving on a chain of $L$ sites ($N$ and $L$ are assumed even, and the lattice spacing is 1 in some unit). We denote by $k_F = \frac{\pi}{2}n$ the Fermi momentum of the non-interacting system. For the Hubbard model $\alpha_j = \sin k_j$, and $c = \frac{u}{2}$, with $u = \frac{U}{t}$, see Lecture 2. Therefore the Bethe-

Ansatz equations take the form:

$$e^{ik_j L} = \prod_{\delta=1}^{M} \frac{\Lambda_\delta - \sin k_j - i\frac{u}{4}}{\Lambda_\delta - \sin k_j + i\frac{u}{4}} \qquad (1)$$

and

$$-\prod_{\delta=1}^{M} \frac{\Lambda_\delta - \Lambda_\gamma + i\frac{u}{2}}{\Lambda_\delta - \Lambda_\gamma - i\frac{u}{2}} = \prod_{i=1}^{N} \frac{\Lambda_\gamma - \sin k_i - i\frac{u}{4}}{\Lambda_\gamma - \sin k_i + i\frac{u}{4}}, \qquad (2)$$

for a state with $M$ down spins and $N - M$ up spins. The quantum number $M$ labels both the total spin and its $z$-component, as discussed earlier, $S_z = \frac{1}{2}(N - 2M) = S$.

Upon taking the logarithm of eqs(1,2) we obtain:

$$Lk_j = 2\pi n_j + \sum_{\delta=1}^{M} \Theta(2\sin k_j - 2\Lambda_\delta), \quad j = 1,...,N \qquad (3)$$

$$\sum_{j=1}^{N} \Theta(2\Lambda_\gamma - 2\sin k_j) = -2\pi I_\gamma + \sum_{\delta=1}^{M} \Theta(\Lambda_\gamma - \Lambda_\delta), \qquad \gamma = 1,...,M \qquad (4)$$

where

$$\Theta(x) = -2\tan^{-1}(2x/u), \qquad -\pi \leq \Theta < \pi \qquad (5)$$

The quantum numbers $\{n_j, I_\gamma\}$ label the states, and are integers or half odd integers. The charge quantum numbers $\{n_j\}$ are integers if $M$ is even and half odd integers if $M$ is odd. They are defined modulo $L$, and will take values between upper and lower bounds $\mathcal{N}^{\pm} = \pm(L-1)/2$. The spin quantum numbers $\{I_\gamma\}$ are integers when $N - M - 1$ is even, and half odd integers when it is odd, and are subject to the same restriction as before,

$$I^-(N, M) = -(N - M - 1)/2 \leq I_\gamma \leq (N - M - 1)/2 = I^+(N, M) \qquad (6)$$

After solving the equations for an (allowed) configuration $\{n_j, I_\gamma\}$ the energy and momentum of the that state are

$$E = \sum_{j=1}^{N} -2t \cos k_j, \qquad (7)$$

$$P = \sum_{j=1}^{N} k_j = \frac{2\pi}{L}(\sum_j n_j + \sum_\gamma I_\gamma), \qquad (8)$$

where the last expression for the momentum is obtained from eq(3)and eq(4).

The solutions, $\{k_j, \Lambda_\gamma\}$, of eqs(1,2) fall, in the thermodynamic limit, into patterns described by the *string hypothesis* [5]:

1. Real $k_j$

2. $\Lambda$ $n$-string, of the form $\Lambda_j^{(n)} = \Lambda^{(n)} + i\frac{c}{2}(n + 1 - 2j), i = 1, 2, ..., n$.

3. $k - \Lambda$ $n$-string, consisting of a $\Lambda$ $n$-string and $n$ complex $k$-pairs $k_j^{(n),\pm}$, each centered around $\Lambda$ string position, $\sin k_j^{(n),\pm} = \Lambda_j^{(n)} \pm i\frac{u}{4}$.

In the Appendix we present a more detailed discussion of some of the solutions listed.

These states form a complete set of $4^L$ eigenstates [11]. Their degeneracy must be properly taken into account. We saw previously that the Bethe-states are highest weight states of the spin $SU(2)$ symmetry group, and the rest of the multiplet can be obtaines by repeated action of the spin lowering operator $S^-$. Less obvious but still true is the fact that the Bethe states are also the highest weight states of the charge $SU(2)$ group and therefore the rest of the multiplet can be obtained by the action of the $C^-$ operator.

We shall be mainly interested in the low lying excitations; the general structure of the solutions will interest us only towards the end of the lecture, when we wish to sum over all eigenstates to obtain the free energy.

We turn now to determine which of the possible solutions corresponds to the lowest energy eigenvalue. We shall find that the answer depends on the sign of the coupling: The ground state is composed of 1-strings (real $k_j$) in the repulsive case, and of $k$ 2-strings (rather $k-\Lambda$ 1-strings in the terminology of the string hypothesis) in the attractive case. Still, the excitations above these ground states so differing in structure are spinons and holons carrying the same quantum numbers and, in particular at half filling, having the same dispersion laws, and interacting via the same S-matrices.

We consider now each case separately.

### The repulsive Hubbard model.

The ground state solution consists of real $k$'s and real $\Lambda$'s. In the thermodynamic limit we shall be interested in the (real) solution-densities $\rho(k)$ and $\sigma(\Lambda)$, defined by;

$$\begin{aligned} L\rho(k_j) &= 1/(k_{j+1} - k_j) \\ L\sigma(\Lambda_\gamma) &= 1/(\Lambda_{\gamma+1} - \Lambda_\gamma) \end{aligned}$$

The $k$-solutions are distributed between the *pseudo* charge Fermi-momenta $-Q$ and $Q$ with $Q \leq \pi$, since they are defined modulo $2\pi$, while the (real) $\Lambda$ solutions will be distributed between the *pseudo* spin Fermi momenta $-B$ and $B$, $B \leq \infty$. The integration limits are to be determined from

$$\int_{-Q}^{Q} \rho(k)dk = N/L \tag{9}$$

$$\int_{-B}^{B} \sigma(\Lambda) = M/L \tag{10}$$

and can vary with the state, even as the number of electrons is held fixed.

In terms of the densities the energy and momentum become

$$E = -2tL \int_{-Q}^{Q} \rho(k) \cos k dk \tag{11}$$

$$P = L \int_{-Q}^{Q} \rho(k) k dk \tag{12}$$

The ground state.

We begin by identifying the ground state configuration $\{n_j^0, I_\gamma^0\}$. We shall find that the ground state is a singlet, $M_o = N/2$, given by a configuration symmetrically arranged around zero and as closely packed as possible:

$$n_{j+1}^o - n_j^o = 1$$
$$I_{\gamma+1}^o - I_\gamma^o = 1$$

with the $n_j$ and $I_\gamma$ filling all the slots from $n^+$ to $n^-$ and from $I^+$ to $I^-$ respectively,

$$I^- \leq I_\gamma \leq I^+, \qquad I^\pm = \pm(N/2 - 1)/2.$$
$$n^- \leq n_j \leq n^+, \qquad n^\pm = \pm(N - 1)/2.$$

We have tacitly made the choice that $N/2$ is an odd integer.

The densities corresponding to this configuration of quantum numbers satisfy the integral equations (their derivation is the same as the one outlined

in Lecture 3)

$$\rho_o(k) = \frac{1}{2\pi} + \cos k \int_{-B_o}^{B_o} d\Lambda \sigma_o(\Lambda) K_1(\sin k - \Lambda) \tag{13}$$

$$\sigma_o(\Lambda) = \int_{-Q_o}^{Q_o} dk \rho_o(k) K_1(\sin k - \Lambda) - \int_{-B_o}^{B_o} d\Lambda' \sigma_o(\Lambda') K_2(\Lambda - \Lambda') \tag{14}$$

Here $\rho_o(k)$, $\sigma_o(\Lambda)$, $Q_o$ and $B_o$ are respectively the densities and integration limits for the ground state, and we encounter the kernels from previous lecture,

$$K_n(x) = \frac{1}{\pi} \frac{n\frac{u}{4}}{(n\frac{u}{4})^2 + x^2} = -\frac{1}{\pi} \frac{d}{dx} \Theta\left(\frac{2}{n}x\right), \qquad n = 1, 2... \tag{15}$$

with Fourier transform $\tilde{K}_n(p) = e^{-n\frac{u}{4}|p|}$.

The structure of the equations (13,14) will recur often, and it is convenient to introduce the notation,

$$\Phi_o = \varphi_o + \mathcal{M}(Q_o, B_o)\Phi_o \tag{16}$$

where we denote

$$\Phi(k, \Lambda) = \begin{pmatrix} \rho(k) \\ \sigma(\Lambda) \end{pmatrix}, \qquad \varphi_o = \begin{pmatrix} \frac{1}{2\pi} \\ 0 \end{pmatrix}$$

and the matrix of integral operators,

$$\mathcal{M}\Phi = \begin{pmatrix} 0 & \cos k \, K_1 \\ K_1 & -K_2 \end{pmatrix} \begin{pmatrix} \rho \\ \sigma \end{pmatrix} \tag{17}$$

given in more detail,

$$\mathcal{M}(Q, B)\Phi(k, \Lambda) = \begin{pmatrix} \cos k \int_{-B}^{B} K_1(\sin k - \Lambda)\sigma(\Lambda) d\Lambda \\ \int_{-Q}^{Q} K_1(\sin k - \Lambda)\rho(k) dk - \int_{-B}^{B} K_2(\Lambda - \Lambda')\sigma(\Lambda') d\Lambda' \end{pmatrix}.$$

To proceed we need to determine the Fermi-levels. We shall find that to minimize the energy we have to set $B_0 = \infty$ irrespective of the density of electrons $n$ (which determines the value of $Q_0$.) The state is a singlet as follows by intergating eq(14) over $\Lambda$, $\int_{-\infty}^{\infty} \sigma_0(\Lambda) d\Lambda = \frac{M_0}{L} = \frac{1}{2}(\frac{N}{L}) = \frac{1}{2}\int_{-Q_o}^{Q_o} \rho_o(k) dk$.

Finite values of $B$ correspond to non-vanishing ground state spin and and will occur in the presence of a magnetic field. We shall return to this point at the end of the lecture.

As $B_0 = \infty$ we may solve the spin-equation, eq(14), for any filling, by means of the Fourier transform,

$$\sigma_0(\Lambda) = \frac{1}{u} \int_{-Q_0}^{Q_0} dk \rho_0(k) \text{sech}(\frac{2\pi}{u}(\Lambda - \sin k)). \tag{18}$$

and derive an integral equation for $\rho_0$ substituting (18) in (13),

$$\rho_0(k) = \frac{1}{2\pi} + \cos k \frac{4}{u} \int_{-Q_0}^{Q_0} dk' \rho_0(k') R\left(\frac{4}{u}(\sin k - \sin k')\right). \tag{19}$$

The function $R$, the resolvent of $K_2$, $(1+R)(1+K_2) = 1$, is given by[4],

$$\begin{aligned}
R(x) &\equiv \int_{-\infty}^{\infty} \frac{dp}{2\pi} \frac{e^{ipx}}{1+e^{2|p|}} = \frac{1}{4\pi} \int_{-\infty}^{\infty} dt \frac{\text{sech}\left(\frac{1}{2}\pi t\right)}{1+(x-t)^2}, \\
&= \frac{1}{\pi} \sum_{n=1}^{\infty} (-1)^{n+1} \frac{2n}{x^2+(2n)^2}.
\end{aligned} \tag{20}$$

It is convenient to introduce the integral operator

$$\mathcal{K}(Q_0)\rho(k) = \int_{-Q_0}^{Q_0} dk' \left\{\delta(k-k') - \cos k \frac{4}{u} R\left(\frac{4}{u}(\sin k - \sin k')\right)\right\} \rho(k'), \tag{21}$$

in terms of which (19) becomes, suppressing the $Q_0$ dependence in the operator,

$$\mathcal{K}\rho_0(k) = \frac{1}{2\pi}. \tag{22}$$

We need to determine $Q_0$. Consider first the ground state at *half filling*, ($N = L$). The quantum numbers $n_j$ assume all allowed values, in other words, $n^\pm = \mathcal{N}^\pm$ and therefore, also the $k$-solutions are distributed over the maximal range: $Q_o = \pi$, as can be immediately verified in the thermodynamic limit by integrating eq(13) over $k$. In this case we can derive explicit expressions for $\rho_o(k)$ and $\sigma_o(\Lambda)$,

$$\rho_0(k) = \frac{\cos k}{\pi} \int_0^\infty dp \frac{J_0(p) \cos(p \sin k)}{1 + e^{\frac{u}{2}|p|}}, \tag{23}$$

$$\sigma_0(\Lambda) = \frac{1}{2\pi} \int_0^\infty dp \frac{J_0(p) \cos(p\Lambda)}{\cosh(\frac{u}{4}p)} = \frac{1}{2\pi u} \int_{-\pi}^{\pi} dk \ \text{sech}\left(\frac{2\pi}{u}(\Lambda - \sin k)\right) \tag{24}$$

Hence the energy can be calculated [1]

$$\frac{E_0}{L} = -2t \int_{-\pi}^{\pi} dk \rho_0(k) \cos k = -4t \int_0^{\infty} dp \frac{J_0(p) J_1(p)}{p(1 + e^{\frac{u}{2}|p|})}. \tag{25}$$

The ground state away from half-filling is not accessible in closed form, except in some limits such as weak coupling, strong coupling or low density [4] [12]. In particular, in the strong coupling limit, $u \gg 1$ we have

$$2\pi \rho_o(k) = 1 + \frac{4}{u} \ln 2 \frac{Q_o}{\pi} \cos k, \tag{26}$$

from which the relation of $n = N/L$ and $Q_o$ can be determined

$$\pi n = Q_o[1 + \frac{4}{u\pi} \ln 2 \sin Q_o], \tag{27}$$

and the energy

$$E_o/L = -2t \left[ \frac{1}{\pi} \sin(\pi n) + \frac{2 \ln 2}{u} \left(\frac{\pi n}{\pi}\right)^2 \left(1 - \frac{\sin(2\pi n)}{2\pi n}\right) \right]. \tag{28}$$

In the half filled case, $n = 1$, this equation reduces to the ground state energy of the Heisenberg antiferromagnet with exchange coupling $J = 2t^2/U$. This relation between couplings can be obtained perturbatively.

In the general case one needs to solve the equations numerically [4]. The results clearly demonstrate the effects of correlations, which become most pronounced at strong coupling, or as $n = N/L$ tends to 1.

### Elementary excitations

Let us consider now excitations above the ground state. Excitations are created by varying the quantum numbers from their ground state configuration; *spinons* - by varying only the spin quantum numbers, $\{I_\gamma\}$, and *holons* - by varying only the charge quantum numbers $\{n_j\}$. We shall find that holons and spinons are interacting, and will compute their scattering matrix. In the low energy limit they decouple in a manner similar to the decoupling of holons and spinons in the Kondo model. We shall see that the low lying spectrum is linear in the momentum and therefore can be captured by a continuum effective hamiltonian: the g-ology model [13] (which is also integrable [14].)

**Spin excitations.**

These are obtained by varying the $\{I_\gamma^o\}$ sequence from its ground state configuartion (the attentive reader will note that the arguments, and words parallel those in Lecture 3).

*The triplet.* The simplest spin excitation (keeping $N$ fixed) is obtained by placing "holes" in the $\{I_\gamma\}$ sequence. Consider a state with $M = N/2 - 1$, namely a spin-1 state. As before, this corresponds to two "holes" in the sequence since the bound $I^\pm(N, M) = \pm(N - M - 1)/2$ takes the value $I^\pm = \pm N/4$. Hence there are $N/2 + 1$ slots for the spin quantum numbers $I_\gamma$ and $M = N/2 - 1$ of them, leaving two "holes". We shall see again that each "hole" corresponds to an excitation, a spinon, carrying spin-$\frac{1}{2}$.

We consider the configuration $\{I_\gamma\}$

$$I_{\gamma+1} - I_\gamma = 1 + \delta_{\gamma,\gamma_1} + \delta_{\gamma,\gamma_2}, \tag{29}$$

leaving the $\{n_j\}$ configuration unchanged. (Actually, the $n_j$ quantum numbers change from half-odd-integers to integers. This change generates extra terms in the energy and momentum which, after a careful analysis, turn out not to be relevant.)

The triplet configuration, eq(29), leads, by methods previously discussed, to the following equations for the densities $\rho(k)$ and $\sigma(k)$,

$$\Phi^t = \varphi^t + \mathcal{M}(Q, B)\Phi^t \tag{30}$$

where we denote

$$\Phi^t(k, \Lambda) = \begin{pmatrix} \rho(k) \\ \sigma(\Lambda) \end{pmatrix}, \quad \varphi^t = \begin{pmatrix} \frac{1}{2\pi} \\ -\frac{1}{L}\sigma^h(\Lambda) \end{pmatrix}$$

with

$$\sigma^h(\Lambda) = \delta(\Lambda - \Lambda_1^h) + \delta(\Lambda - \Lambda_2^h) \tag{31}$$

or expilicitly,

$$\rho(k) = \frac{1}{2\pi} + \cos k \int_{-\infty}^{\infty} d\Lambda \sigma(\Lambda) K_1(\sin k - \Lambda) \tag{32}$$

$$\sigma(\Lambda) = -\frac{1}{L}\sigma^h(\Lambda) + \int_{-Q}^{Q} dk \rho(k) K_1(\sin k - \Lambda) - \int_{-\infty}^{\infty} d\Lambda' \sigma(\Lambda') K_2(\Lambda - \Lambda') \tag{33}$$

We denoted by $\Lambda_1^h, \Lambda_2^h$ the hole positions corresponding to the quantum numbers omitted from the sequence. To determine them we use the same line of argument as in previous lecture; from the prescribed configurations $\{n_j, I_\gamma\}$ one finds the corresponding solutions $k_j, \Lambda_\gamma$, and defines the counting functions,

$$\nu(\Lambda) = \frac{1}{2\pi}\left(\sum_{\delta=1}^{M}\Theta(\Lambda - \Lambda_\delta) - \sum_{j=1}^{N}\Theta(2\Lambda - 2\sin k_j)\right) \quad \gamma = 1,...,M$$

$$\omega(k) = \frac{1}{2\pi}\left(Lk - \sum_{\delta=1}^{M}\Theta(2\sin k - 2\Lambda_\delta)\right) \qquad j = 1,...,N$$

The hole positions in the triplet state then satisfy

$$\nu(\Lambda_{1,2}^h) = I_{1,2}^h. \tag{34}$$

Note that the densities $\rho(k) = \rho(k; \Lambda_1^h, \Lambda_2^h, Q)$ and $\sigma(\Lambda) = \sigma(\Lambda; \Lambda_1^h, \Lambda_2^h, Q)$ depend also on the hole positions and on the charge pseudo Fermi-momentum, $Q$, which is determined from the condition $\int_{-Q}^{Q} dk\rho(k) = \frac{N}{L}$ and may differ from its ground state value $Q_o$. The $\Lambda$ integration limit stays at infinity since the corrections are of order $1/L$. (We shall usually spell out only the variables we need.)

Eqs(32,33) describe a spin-1 state, as can be deduced by integrating with respect to $\Lambda$. One finds, $\frac{N}{L} = \frac{M}{L} + \frac{M}{L} + \frac{1}{L}(1+1)$, that is $M = N/2 - 1$. Again, this consideration is just the counterpart in the thermodynamic limit of the counting argument presented above.

Since the holes contribute to the distribution equations terms of order $\frac{1}{L}$, we may write $\rho(k)$ and $\sigma(\Lambda)$ as

$$\rho(k) = \rho_o(k) + \frac{1}{L}\rho_1(k)$$

$$\sigma(\Lambda) = \sigma_o(\Lambda) + \frac{1}{L}\sigma_1(k)$$

with $\rho_o(k) = \rho_o(k; Q)$ and $\sigma_o(\Lambda) = \sigma_o(\Lambda; Q)$ satisfying the ground-state equations (13,14) for a given value of $Q$, while $\rho_1(k) = \rho_1(k; \Lambda_1^h, \Lambda_2^h, Q)$ and $\sigma_1(\Lambda) = \sigma(\Lambda; \Lambda_1^h, \Lambda_2^h, Q)$ satisfy the equations

$$\rho_1(k) = \cos k \int_{-\infty}^{\infty} d\Lambda \sigma_1(\Lambda) K_1(\sin k - \Lambda) \tag{35}$$

$$\sigma_1(\Lambda) = -\sigma^h(\Lambda) + \int_{-Q}^{Q} dk \rho_1(k) K_1(\sin k - \Lambda) - \int_{-\infty}^{\infty} d\Lambda' \sigma_1(\Lambda') K_2(\Lambda - \Lambda') \tag{36}$$

The structure of the equations is as before, $\Phi = \varphi + \mathcal{M}(Q,B)\Phi$, but now the inhomogeneous term is

$$\varphi = \varphi_o + \frac{1}{L}\begin{pmatrix} 0 \\ -\sigma^h(\Lambda) \end{pmatrix} = \varphi_o + \frac{1}{L}\varphi_1 \qquad (37)$$

and writing $\Phi = \Phi_o + \frac{1}{L}\Phi_1$ we derived a reduced equation for $\Phi_1$

$$\Phi_1 = \varphi_1 + \mathcal{M}(Q,B)\Phi_1. \qquad (38)$$

All excitations will be determined from this equation with only the inhomogeneous term $\varphi_1$ and the integration bounds $Q$ and $B$ varying from case to case.

As before, we solve for $\sigma_1$ by taking the Fourier transform with respect to $\Lambda$ of eq(45),

$$\sigma_1(\Lambda) = \frac{1}{u}\int_{-Q}^{Q} dk \rho_1(k)\text{sech}\frac{2\pi}{u}(\Lambda - \sin k) - \frac{1}{2\pi}\int_{-\infty}^{\infty} dp \frac{e^{ip(\Lambda-\Lambda_1^h)} + e^{ip(\Lambda-\Lambda_2^h)}}{1 + e^{-\frac{u}{2}|p|}} \qquad (39)$$

The second term is identical to the corresponding spin contribution in the Kondo model, while the first term represents the rearrangement in the charge sector that takes place due to its coupling to the spin sector.

Feeding eq(39) into eq(35) one has,

$$\rho_1(k) \equiv \rho_1(\Lambda_1^h, \Lambda_2^h) = \rho_1^s(k;\Lambda_1^h) + \rho_1^s(k;\Lambda_2^h), \qquad (40)$$

where each term satisfies,

$$\mathcal{K}\rho_1^s(k;\Lambda^h) = -\frac{\cos k}{u}\text{sech}\frac{2\pi}{u}(\sin k - \Lambda^h). \qquad (41)$$

We may take the integration limit implicit in the equation to be $Q_0$ rather than $Q$, since the correction are of a higher order in $1/L$.

From $\rho_1$ we compute the excitation energy and momentum. The total energy eigenvalue is given by

$$E = -2tL\int_{-Q}^{Q}\rho(k)\cos k = E_o(Q) - 2t\int_{-Q}^{Q}\rho_1(k)\cos k,$$

hence the excitation energy, $\Delta E = E - E_0$,

$$\Delta E(\Lambda_1^h, \Lambda_2^h) = -2t\int_{-Q}^{Q}\rho_1(k)\cos k\, dk + \Big(E_o(Q) - E_o(Q_o)\Big), \qquad (42)$$

where we define

$$E_o(Q) = -2tL \int_{-Q}^{Q} \rho_o(k,Q) \cos k\, dk. \tag{43}$$

with $\rho_o(k,Q)$ being the solution of the ground-state equation eq(13) with integration limit $Q$, which we now turn to determine. Denoting $N_o(Q) = L \int_{-Q}^{Q} \rho_o(k,Q)dk$ allows us to rewrite the condition $L \int_{-Q}^{Q} \rho(k,Q)dk = N$ as $N_o(Q) + \int_{-Q}^{Q} \rho_1(k,Q)dk = N_o(Q_o)$, yielding to order $O(\frac{1}{L})$,

$$Q - Q_o = -\int_{-Q}^{Q} \rho_1(k)dk \left(\frac{\partial N_o(Q)}{\partial Q}\right)^{-1}_{Q=Q_0}. \tag{44}$$

Thus to order $O(\frac{1}{L})$ the excitation energy becomes,

$$\begin{aligned}\Delta E(\Lambda_1^h, \Lambda_2^h) &= -2t \int_{-Q_o}^{Q_o} dk\, \rho_1(k) \cos k + (Q - Q_o)\frac{\partial E_o(Q_o)}{\partial Q_o} \\ &= -2t \int_{-Q_o}^{Q_o} dk\, \rho_1(k) \cos k - \mu \int_{-Q_o}^{Q_o} dk\, \rho_1(k)\end{aligned}$$

where the chemical potential, $\mu$, is given by

$$\mu = \frac{dE_0}{dN} = \left(\frac{\partial E_o(Q_o)}{\partial Q_o}\right)\left(\frac{\partial N_o(Q_o)}{\partial Q_o}\right)^{-1} = -2t\frac{\cos Q_o + \int_{-Q_o}^{Q_o} \phi(k,Q_o) \cos k\, dk}{1 + \int_{-Q_o}^{Q_o} \phi(k,Q_o) dk}. \tag{45}$$

We denoted

$$\phi(k,Q) = \frac{1}{2\rho_o(Q,Q)}\frac{\partial \rho_o(k,Q)}{\partial Q}, \tag{46}$$

and used the fact that $\rho_o(k,Q) = \rho_o(-k,Q)$.

Again as in Lecture 3 we find that the triplet excitation energy is composed of two terms $\Delta E = \epsilon(\Lambda_1^h) + \epsilon(\Lambda_2^h)$ each of which we identify as a spinon excitation energy,

$$\epsilon_s(\Lambda^h) = -\int_{-Q_o}^{Q_o} dk\, \rho_1^s(k, \Lambda^h)[2t \cos k + \mu] \tag{47}$$

Now to the excitation momentum $\Delta P(\Lambda_1^h, \Lambda_2^h)$. The total momentum of a state characterized by a configuration $\{n_j, I_\gamma\}$ is, as we saw earlier, $P = \frac{2\pi}{L}\left(\sum_{j=1}^{N} n_j + \sum_{\gamma=1}^{M} I_\gamma\right)$. Here we are considering a configuration with two "holes" in the spin sequence, $I_1^h$ and $I_2^h$, hence the momentum of the excitation is

$$\Delta P(\Lambda_1^h, \Lambda_2^h) = -\frac{2\pi}{L}(I_1^h + I_2^h) = -\frac{2\pi}{L}(\nu(\Lambda_1^h) + \nu(\Lambda_2^h)) \qquad (48)$$

We used $P_0 = 0$ and dropped a term arising upon the shift of all quantum numbers $n_j$ from half-odd-integers to integers when $M$ changes by one. This shift has no effect on the momentum which is periodic in $\pi n$.

We wish to identify the momentum of a single spinon in the sum eq(48). It must depend on a *single* parameter $\Lambda^h$ and have a finite limit as $L \to \infty$. The function $\nu(\Lambda^h)$ implicitly depends on both $\Lambda_1^h$ and $\Lambda_2^h$. However, we may write

$$\begin{aligned}\nu(\Lambda_1^h) &= \int_{-Q}^{Q} dk \Theta(2\sin k - 2\Lambda_1^h) - \int d\Lambda' \sigma(\Lambda')\Theta(\Lambda' - \Lambda_1^h) \\ &= \nu_0(\Lambda_1^h) + \frac{1}{L}\delta(\Lambda_1^h, \Lambda_2^h)\end{aligned} \qquad (49)$$

(and a similar expression for $\Lambda_2^h$) where,

$$\frac{1}{L}\delta(\Lambda_1^h, \Lambda_2^h) = (Q - Q_o)\frac{d\nu_0}{dQ_o} + \frac{1}{L}\int_{-Q_0}^{Q_0} dk \rho_1(k; \Lambda_1^h, \Lambda_2^h)\Theta(2\sin k - 2\Lambda_j^h) \qquad (50)$$
$$- \frac{1}{L}\int d\Lambda' \sigma_1(\Lambda'; \Lambda_1^h, \Lambda_2^h)\Theta(\Lambda' - \Lambda_j^h)$$

and $\nu_0$ is the ground state counting function. Hence, it is clear that the spinon momentum can be identified as

$$p_s(\Lambda^h) = -\frac{2\pi}{L}\nu_0(\Lambda^h) = -\frac{\pi}{2}n + 2\int_{-Q_o}^{Q_0} dk \rho_o(k)\tan^{-1}\exp\frac{2\pi}{u}(\Lambda^h - \sin k), (51)$$

yielding, together with the expression for the spinon energy eq(47), a parametric representation of the spinon dispersion.

The shift of the momentum from a free value, $\delta(\Lambda_1^h, \Lambda_2^h) = L(\nu(\Lambda_1^h) - \nu_0(\Lambda_1^h))$, is the scattering phase shift of spinon 1 off spinon 2, according to the method of momentum shifts discussed in Lecture 3 [15]. A general expression for it is given in eq(50), shall calculate it more explicitly soon.

We can see immediately that the spinons are gapless; consider $\epsilon(\Lambda^h)$ in the limit $\Lambda^h \to \pm\infty$. The form of $\rho_1(k, \Lambda)$ in this case can be explicitly found

from eq(41) [3]
$$\rho_1^s(k, \Lambda^h) = -\frac{2}{u} \cos k \, e^{\mp \frac{2\pi}{u} \Lambda^h} \psi(k), \tag{52}$$

with $\psi(k)$ satisfying [4],

$$\psi(k) = e^{\frac{2\pi}{u} \sin k} + \frac{4}{u} \int_{-Q_o}^{Q_o} dk' \cos k' R(\frac{4}{u}(\sin k - \sin k')) \psi(k'). \tag{53}$$

Hence,

$$\epsilon_s(\Lambda^h) \to 8\pi \frac{t}{u} [C^{(2)} + \frac{\mu}{2t} C^{(1)}] e^{\mp \frac{2\pi}{u} \Lambda^h}$$

$$p_s(\Lambda^h) \to \mp[\frac{\pi}{2} n - 4\pi e^{\pm \frac{2\pi}{u} \Lambda^h} C^{(0)}]$$

with

$$C^{(n)} = C^{(n)}(u, Q_o) = \frac{1}{2\pi} \int_{-Q_o}^{Q_o} dk \psi(k) . \cos^n(k) \tag{54}$$

It follows that the energy vanishes linearly with the momentum measured with respect to the *spin Fermi-momentum* $k_F^s = \frac{\pi}{2} n = k_F$,

$$\epsilon_s(p) = \pm v_s(p \pm k_F^s) \tag{55}$$

with the spin-velocity

$$v_s = \frac{2t}{u} \left[ \frac{C^{(2)}}{C^{(0)}} + \frac{\mu}{2t} \frac{C^{(1)}}{C^{(0)}} \right] \tag{56}$$

These spin-excitations, surviving in the low-enrgy limit, lead to a power law behaviour in the spin density correlation function for momentum transfer $q \approx 2k_F$.

The spinons can be excited (in pairs) without exciting charged modes. However, due to the coupling of the charge-sector to the spin-sector, the charge Fermi sea rearranges when spin modes are excited and modifies their energy. This can be seen, in particular, when we study the spinon scattering phase shifts.

To compute the scattering phase shifts we consider the shift of the one spinon momentum from its free value $\frac{2\pi}{L} I^h$, due to the presence of the other spinon, $-\frac{2\pi}{L}I^h = -\frac{2\pi}{L}\nu(\Lambda^h) = p_s(\Lambda^h) + \frac{\delta(\Lambda_1^h,\Lambda_2^h)}{L}$. Thus, scattering phase shift for spinons in the triplet state,

$$\begin{aligned}
\delta^{trip}(\Lambda_1^h, \Lambda_2^h) &= -2\pi(\nu(\Lambda_1^h) - \nu_0(\Lambda_1^h)) \\
&= -\pi + \int_{-Q_0}^{Q_0} dk \rho_1(k) \left\{ \frac{dp_s(\Lambda_1^h)}{dn} - \Theta(2(\sin k - \Lambda_1^h)) \right\} \\
&\quad + \int_{-\infty}^{\infty} d\Lambda' \sigma_1(\Lambda')\Theta(\Lambda' - \Lambda_1^h).
\end{aligned} \tag{57}$$

The constant $-\pi$ appears because the set $\{I_\alpha\}$ is shifted by $\frac{1}{2}$ with respect to the ground state set. We have

$$\delta^{trip}(\Lambda_1^h, \Lambda_2^h) = -\pi + \frac{1}{i}\log\left\{\frac{\Gamma\left(\frac{1}{2} + i\frac{\Lambda_1^h - \Lambda_2^h}{u}\right)\Gamma\left(1 - i\frac{\Lambda_1^h - \Lambda_2^h}{u}\right)}{\Gamma\left(\frac{1}{2} - i\frac{\Lambda_1^h - \Lambda_2^h}{u}\right)\Gamma\left(1 + i\frac{\Lambda_1^h - \Lambda_2^h}{u}\right)}\right\} \tag{58}$$

$$+ \int_{-Q_0}^{Q_0} dk \rho_1(k; \Lambda_1^h, \Lambda_2^h) \left\{ 2\arctan\left(e^{\frac{2\pi}{u}(\Lambda_1^h - \sin k)}\right) + \frac{\pi}{2} + \frac{dp_s(\Lambda_1^h)}{dn} \right\},$$

with

$$\frac{dp_s(\Lambda)}{dn} = \frac{\pi}{2} + \frac{\arctan\left(\frac{\cosh\left(\frac{2\pi}{u}\sin Q_0\right)}{\sinh\left(\frac{2\pi}{u}\Lambda\right)}\right) - 2\int_{-Q_0}^{Q_0} \arctan\left(e^{\frac{2\pi}{u}(\Lambda - \sin k)}\right)\phi(k, Q_0)dk}{1 + \int_{-Q_0}^{Q_0} dk \phi(k, Q_o)}$$

The first two terms in (58) correspond to the pure spin scattering and have already appeared in the spinon scattering in the Kondo model. The last term corresponds to the interaction of the spinon with the change in the charge distribution produced by the presence of the triplet. As $Q_0$ tends to $\pi$, the redistribution of the charge degrees of freedom decreases and so does its contribution to the phase shift.

At half filling the expressions simplify. The $Q$-level does not shift from its ground state value $Q_o = \pi$, so $\mu = 0$, and the solution of eq(41) is straightforward,

$$\rho_1(k) = -\cos k \left[\frac{1}{u}\operatorname{sech}\frac{2\pi}{u}(\sin k - \Lambda_1^h) + \frac{1}{u}\operatorname{sech}\frac{2\pi}{u}(\sin k - \Lambda_2^h)\right].$$

Hence the spinon excitation energy and momentum are

$$\epsilon_s^\pi(\Lambda^h) = \frac{2t}{u}\int_{-\pi}^{\pi}\cos^2 k\ \text{sech}\frac{2\pi}{u}(\sin k - \Lambda^h)dk = 2t\int_0^\infty dp\frac{J_1(p)\cos p\Lambda^h}{p\cosh\frac{u}{4}p}$$

$$p_s^\pi(\Lambda^h) = -\frac{\pi}{2} + \frac{1}{\pi}\int_{-\pi}^{\pi}\tan^{-1}\exp\frac{2\pi}{u}(\Lambda^h - \sin k) = -\frac{\pi}{2} + \int_0^\infty dp\frac{J_0(p)\sin p\Lambda^h}{p\cosh\frac{u}{4}p},$$

and the spin velocity at half filling becomes

$$v_s = 2t\frac{I_1(\frac{2\pi}{u})}{I_0(\frac{2\pi}{u})}, \tag{59}$$

with $I_n$ being Bessel functions with imaginary argument. The spinon velocity varies from $v_s = 2t$ when $U = 0$ to $v_s = 0$ when $U = \infty$, corresponding to the excitation spectrum $\epsilon_s(p) = 2t\cos p$ and $\epsilon_s(p) = 0$, respectively.

The spinon-spinon scattering matrix becomes identical to the Kondo spinon scttering matrix (and to the pure spin S-matrix in any SU(2) model soluble with the R-matrix given in lecture 2)

$$S_\pi^{(trip)}(\Lambda_1^h, \Lambda_2^h) = \frac{\Gamma(1 - \frac{2i}{u}(\Lambda_1^h - \Lambda_2^h))\Gamma(\frac{1}{2} + \frac{2i}{u}(\Lambda_1^h - \Lambda_2^h))}{\Gamma(1 + \frac{2i}{u}(\Lambda_1^h - \Lambda_2^h))\Gamma(\frac{1}{2} - \frac{2i}{u}(\Lambda_1^h - \Lambda_2^h))}. \tag{60}$$

To validate the claim that the spin-1 state considered thus far consists of two spin-$\frac{1}{2}$ excitations we must show that a state exists in the spectrum, degenerate in energy with the triplet state, with these spin excitations coupled to form a spin-0 eigenstate.

*The singlet.* The spinons can be coupled antisymmetrically to form a singlet by adding to a $\Lambda$-configuration with "holes" at $\Lambda_1^h$ and $\Lambda_2^h$ a 2-string $\Lambda^\pm = \Lambda_o \pm iu/4$, $\Lambda_o = \frac{1}{2}(\Lambda_1^h + \Lambda_2^h)$. This is, indeed a solution corresponding to the choice of $I_\gamma$-quantum numbers with two unfilled slots at $I_1^h$ and $I_2^h$ and an additional $I^{(2)}$ quantum number related to the 2-string position.

The equations for the (reduced) densities $\rho_1, \sigma_1$ satisfy the reduced equation with inhomogeneous term

$$\varphi_1 = \begin{pmatrix} \cos k\ K_2(\sin k - \Lambda_o) \\ -\sigma^h(\Lambda) - \sigma^{st}(\Lambda) \end{pmatrix} \tag{61}$$

where
$$\sigma^h(\Lambda) = \delta(\Lambda - \Lambda_1^h) + \delta(\Lambda - \Lambda_2^h)$$
$$\sigma^{st}(\Lambda) = K_1(\Lambda - \Lambda_o) + K_3(\Lambda - \Lambda_o).$$

We find for the spin density

$$\tilde{\sigma}_1(p) = \int_{-Q}^{Q} dk \rho_1(k) \frac{e^{-ip\sin k}}{2\cosh(\frac{u}{4}p)} - \frac{\left(e^{-ip\Lambda_1^h} + e^{-ip\Lambda_2^h}\right)}{1 + e^{-\frac{u}{2}|p|}} - e^{-\frac{u}{4}|p|} e^{-i\Lambda_o p}$$

leading to an equation for $\rho_1$ which is identical to the equation determining $\rho_1(k)$ in the triplet case, (the 2-string contributions, although modifying the the spin-sector, cancel in the equation for $\rho_1$ leaving the charge-sector unchanged!) Hence the triplet and singlet are degenerate in energy (in the limit $L \to \infty$), confirming the physical picture, totally analogous to the one we have found in the Kondo model, of these states describing two spin-$\frac{1}{2}$ objects coupled symmetrically in one case and antisymmetrically in the other.

The interaction of the spinons, however, depends on the spin-state. Following the procedure previously outlined we find that the spinons, when in the singlet state, undergo scattering with the phase shift

$$\delta^{sing}(\Lambda_1^h, \Lambda_2^h) = \tag{62}$$
$$= \frac{1}{i} \log \left\{ \frac{1 + i\frac{2}{u}(\Lambda_1^h - \Lambda_2^h)}{1 - i\frac{2}{u}(\Lambda_1^h - \Lambda_2^h)} \right\} + \frac{1}{i} \log \left\{ \frac{\Gamma\left(1 - i\frac{\Lambda_1^h - \Lambda_2^h}{u}\right) \Gamma\left(\frac{1}{2} + i\frac{\Lambda_1^h - \Lambda_2^h}{u}\right)}{\Gamma\left(1 + i\frac{\Lambda_1^h - \Lambda_2^h}{u}\right) \Gamma\left(\frac{1}{2} - i\frac{\Lambda_1^h - \Lambda_2^h}{u}\right)} \right\}$$
$$+ \int_{-Q_0}^{Q_0} dk \rho_1(k; \Lambda_1^h, \Lambda_2^h) \left\{ \arctan(e^{\frac{2\pi}{u}(\Lambda_1^h - \sin k)}) - \frac{\pi}{2} - \frac{dp_s(\Lambda_1^h)}{dn} \right\},$$

consisting of a charge contribution identical to the one we encountered discussing the triplet scattering, and a spin contribution that is modified in a manner similar to the modification we found for the Kondo-spinons scattering in the singlet state.

Again, in the case of half filling the charge sector does not contribute to the spinon scattering, and the singlet S-matrix is found from the first two terms on the right hand side of eq(62)

$$S_\pi^{(sing)}(\Lambda_1^h, \Lambda_2^h) = \frac{1 + \frac{2i}{u}(\Lambda_1^h - \Lambda_2^h)}{1 - \frac{2i}{u}(\Lambda_1^h - \Lambda_2^h)} \frac{\Gamma(1 - \frac{i}{u}(\Lambda_1^h - \Lambda_2^h))\Gamma(\frac{1}{2} + \frac{i}{u}(\Lambda_1^h - \Lambda_2^h))}{\Gamma(1 + \frac{i}{u}(\Lambda_1^h - \Lambda_2^h))\Gamma(\frac{1}{2} - \frac{i}{u}(\Lambda_1^h - \Lambda_2^h))}. \tag{63}$$

We can combine (60) and (63) to get the scattering matrix for spinons at half filling

$$S^{spin}_\pi(\Lambda_1^h, \Lambda_2^h) = -\frac{\Gamma(1 - \frac{i}{u}(\Lambda_1^h - \Lambda_2^h))\Gamma(\frac{1}{2} + \frac{i}{u}(\Lambda_1^h - \Lambda_2^h))}{\Gamma(1 + \frac{i}{u}(\Lambda_1^h - \Lambda_2^h))\Gamma(\frac{1}{2} - \frac{i}{u}(\Lambda_1^h - \Lambda_2^h))} \left\{ \frac{(\Lambda_2^h - \Lambda_1^h)I^{12} + i\frac{u}{2}P^{12}}{(\Lambda_2^h - \Lambda_1^h) + i\frac{u}{2}} \right\}, \quad (64)$$

where $I^{12}$ and $P^{12}$ are the identity and the exchange operator in spin space, respectively. This expression satisfies the Yang-Baxter equation guaranteeing that the physical, dressed S-matrices which describe the scattering of the actual quasi-particles do factorize consistently in the same manner as do the bare S-matrices discussed in Lecture 1. As a consequence we conclude that the excitations, though interacting, never decay! This is more remarkable still since gapless excitations are present, and is due to the dynamical conservation laws (briefly mentioned in Lecture 1) which protect the excitations.

**Charge excitations.**

Still keeping the number of electrons fixed, we now vary the $\{n_j\}$ sequence from its vacuum configuration, leaving the spin configuration $\{I_\gamma\}$ unchanged.

*The holon-antiholon.* We consider first the case $N < L$, where the following charge configuration is allowed; remove the level $n_A$ from the ground state sequence (creating a hole at the corresponding $k_A \leq Q_0$), and add a level outside the charge-sea, $n_B$ (creating a particle at $k_B \geq Q_0$). Obviously, this excitation is *not* present at half filling.

The $\{n_j\}$ configuration we consider, $n_B > n^+$, $n_{j+1} - n_j = 1 + \delta_{j,A}$, leads to the equations (here $-Q \leq k \leq Q$)

$$\rho(k) = \frac{1}{2\pi} - \frac{1}{L}\delta(k - k_A) + \cos k \int_{-\infty}^\infty d\Lambda \sigma(\Lambda) K_1(\sin k - \Lambda)$$

$$\sigma(\Lambda) = \frac{1}{L}K_1(\sin k_B - \Lambda) + \int_{-Q}^Q dk \rho(k) K_1(\sin k - \Lambda) - \int_{-\infty}^\infty d\Lambda' \sigma(\Lambda') K_2(\Lambda - \Lambda'),$$

with the Q-level set by the requirement $\int_{-Q}^Q \rho(k)dk = \frac{N-1}{L}$. It can be easily checked integrating over $\Lambda$ that $M = \frac{1}{2}N$ as expected since we did not change the spin-sequence $\{I_\gamma^o\}$.

Again, $\rho_1(k)$ and $\sigma_1(\Lambda)$ satisfy the reduced equation with the inhomogeneous term

$$\varphi_1 = \begin{pmatrix} -\delta(k - k_A) \\ K_1(\sin k_B - \Lambda) \end{pmatrix}. \tag{65}$$

Hence immediately,

$$\tilde{\sigma}_1(p) = \int_{-Q}^{Q} dk \rho_1(k) \frac{e^{-ip\sin k}}{2\cosh(\frac{u}{4}p)} - \frac{e^{-ip\sin k_B}}{2\cosh(\frac{u}{4}p)},$$

and the equation for $\rho_1(k)$ follows. It is convenient to introduce a smooth density $\rho'_1(k)$,

$$\rho'_1(k) = \rho_1(k) + \delta(k - k_A) \tag{66}$$

and one has

$$\rho'_1(k) \equiv \rho'_1(k; k_A, k_B) = \rho^c_1(k; k_A) - \rho^c_1(k; k_B). \tag{67}$$

with

$$\mathcal{K}\rho^c_1(k; k_j) = -\frac{4}{u}\cos k \; R\left(\frac{4}{u}(\sin k - \sin k_j)\right) \quad j = A, B. \tag{68}$$

The calculation of the energy and momentum proceeds as before; the energy is given by

$$\begin{aligned} E &= -2t\cos k_B - 2tL\int_{-Q}^{Q} dk \rho(k) \cos k \\ &= -2t\cos k_B - 2t\int_{-Q}^{Q} dk(\rho'_1(k) - \delta(k - k_A))\cos k + E_o(Q) \end{aligned}$$

and the excitation energy by

$$\begin{aligned} \Delta E(k_A, k_B) &= 2t\cos k_A - 2t\cos k_B - 2t\int_{-Q}^{Q} dk \rho'_1(k; k_A, k_B) \cos k + E_o(Q) - E_o(Q_o) \\ &= -\epsilon_c(k_A) + \epsilon_c(k_B), \end{aligned} \tag{69}$$

where we introduced the holon energy

$$\epsilon_c(k_A) = -2t \cos k_A + \int_{-Q_o}^{Q_o} dk \rho_1^c(k, k_A)[2t \cos k + \mu]. \tag{70}$$

This function is monotonically decreasing to

$$\epsilon_c(Q_o) = \mu \tag{71}$$

as $k_A$ tends to $Q_0$.

The momentum of the excitation is

$$P - P_0 = \frac{2\pi}{L} \sum_j (n_j - n_j^0) + \frac{2\pi}{L} n_B = \frac{2\pi}{L}(n_B - n_A) = -p_c(k_A) + p_c(k_B), \tag{72}$$

with the holon momentum, following previous arguments,

$$\begin{aligned}
p_c(k_A) &= \frac{2\pi}{L}\omega_0(k_A) = 2\pi \int_0^{k_A} \rho_o(k)dk \\
&= k_A - i \int_{-Q_0}^{Q_0} dk \rho_0(k) \log \left\{ \frac{\Gamma\left(1 + i\frac{\sin k_A - \sin k}{u}\right) \Gamma\left(\frac{1}{2} - i\frac{\sin k_A - \sin k}{u}\right)}{\Gamma\left(1 - i\frac{\sin k_A - \sin k}{u}\right) \Gamma\left(\frac{1}{2} + i\frac{\sin k_A - \sin k}{u}\right)} \right\}.
\end{aligned} \tag{73}$$

Both forms for the momentum will be useful. Together with the expression for the holon energy we have a parametric representation of the holon dispersion.

It is obvious that the excitation is gapless when $k_A = k_B = Q_0$ (recall that $-\pi \leq -Q_0 \leq k_A \leq Q_0 \leq k_B \leq \pi$.) The holon momentum in this limit tends to the charge Fermi momentum $k_F^c = \pi n$, $p_c(k_A) = 2\pi \int_0^{k_A} \rho_o(k)dk \to \pm \pi n$, hence the vanishing of the excitation energy will lead to power law behaviour for charge density correlation functions at momentum transfer $q \approx 2k_F^c = 4k_F$. In a few paragraphs we shall find that the holon-holon excitation, on the other hand, is gapful.

Once more we are able to describe the excitation as a combination of two objects. We will call holon the one with energy $\epsilon_c(k_A)$ and momentum $p_c(k_A)$ associated with the hole in the sequence, and antiholon the one created by adding an electron to the system and characterized by $-\epsilon_c(k_B)$ and $-p_c(k_B)$.

The holon-antiholon phase shift $\delta^{h,\bar{h}} = -2\pi(\omega(k_B) - \omega_0(k_B))$ is given by

$$\delta^{h,\bar{h}} = \frac{1}{i} \log \left\{ \frac{\Gamma\left(1 + i\frac{\sin k_B - \sin k_A}{u}\right) \Gamma\left(\frac{1}{2} - i\frac{\sin k_B - \sin k_A}{u}\right)}{\Gamma\left(1 - i\frac{\sin k_B - \sin k_A}{u}\right) \Gamma\left(\frac{1}{2} + i\frac{\sin k_B - \sin k_A}{u}\right)} \right\} \tag{74}$$

$$-\int_{-Q_0}^{Q_0} dk \rho_1'(k;k_A,k_B) \left[\frac{1}{i}\log\left\{\frac{\Gamma\left(1+i\frac{\sin k_B-\sin k}{u}\right)\Gamma\left(\frac{1}{2}-i\frac{\sin k_B-\sin k}{u}\right)}{\Gamma\left(1-i\frac{\sin k_B-\sin k}{u}\right)\Gamma\left(\frac{1}{2}+i\frac{\sin k_B-\sin k}{u}\right)}\right\} - \frac{dp_c(k_B)}{dn}\right].$$

As before there are two contributions to the phase shift: the first line corresponds to direct holon-antiholon interaction; the rest describes the effect of the interaction between the excitations and the redistribution of charge they produce. Since the interaction is between charge degrees of freedom, the integrand is of the same form as the first term of the phase shift. At half filling it vanishes and the first term would describe the holon-antiholon scattering had they existed.

*The holon-holon excitation.* We were considering configurations with only real $k$-momenta and holes in them. To discuss states with double occupancy, however, we need to consider solutions involving complex $k$-pairs. As a result, a gap of order $U$ opens, and the excitations lie in the "upper Hubbard band". These gapful excitations are present both at half filling and away from it, but are the only ones surviving at half filling ( keeping the number of electrons fixed.) The disappearance of the gapless holon-antihilon excitation from the spectrum at half filling is the origin of the charge gap and the concomitant metal-insulator transition.

Consider then a configuration $\{n_j\}$ leading to a complex $k$−pair $k^\pm = \kappa \pm i\chi$ and two "holes" in the real $k$-sea (the number of electrons is kept fixed). In the notation of the Appendix we have $M' = 1$ complex $k$-pairs (i.e. a $k - \Lambda$ 1-string) with the associated quantum number $I'$. The number of real $k$-momenta is $N - 2$ and $M = N/2 - 1$. The number of real $\Lambda$ decreased by one since one of them becomes the parameter $\lambda$ to describe the complex pair. From the counting arguments we have $I' = 0$, the set $\{I_\gamma\}$ splits around zero so that $\sum_\gamma I_\gamma = 0$, but there is a $\pi$ contribution to the phase shift.

The Bethe Ansatz equations now take the form

$$L(\kappa \pm i\chi) = 2\pi n_j + \sum_{\delta=1}^{M-1} \Theta[2\sin(\kappa \pm i\chi) - 2\Lambda_\delta] \qquad (75)$$

The right-hand side can be converted to an integral and evaluated to order $\frac{1}{L}$ by means of the ground state density $\sigma_o(\Lambda)$. One finds the complex is driven to string positions centered around a particular $\Lambda$ solution. Denoting this particular solution by $\lambda$, we find that it must satisfy (to order $O(e^{-\eta L})$ with $\eta$ a number of order 1)

$$\sin(\kappa \pm i\chi) = \lambda \mp iu/4 + O(e^{\eta L}) \tag{76}$$

namely,

$$k_{\pm} = \pi - \arcsin(\lambda \pm i\frac{u}{4}) \tag{77}$$

with $\lambda$ real.

Substituting (77) in the eigenvalue equations for real $k$,

$$e^{ik_j L} = \left(\prod_{\gamma=1}^{M-1} \frac{\Lambda_\gamma - \sin k_j - i\frac{u}{4}}{\Lambda_\gamma - \sin k_j + i\frac{u}{4}}\right) \frac{\lambda - \sin k_j - i\frac{u}{4}}{\lambda - \sin k_j + i\frac{u}{4}}$$

it follows that the real $k$'s (the 1-strings) satisfy

$$Lk_j = 2\pi n_j + \sum_{\delta=1}^{M-1} \Theta(2\sin k_j - 2\Lambda_\delta) + \Theta(2\sin k_j - 2\lambda) \tag{78}$$

where the $\{n_j\}$ set has two "holes" in it, at $n_1^h$ and $n_2^h$, corresponding to omitted momenta $k_1^h$ and $k_2^h$, respectively.

Similarly, the eigenvalue equation for real-$\Lambda$, becomes

$$-\left(\prod_{\delta=1}^{M-1} \frac{\Lambda_\delta - \Lambda_\gamma + i\frac{u}{2}}{\Lambda_\delta - \Lambda_\gamma - i\frac{u}{2}}\right) \frac{\lambda - \Lambda_\gamma + i\frac{u}{2}}{\lambda - \Lambda_\gamma - i\frac{u}{2}} = \left(\prod_{j=1}^{N-2} \frac{\Lambda_\gamma - \sin k_j - i\frac{u}{4}}{\Lambda_\gamma - \sin k_j + i\frac{u}{4}}\right) \frac{\Lambda_\gamma - \lambda - i\frac{u}{2}}{\Lambda_\gamma - \lambda + i\frac{u}{2}}$$

namely,

$$\sum_{j=1}^{N-2} \Theta(2\Lambda_\gamma - 2\sin k_j) = -2\pi I_\gamma + \sum_{\delta=1}^{M-1} \Theta(\Lambda_\gamma - \Lambda_\delta), \quad \gamma = 1, ..., M-1 \tag{79}$$

which has the same form as eq(4) but for $N-2$ particles.

Solving the equations for real $k$ and $\Lambda$ and evaluating eq(75) one has

$$\lambda = \frac{1}{2}(\sin k_1^h + \sin k_2^h) \tag{80}$$

similar to the result we encountered discussing singlet spin excitations.

We convert (78) and (79), to the integral equations for $\rho(k)$ and $\sigma(\Lambda)$, then defining $\rho_1$ and $\sigma_1$, and finally introducing

$$\rho_1'(k) = \rho_1(k) + \frac{1}{L}\delta(k - k_1^h) + \frac{1}{L}\delta(k - k_2^h) \tag{81}$$

we find that the densities $\sigma_1$ and $\rho_1'$ satsfy the usual reduced equation with the following inhomogeneous term

$$\varphi_1 = \begin{pmatrix} \cos k\, K_1(\sin k - \lambda) \\ -K_1(\sin k_1^h - \Lambda) - K_1(\sin k_2^h - \Lambda) \end{pmatrix} \tag{82}$$

and with the $Q$-level set by the condition $\int_{-Q}^{Q} \rho(k)dk = \frac{N-2}{L}$.

Hence,

$$\tilde{\sigma}_1(p) = \int_{-Q}^{Q} \rho_1'(k) \frac{e^{-ip\sin k}}{2\cosh(\frac{u}{4}p)} - \frac{e^{-ip\sin k_1^h} + e^{-ip\sin k_2^h}}{2\cosh(\frac{u}{4}p)} \tag{83}$$

and

$$\rho_1'(k) \equiv \rho_1'(k; k_1^h, k_2^h) = \rho^c(k; k_1^h) + \rho^c(k; k_2^h) + \rho^{st}(k; \lambda) \tag{84}$$

where $\rho^c(k; k_j^h)$ satisfy eq(68), and $\rho^{st}(k; \lambda)$ is the solution of

$$\mathcal{K}\rho^{st}(k; \lambda) = \cos k\, K_1(\sin k - \lambda). \tag{85}$$

The energy of the state is given by

$$E = -2t(\cos k_+ + \cos k_-) - 2tL \int_{-Q}^{Q} dk \rho(k) \cos k \tag{86}$$

with the first term being the contribution of the string. Using some identities: $\cos k_+ + \cos k_- = 2Re\left(\sqrt{1 - (\lambda + i\frac{u}{4})^2}\right) = \frac{u}{2} + \frac{1}{\pi}\int_{-\pi}^{\pi} dk \cos^2 k K_1(\sin k - \Lambda)$, we find for the excitation energy

$$\begin{aligned}\Delta E(k_1^h, k_2^h) &= U + \sum_{j=1}^{2} \epsilon_c(k_j^h) + 2t \int_{-\pi}^{\pi} dk \cos^2 k K_1(\sin k - \lambda) \\ &\quad -2t \int_{-Q_0}^{Q_0} dk \rho_1^{st}(k; \lambda) \left\{\cos k + \frac{\mu}{2t}\right\} \\ &= \Delta_g(U, Q_0) + \sum_{j=1}^{2} \epsilon_c(k_j^h) \end{aligned} \tag{87}$$

with the $\epsilon_c(k)$ same as defined for the holon-antiholon case, eq(70). The holon-holon energy gap $\Delta_g(U, Q_0) + 2\epsilon_c(Q_0)$ does not vanish for any filling. Away from half filling, however, the holon-antiholon gapless modes are available to carry the charge.

We evaluate now the excitation momentum. Since there are no holes in the spin quantum numbers and we are taking $I'_\alpha = 0$, the only contribution to the change in momentum comes from the holes in the charge distribution. Then

$$\Delta P = -\frac{2\pi}{L}(n_1^h + n_2^h) \tag{88}$$

$$= k_1^h + k_2^h - \frac{1}{L}\sum_{\beta=1}^{\frac{N}{2}-1}\left\{\Theta(2(\sin k_1^h - \Lambda_\beta)) + \Theta(2(\sin k_2^h - \Lambda_\beta))\right\}$$

$$+ \frac{1}{L}\left\{\Theta(2(\sin k_1^h - \lambda)) + \Theta(2(\sin k_2^h - \lambda))\right\}$$

$$= p_c(k_1^h) + p_c(k_2^h). \tag{89}$$

with $p_c(k)$ defined in (73).

The holon-holon phase shift is

$$\delta^{h,h}(k_1^h, k_2^h) = -\pi - \left\{\frac{\int_{-Q_0}^{Q_0} dk \rho_1'(k)}{2\rho_0(Q_0) + \int_{-Q_0}^{Q_0} \frac{d\rho_0(k)}{dQ_0}}\right\}\frac{dp_c(k_1)}{dQ_0} \tag{90}$$

$$+ \int_{-\infty}^{\infty} d\Lambda \sigma_1(\Lambda)\Theta(\sin k_1 - \Lambda) - \Theta(2(\sin k_1 - \lambda)),$$

and can be written more explicitly as

$$\delta^{h,h}(k_1^h, k_2^h) = \pi + \frac{1}{i}\log\left\{\frac{1 + i\frac{2}{u}(\sin k_1^h - \sin k_2^h)}{1 - i\frac{2}{u}(\sin k_1^h - \sin k_2^h)}\frac{\Gamma\left(1 - i\frac{\sin k_1^h - \sin k_2^h}{u}\right)\Gamma\left(\frac{1}{2} + i\frac{\sin k_1^h - \sin k_2^h}{u}\right)}{\Gamma\left(1 + i\frac{\sin k_1^h - \sin k_2^h}{u}\right)\Gamma\left(\frac{1}{2} - i\frac{\sin k_1^h - \sin k_2^h}{u}\right)}\right\}$$

$$+ \int_{-Q_0}^{Q_0} dk \rho_1'(k; k_1^h, k_2^h)\left[\frac{1}{i}\log\frac{\Gamma\left(1 + i\frac{\sin k_1^h - \sin k}{u}\right)\Gamma\left(\frac{1}{2} - i\frac{\sin k_1^h - \sin k}{u}\right)}{\Gamma\left(1 - i\frac{\sin k_1^h - \sin k}{u}\right)\Gamma\left(\frac{1}{2} + i\frac{\sin k_1^h - \sin k}{u}\right)} - \frac{dp_c(k_1^h)}{dn}\right] \tag{91}$$

It is worth comparing (74) and (91), specially the terms that do not depend on the charge densities. These, as we shall see, are the only terms surviving at half filling. If both excitations were present at half filling, they would have the singlet-triplet relation that we found for the spin excitations.

At half filling matters simplify again; we have explicit solutions

$$\rho^c(k; k_j^h) = -\cos k \frac{4}{u} R(\frac{4}{u}(\sin k - \sin k_j^h)), \quad \rho^{st}(k; \lambda) = \cos k K_1(\sin k - \lambda),$$

leading to the holon-holon excitation energy

$$\Delta E^{h,h}(k_1^h, k_2^h) = U + \sum_{j=1}^{2} \epsilon_c(k_j^h), \tag{92}$$

with the holon energy explicitly given by

$$\epsilon_c^\pi(k^h) = -2t \cos k^h + 4t \int_0^\infty dp \frac{J_1(p) \cos(p \sin k^h)}{p(1 + e^{\frac{u}{2}|p|})}, \tag{93}$$

and the gap by

$$\Delta_g = U + 2\epsilon_c^\pi(\pi). \tag{94}$$

Likewise, the momentum of the holon can be explicitly computed

$$\begin{aligned}
p_c(k^h) &= 2\pi \int_0^{k^h} \rho_o(k) dk = k^h + \int_0^\infty dp \frac{e^{-\frac{u}{4}p} J_0(p)}{p \cosh \frac{u}{4} p} \sin(p \sin k^h) \\
&= k^h + \frac{1}{2\pi i} \int_{-\pi}^{\pi} dk' \log \left\{ \frac{\Gamma\left(1 + i\frac{\sin k^h - \sin k'}{u}\right) \Gamma\left(\frac{1}{2} - i\frac{\sin k^h - \sin k'}{u}\right)}{\Gamma\left(1 - i\frac{\sin k^h - \sin k'}{u}\right) \Gamma\left(\frac{1}{2} + i\frac{\sin k^h - \sin k'}{u}\right)} \right\}
\end{aligned}$$

and the phase shift,

$$\delta_\pi^{h,h} = -\pi + \frac{1}{i} \log \left\{ \frac{1 - i\frac{2}{u}(\sin k_2^h - \sin k_1^h)}{1 + i\frac{2}{u}(\sin k_2^h - \sin k_1^h)} \frac{\Gamma\left(1 - i\frac{\sin k_2^h - \sin k_1^h}{u}\right) \Gamma\left(\frac{1}{2} + i\frac{\sin k_2^h - \sin k_1^h}{u}\right)}{\Gamma\left(1 + i\frac{\sin k_2^h - \sin k_1^h}{u}\right) \Gamma\left(\frac{1}{2} - i\frac{\sin k_2^h - \sin k_1^h}{u}\right)} \right\} \tag{95}$$

Note the similarity with the singlet scattering shift, eq(63).

*Charge-spin excitations.*

We studied thus far pure spin excitations, as well as pure charge excitations. The latter fall into two categories; the gapless holon-antiholon and the gapful holon-holon excitations. In the half filled band only the latter exists and a charge gap opens. Similarly the simplest spin excitation are the triplet and the singlet composed of two spin-1/2 spinons.

A single holon cannot be excited if we keep the the number of electrons fixed. Neither can we excite a single spinon. To do so we need to change the number of particles in the system. When we add an electron we modify both the spin and charge configuration, creating a hole in the former and adding a level in the latter, thus exciting a spinon and an antiholon excitation (rather a coherent superposition of them). When we remove an electron we create holes both in the spin and the charge sequences, exciting a spinon and a holon. These considerations are similar to the corresponding ones in the Kondo model. However, unlike the situation there, the spinon and holon do not decouple except in the low-energy limit.

Now that we are considering a change in the number of particles it is convenient to change the form of the hamiltonian to make it more symmetric. We shall replace the interaction term $H_I = U \sum_j n_{j\uparrow} n_{j\downarrow}$ by $H_I' = \frac{1}{2} U \sum_j (n_{j\uparrow} + n_{j\downarrow} - 1)^2$. The difference, $H_I' - H_I = -\frac{1}{2} U \hat{N} = -\frac{1}{2} U \sum_j (n_{j\uparrow} + n_{j\downarrow})$ commutes with the hamiltonian and only modifies the energetics. We have, in other words, added a chemical potential, $A = -\frac{1}{2} U$, to the system to make it particle-hole symmetric.

We begin by *removing* an electron, creating a charge hole at $k^h$ and a spin hole at $\Lambda^h$ corresponding to the omitted quantum numbers $n^h$, $I^h$. The $n_j$ quantum numbers change from half-odd-integers to integers, while the $I_\gamma$ stay half-odd-integers.

Following the well trodden path we introduce the densities $\rho_1(k)$ and $\sigma_1(\Lambda)$, then $\rho_1'(k) = \rho_1(k) + \delta(k - k^h)$ obeying

$$\mathcal{K} \rho_1'(k) = -\cos k \frac{1}{u} \mathrm{sech}(\frac{2\pi}{u}(\sin k - \Lambda^h)) - \cos k \frac{4}{u} R\left(\frac{4}{u}(\sin k - \sin k^h)\right),$$

with the solution given by

$$\rho_1'(k) = \rho_1^s(k; \Lambda^h) + \rho_1^c(k; k^h).$$

Hence the excitation energy

$$E(N-1; k^h, \Lambda^h) - E_0(N) = \Delta E_-(k^h, \Lambda^h) = \frac{U}{2} + \epsilon_s(\Lambda^h) + \epsilon_c(k^h). \quad (96)$$

and momentum

$$\Delta P = -\frac{2\pi}{L}\left(I^h + n^h\right) = p_c(k^h) + p_s(\Lambda^h), \quad (97)$$

Therefore, this excitation is composed of a holon and a spinon. They interact and the phase shift can be calculated, for instance, from the spin contribution to be,

$$\delta^{h,s} = \frac{\pi}{2} - 2\arctan(e^{\frac{2\pi}{u}(\Lambda^h - \sin k^h)})$$
$$- \int_{-Q_0}^{Q_0} dk \rho_1'(k) \left\{ 2\arctan(e^{\frac{2\pi}{u}(\Lambda^h - \sin k)}) - \frac{\pi}{2} - \frac{dp_s}{dn} \right\}. \quad (98)$$

When we remove an electron in a definite momentum state $p$ forming the state $c_{p,a}|\Omega>$ we end up with a coherent superposition of the eigenstates just described, subject to $p = p_c(k^h) + p_s(\Lambda^h)$. The associated spread in energies gives a measure of the life time of an electron (or electron hole) in the system.

At half-filling the excitation energy and momentum are

$$\Delta E_-(k^h, \Lambda^h) = \frac{U}{2} + 2t\cos k^h + 4t\int_0^\infty dp \frac{J_1(p)\cos(p\sin k^h)}{p(1 + e^{\frac{u}{2}|p|})} + 2t\int_0^\infty dp \frac{J_1(p)\cos p\Lambda^h}{p\cosh\frac{u}{4}p}$$

$$\Delta P(k^h, \Lambda^h) = \frac{\pi}{2} - \int_0^\infty dp \frac{J_0(p)\sin(p\Lambda^h)}{p\cosh\frac{u}{4}p}$$
$$+ k^h + \frac{1}{2\pi i}\int_{-\pi}^{\pi} dk' \log \left\{ \frac{\Gamma\left(1 + i\frac{\sin k^h - \sin k'}{u}\right)\Gamma\left(\frac{1}{2} - i\frac{\sin k^h - \sin k'}{u}\right)}{\Gamma\left(1 - i\frac{\sin k^h - \sin k'}{u}\right)\Gamma\left(\frac{1}{2} + i\frac{\sin k^h - \sin k'}{u}\right)} \right\}$$
$$= \int_0^\infty dp \frac{J_0(p)\sin p\Lambda^h}{p\cosh\frac{u}{4}p} + k^h + \int_0^\infty dp \frac{e^{-p\frac{u}{4}}J_0(p)}{p\cosh p\frac{u}{4}}\sin(p\sin k^h).$$

The minimum energy to remove a particle from a half filled band occurs at $k^h = \pi$, $\Lambda^h = \infty$,

$$\Delta E_- = E_o(N-1) - E_o(N) = \frac{1}{2}U - 2t + 4t\int_0^\infty dp \frac{J_1(p)}{p\left(1 + e^{\frac{u}{2}|p|}\right)} = \frac{1}{2}U + \epsilon_c(\pi) = \frac{1}{2}\Delta_g(\pi).$$
$$(99)$$

We shall compare it to the energy required to add a particle at half filling in the next subsection.

The phase shift is just the first term in (98)

$$\delta^{h,s} = \frac{\pi}{2} - 2\arctan(e^{\frac{2\pi}{u}(\Lambda^h - \sin k^h)}), \qquad (100)$$

and is similar to the spinon-impurity scattering phase shift we encounterd in the Kondo model. The reason for the similarity is that in the half filled case in the Hubbard model the charge distribution is locked and does not interact directly, while in the Kondo model it decouples completely. We conclude that spinon-holon scattering matrix

$$S_\pi^{s,h}(k^h, \Lambda^h) = i\frac{e^{\frac{2\pi}{u}\sin k^h} + ie^{\frac{2\pi}{u}\Lambda^h}}{e^{\frac{2\pi}{u}\sin k^h} - ie^{\frac{2\pi}{u}\Lambda^h}}, \qquad (101)$$

indicates that the low-energy spinon sector ($\Lambda \to \pm\infty$) completely decouples from the charge sector. This decoupling is captured by the g-ology model, describing the low-energy physics of the model. As discussed earlier, in the latter model spinon sector and holon sector belong to different Hilbert spaces in analogy to the situation in the Lecture 3. It was in this context that the first spinon S-matrix was calculated [15].

Consider now *adding* a particle to the system. This can be done in two ways; creating a state with double occupancy, described in the Bethe-Ansatz language by configuration containing complex $k$-pairs, or creating a state with all momenta real. At half filling only the first possibility exists.

We begin by discussing this situation. We consider a configuration with one hole in the real $k$-sea and a 2-string. A hole also opens in the $\Lambda$-sequence since $N$ is increased by one while $M$ is unchanged. Placing the charge-hole at $k^{\bar h}$, the charge-string at $k^\pm$, where $\sin k^\pm = \lambda \pm i\frac{u}{4}$, and the spin-hole at $\Lambda^h$, we find upon solving for the density of real-$\Lambda$ and real-$k$ that $\lambda = \sin k^{\bar h}$. The relevant integral equation is

$$\mathcal{K}\rho_1'(k) = \cos k\left\{K_1(\sin k - \lambda) - \frac{1}{u}\mathrm{sech}(\frac{2\pi}{u}(\sin k - \Lambda^h)) - \frac{4}{u}R\left(\frac{4}{u}(\sin k - \sin k^{\bar h})\right)\right\}.$$

leading to the excitation energy, momentum and phase-shift

$$\Delta E_+^{(2)}(k^{\bar h}, \Lambda^h) = E^{(2)}(N+1; k^{\bar h}, \Lambda^h) - E_o(N)$$

$$\begin{align}
&= \Delta_g(U, Q_0) + \epsilon_s(\Lambda^h) + \epsilon_c(k^{\bar{h}}), \tag{102}\\
\Delta P(k^{\bar{h}}, \Lambda^h) &= -\frac{2\pi}{L}(I^h + n^h) = p_c(k^{\bar{h}}) + p_s(\Lambda^h), \tag{103}\\
\delta^{s,\bar{h}(2)}(k^{\bar{h}}, \Lambda^h) &= \frac{\pi}{2} - 2\arctan(e^{\frac{2\pi}{u}(\Lambda_h - \sin k_h)}) \\
&\quad - \int_{-Q_0}^{Q_0} dk \rho_1'(k) \left\{ 2\arctan(e^{\frac{2\pi}{u}(\Lambda_h - \sin k^{\bar{h}})}) - \frac{\pi}{2} - \frac{dp_s}{dn} \right\} \tag{104}
\end{align}$$

At half filling we have explicit results, and it can be easily seen that they are the same as when we were removing an electron,

$$\Delta E_+^{(2)}(k^{\bar{h}}, \Lambda^h) = \frac{1}{2}U + 2t\cos k^{\bar{h}} + 2t\int_0^\infty dp \frac{J_1(p)\cos p\Lambda^h}{p\cosh\frac{u}{4}p} + 4t\int_0^\infty dp \frac{J_1(p)\cos(p\sin k^{\bar{h}})}{p\left(1 + e^{\frac{u}{2}|p|}\right)}$$

where $\frac{1}{2}U$ is composed of two contributions: $\frac{1}{2}U$ — the chemical potential, and $U$ — string contribution arising from double occupancy. Hence the minimum energy required to add an electron,

$$E_0(N+1) - E_o(N) = \frac{1}{2}U + \epsilon_c(\pi) \tag{105}$$

is the same as to remove one (due to our choice of a particle-hole symmetric hamiltonian).

Away from half filling we can add a particle without creating double occupancy. There are spaces available for an extra $k = k^{\bar{h}}$, but as before a hole opens in the $\{I_\gamma\}$ sequence at position $\Lambda^h$.

The excitation energy is

$$\begin{align}
\Delta E_+^{(1)}(k^{\bar{h}}, \Lambda^h) &= -2t\cos k^{\bar{h}} - \int_{-Q_0}^{Q_0} dk \rho_1(k)\{2t\cos k - \mu\} - \frac{U}{2}, \\
&= -\frac{U}{2} + \epsilon_c(\Lambda^h) + \epsilon_s(k^{\bar{h}}), \tag{106}
\end{align}$$

and the momentum

$$\Delta P(k^{\bar{h}}, \Lambda^h) = \frac{2\pi}{L}(n^{\bar{h}} - I^h) = p_c(k^{\bar{h}}) + p_s(\Lambda^h). \tag{107}$$

The spinon-antiholon phase-shift,

$$\delta^{s,\bar{h}(1)} = \frac{\pi}{2} + 2\arctan(e^{\frac{2\pi}{u}(\Lambda^h - \sin k^{\bar{h}})}) - \int_{-Q_0}^{Q_0} dk \rho_1(k) \left\{ \arctan(e^{\frac{2\pi}{u}(\Lambda^h - \sin k)}) - \frac{\pi}{2} - \frac{dp_s}{dn} \right\}$$

is conjugate to the spinon-holon phase-shift.

To summarize our conclusions for the repulsive interaction: the spin excitations are gapless for any value of $n$, while gapless charge excitations exist only away from half filling. When the density reaches this critical point, all gapless charge carrying modes disappear from the spectrum (gapful modes are always present), and the system becomes an insulator. This is the mechanism underlying the Mott transition in this model.

The structure of the excitations is rather complex. Away from half filling, there is an incoherent charge background that is modified in the excited states. This is particularly manifest in the phase shifts. We see that, for a given excited state, the phase shift of one of the elemantary excitations consists of a term due to the interaction with the other excitation, and a term that describes the interaction with the change from the ground state of the charge incoherent background.

At half filling, the charge distribution becomes *rigid*, and we only see interactions between the elementary excitations.

When a magnetic field is turned on similar comments will apply to a spin background. We will see the same behavior in the attractive case.

The low energy physics can be captured by an effective fixed point hamiltonian, calculated by repeatedly integrating out higher energy degrees of freedom. This effective hamiltonian in our case will describe excitations with linear dispersion and hence will be conformally invariant. One can, then, from the exact solution calculate the parameters specifying the conformal hamiltonian directly without carrying out RG transformations [20][21].

## The attractive Hubbard model

The structure of the wave functions of the ground state and the excitations changes completely; it is advantageous to have doubly occupied sites to lower the energy. Therefore the ground state is composed entirely of $k$ 2-strings, and the excitations involve either the breaking or displacing of 2-strings.

### The ground state

The ground state configuration for $u < 0$ consist of a *sea* of $k$ 2-strings. To minimize the energy we choose the configuration of quantum numbers $\{I'_\alpha\}$ to consists of consecutive integers centered around 0, filling all slots between $I^-$ and $I^-$. The $k$ 2-strings satisfy the *string hypothesis*

$$k_{\delta,\pm} = \arcsin(\Lambda_\delta \pm i\frac{u}{4}) \qquad \delta = 1,\cdots,M, \tag{108}$$

where $\Lambda_\delta$ are real solutions of the eigenvalue equations. It is easy to see that this choice satisfies the $\Lambda$-equations (2) trivially, whereas the $k$-equations become

$$e^{i2Re\left(\arcsin(\Lambda_\delta + i\frac{u}{4})\right)L} = \left(\prod_{\gamma=1}^{M} \frac{\Lambda_\gamma - \Lambda_\delta - i\frac{u}{2}}{\Lambda_\gamma - \Lambda_\delta + i\frac{u}{2}}\right) \tag{109}$$

whose logarithmic version is

$$2LRe\{\arcsin(\Lambda_\alpha + i\frac{|u|}{4})\} = 2\pi I'_\alpha - \sum_{\beta=1}^{M} \Theta(\Lambda_\alpha - \Lambda_\beta)), \tag{110}$$

with $I'_\alpha$ integer (h.o.i) if $L - M$ is odd (even). The integral equation satisfied by the density of strings, $\sigma_0(\Lambda)$, for the ground state in the thermodynamic limit is

$$\mathcal{L}\sigma_0(\Lambda) \equiv \sigma_0(\Lambda) + \int_{-B_0}^{B_0} d\Lambda' \sigma_0(\Lambda') K_2(\Lambda - \Lambda') = \frac{1}{\pi} Re\left\{\frac{1}{\sqrt{1 - (\Lambda + i\frac{|u|}{4})^2}}\right\} \tag{111}$$

where now $K_n(x) = \frac{1}{\pi}\frac{n\frac{|u|}{4}}{(n\frac{|u|}{4})^2 + x^2}$, and the integration limit $B_0$ depends on the density $n = N/L$ through $\int_{-B_0}^{B_0} d\Lambda \sigma_0(\Lambda) = \frac{1}{2}N/L$. The integral operator $\mathcal{L}$ introduced in eq(111) plays a role similar to the one played by the operator $\mathcal{K}$ in the repulsive case.

The energy and the momentum of the ground state are given by

$$\begin{aligned}E_0(B_0) &= 2t\sum_{\delta=1}^{M'}(\cos k_{\delta,+} + \cos k_{\delta,-}) = -4tL\int_{-B_0}^{B_0} d\Lambda \sigma(\Lambda) Re\sqrt{1 - (\Lambda - i\frac{|u|}{4})^2}, \\ &= -|U|\frac{N}{2} - 2t\int_{-\pi}^{\pi} dk \cos^2 k \int_{-B_0}^{B_0} d\Lambda \sigma_0(\Lambda) K_1(\sin k - \Lambda), \quad (112) \\ P_0(B_0) &= \frac{2\pi}{L}\sum_{\alpha=1}^{M'} I'_\alpha = 0.\end{aligned}$$

where we explicitly separated the contribution to the energy, $-\frac{|U|}{2}N$, due to double occupancy.

Consider first the case of half filling. It is obtained by setting $B_0 = \infty$, as can be easily verified by integrating eq(111). In this case the equation can be solved by Fourier transform to yield

$$\sigma_0^\infty(\Lambda) = \frac{1}{2\pi}\int_0^\infty dp \frac{J_0(p)\cos(\Lambda p)}{\cosh\left(\frac{|u|}{4}p\right)}, \qquad (113)$$

the same function that described the ground state spin density in the $U > 0$ case, eq(24). From eq(113) and eq(112) we get

$$\frac{E_0^\infty(U)}{L} = \frac{U}{2} - 4t\int_0^\infty dp \frac{J_0(p)J_1(p)}{p\left(1 + e^{\frac{|u|}{2}|p|}\right)}. \qquad (114)$$

Comparing (114) to the results for the $U > 0$ case, we find a relation between the the ground state energies of the attractive and repulsive hamiltonias

$$E_0^{att}(U; B_o = \infty) = \frac{UL}{2} + E_0^{rep}(|U|; Q_0 = \pi)$$

as expected on general grounds [1].

To go away from half-filling one needs $B_0 < \infty$, and it is convenient to use a formalism where the physical quantities at arbitrary density are given by equations with those quantities evaluated at half filling appearing as inhomogenous terms. The method is due to Griffith [16]: eq(111) is of the form

$$\sigma(\Lambda) = f(\Lambda) + \int_{-B_0}^{B_0} d\Lambda'\sigma(\Lambda')K(\Lambda - \Lambda'), \qquad (115)$$

and from the solution we wish to calculate integrals of the form

$$I = \int_{-B_0}^{B_0} d\Lambda \sigma(\Lambda)\varepsilon_0(\Lambda), \qquad (116)$$

where $f(\Lambda)$, $K(\Lambda - \Lambda')$, and $\varepsilon_0(\Lambda)$ are known functions defined in the whole real axis, with Fourier transforms $\tilde{f}(p)$, $\tilde{K}(p)$, $\tilde{\varepsilon}(p)$. In this particular case $f(\Lambda) = \phi_0(\Lambda) \equiv \frac{1}{\pi}Re\left\{1/\sqrt{1 - (\Lambda + i\frac{|u|}{4})^2}\right\}$ and $\varepsilon_0(\Lambda) = Re\sqrt{1 - (\Lambda - i\frac{|u|}{4})^2}$.

The function $\sigma(\Lambda)$ is physically relevant in the interval $[-B_0, B_o]$. However, since $f(\Lambda)$ and $K(\Lambda)$ are defined in the whole real axis, the integral equation (115) defines a continuation of $\sigma(\Lambda)$ to the whole real axis. Therefore, we may manipulate the Fourier transform in (115),

$$\begin{aligned}\tilde{\sigma}(p) &= \tilde{f}(p) + \int_{-B_0}^{B_0} d\Lambda' \sigma(\Lambda') e^{ip\Lambda'} \tilde{K}(p), \\ &= \tilde{f}(p) + \left\{ \int_{-\infty}^{\infty} - \int_{|\Lambda'|>B_0} \right\} d\Lambda' \sigma(\Lambda') e^{ip\Lambda'} \tilde{K}(p), \\ &= \tilde{f}(p) + \tilde{K}(p)\tilde{\sigma}(p) - \int_{|\Lambda'|>B_0} d\Lambda' \sigma(\Lambda') e^{ip\Lambda'} \tilde{K}(p) \end{aligned}$$

to obtain an equivalent equation for $\sigma(\Lambda)$

$$\sigma(\Lambda) = \sigma^\infty(\Lambda) - \int_{|\Lambda'|>B_o} d\Lambda' \sigma(\Lambda') R(\Lambda - \Lambda'),. \tag{117}$$

The function

$$\sigma^\infty(\Lambda) = \frac{1}{2\pi} \int_{-\infty}^{\infty} dp \frac{\tilde{f}(p) e^{ip\Lambda}}{1 + \tilde{K}(p)} \tag{118}$$

is the solution of (115) for $B_0 \to \infty$, and we have often encountered the resolvent $R(\Lambda - \Lambda') = \frac{1}{2\pi} \int_{-\infty}^{\infty} dp \frac{\tilde{K}(p) e^{ip(\Lambda-\Lambda')}}{1+\tilde{K}(p)}$.

We rewrite now the integral (116)

$$\begin{aligned} I &= \int_{-B_0}^{B_0} d\Lambda \sigma(\Lambda) \varepsilon_0(\Lambda) = \left\{ \int_{-\infty}^{\infty} - \int_{|\Lambda|>B_o} \right\} d\Lambda \sigma(\Lambda) \varepsilon_0(\Lambda) \\ &= \int_{-\infty}^{\infty} d\Lambda \sigma^\infty(\Lambda) \varepsilon_0(\Lambda) \\ &\quad - \int_{|\Lambda|>B_0} \int_{-\infty}^{\infty} d\Lambda \sigma(\Lambda') R(\Lambda - \Lambda') \varepsilon_0(\Lambda) - \int_{|\Lambda|>B_0} d\Lambda \sigma(\Lambda) \varepsilon_0(\Lambda) \\ &= I^\infty - \int_{|\Lambda|>B} d\Lambda \sigma(\Lambda) \varepsilon(\Lambda), \end{aligned}$$

where $\varepsilon(\Lambda)$ is given by

$$\varepsilon(\Lambda) = \varepsilon_0(\Lambda) + \int_{-\infty}^{\infty} d\Lambda' R(\Lambda - \Lambda') \varepsilon_0(\Lambda'), \tag{119}$$

and $I^\infty$ is the value of $I$ when $B_0 \to \infty$.

In our particular case we have therefore,

$$\mathcal{F}\sigma_0(\Lambda) \equiv \sigma_0(\Lambda) - \frac{4}{|u|}\int_{|\Lambda|>B_0} d\Lambda' \sigma_0(\Lambda') R\left(\frac{4}{|u|}(\Lambda-\Lambda')\right) = \sigma_0^\infty(\Lambda) \quad (120)$$

where $\sigma_0^\infty$ is the density at half filling and is given below. The integral operator $\mathcal{F}$ will recur in our subsequent discussion. (This type of equation also describes the magnetization of the repulsive Hubbard model, with $B_0$ related to the magnetic field.)

The ground state energy is

$$E_0(B_0)/L = E_0^\infty/L + |U|\delta(B_0)/2 + \int_{|\Lambda|>B_0} d\Lambda \sigma_0(\Lambda') \bar{\epsilon}_c^\infty(\Lambda'), \quad (121)$$

where $\delta = \delta(B_0) = 1 - n(B_0)$ measures the doping, and $\bar{\epsilon}_c^\infty$ is the the function

$$\bar{\epsilon}_c^\infty(\Lambda) = 2t \int_0^\infty dp \frac{J_1(p)\cos(p\Lambda)}{p \cosh(\frac{|u|}{4}p)} \quad (122)$$

we met earlier when discussing spinon excitations of the repulsive model at half filling, $\bar{\epsilon}_c^\infty = \epsilon_s^\pi$. We shall see soon, when we discuss charge excitation in the attractive case, that it also corresponds to the energy associated with the charge excitations, hence the notation. The fact that the same function appears in the ground state energy and in the excitations is due to the structure of the ground state in the attractive case. Eq(121) allows us to view the ground state energy as the energy at half filling plus the energy of the holes that have been made on the charge distribution in order to reach a particular filling.

Close to half filling we find an expansion in $\delta$,

$$E_0/L = E_0^\infty/L + |U|\frac{\delta}{2} + \frac{t}{\pi}\delta^2 \frac{I_1\left(\frac{2\pi}{|u|}\right)}{I_0\left(\frac{2\pi}{|u|}\right)}. \quad (123)$$

From this, the chemical potential is obtained

$$\mu = \frac{dE_0}{dN} = -\frac{|U|}{2} - \pi t \delta \frac{I_1\left(\frac{2\pi}{|u|}\right)}{I_0\left(\frac{2\pi}{|u|}\right)}. \quad (124)$$

(Notice that the chemical potential remains negative for any $|U| \neq 0$.) Finally, we write the charge susceptibility

$$\chi_c = \frac{dn}{d\mu} = \frac{1}{\pi t}\frac{I_0\left(\frac{2\pi}{|u|}\right)}{I_1\left(\frac{2\pi}{|u|}\right)}. \tag{125}$$

Elementary excitations

To obtain excitations one has to consider small variations of the quantum numbers away from the ground state configuration. This can be done with pair-breaking (spin excitations) or pair-rearrangement (charge excitations). We shall discuss both possibilities.

**Charge excitations**

*The Holon-holon (Quartet)*. This excitation exists at half-filling and away from it. It is gapless and resembles the spinon excitation of repulsive model. To create it remove two 2-strings and combine the resulting four momenta into the following configuration,

$$\begin{aligned} k_\alpha^1 &= \arcsin(\Lambda_\alpha'' + i\frac{|u|}{2}), & k_\alpha^2 &= \pi - \arcsin(\Lambda_\alpha''), \\ k_\alpha^3 &= \arcsin(\Lambda_\alpha''), & k_\alpha^4 &= \arcsin(\Lambda_\alpha'' - i\frac{|u|}{2}). \end{aligned} \tag{126}$$

The *quartet* is parameterized by $\Lambda_\alpha''$ and the quantum number associated is $I_\alpha''$.

From the constraints on the quantum numbers we can see that, at half-filling, $I_\alpha'' = 0$, and the set $\{I_\alpha'\}$ is not shifted with respect to the ground state configuration. We will assume that this holds away from half-filling.

The set $\Lambda_\alpha'$ is determined therefore from,

$$2LRe\{\arcsin(\Lambda_\alpha' + i\frac{|u|}{4})\}$$
$$= 2\pi I_\alpha' - \sum_{\beta=1}^{N/2-1} \Theta(\Lambda_\alpha' - \Lambda_\beta') - \left\{\Theta(2(\Lambda_\alpha' - \Lambda'')) + \Theta(\frac{2}{3}(\Lambda_\alpha' - \Lambda''))\right\} \tag{127}$$

which translates in the thermodynamic limit to an integral equation for the $k$ 2-string distribution, $\sigma(\Lambda; B)$,

$$\mathcal{L}\sigma(\Lambda) = \phi_0(\Lambda) - \frac{1}{L}\left(\delta(\Lambda - \Lambda_1^h) + \delta(\Lambda - \Lambda_2^h) + K_1(\Lambda - \Lambda'') + K_3(\Lambda - \Lambda'')\right) \tag{128}$$

with $\Lambda'' = \frac{1}{2}(\Lambda_1^h + \Lambda_2^h)$. The integration limits are set by the condition $L \int_{-B}^{B} d\Lambda \sigma(\Lambda) = N/2 - 1$.

Writing $\sigma(\Lambda; B)$ as

$$\sigma(\Lambda; B) = \sigma_0(\Lambda; B) + \frac{1}{L}\sigma_1'(\Lambda; B_0, \Lambda_1^h, \Lambda_2^h) - \frac{1}{L}(\delta(\Lambda - \Lambda_1^h) + \delta(\Lambda - \Lambda_2^h)) \quad (129)$$

we find that the excitation density is given by

$$\sigma_1'(\Lambda) = \sigma_1^c(\Lambda; \Lambda_1^h) + \sigma_1^c(\Lambda; \Lambda_2^h) - \sigma_1^q(\Lambda) \quad , \tag{130}$$

where

$$\mathcal{L}\sigma_1^c(\Lambda; \Lambda^h) = K_2(\Lambda - \Lambda_j^h), \tag{131}$$

$$\mathcal{L}\sigma_1^q(\Lambda; \Lambda^h) = K_1(\Lambda - \Lambda'') + K_3(\Lambda - \Lambda'') \ . \tag{132}$$

These equations bear similarity to the singlet equations in the repulsive case. Here, as note before, they are defined with respect to finite integration limits $\pm B_0$ while there, as long as we considered excitations out of the ground state, $B_0 = \infty$. Rewriting the equations as,

$$\mathcal{F}\sigma_1^c(\Lambda); \Lambda^h) = \sigma_1^{c,\infty}(\Lambda; \Lambda^h), \quad \mathcal{F}\sigma_1^q(\Lambda; \Lambda_1^h, \Lambda_2^h) = \sigma_1^{q,\infty}(\Lambda; \Lambda''). \tag{133}$$

we find

$$\Delta E^{quartet}(\Lambda_1^h, \Lambda_2^h) = \bar{\epsilon}_c(\Lambda_1^h) + \bar{\epsilon}_c(\Lambda_2^h) - \bar{\epsilon}_q(\Lambda'') \quad . \tag{134}$$

with

$$\bar{\epsilon}_c(\Lambda_j^h) = \bar{\epsilon}_c^\infty(\Lambda_j^h) + \int_{|\Lambda|>B_0} d\Lambda \sigma_1^c(\Lambda; \Lambda_j^h) \left\{ \frac{|U|}{2} + \mu + \bar{\epsilon}_c^\infty(\Lambda) \right\},$$

$$\bar{\epsilon}_q(\Lambda'') = \int_{|\Lambda|>B_0} d\Lambda \sigma_1^q(\Lambda; \Lambda'') \left\{ |U| + 2\mu + \bar{\epsilon}_c^\infty(\Lambda) \right\},$$

where $\bar{\epsilon}_c^\infty(\Lambda^h)$ the charge excitation energy at half filling, to be evaluated shortly.

The chemical potential $\mu$ is

$$\mu = \frac{dE_0}{dN} = -2t \frac{Re\left\{\sqrt{1 - (B_0 + i\frac{|u|}{4})^2}\right\} + \int_{-B_0}^{B_0} d\Lambda \Psi_0(\Lambda) Re\left\{\sqrt{1 - (\Lambda) + i\frac{|u|}{4})^2}\right\}}{1 + \int_{-B_0}^{B_0} d\Lambda \Psi_0(\Lambda)} \tag{135}$$

with $\Psi_0(\Lambda) \equiv \frac{1}{2\sigma_0(B_0)} \frac{d\sigma_0}{dB_0}$ satisfying $\mathcal{L}\{\Psi_0(\Lambda)\} = -\frac{1}{2}\{K_2(\Lambda - B_0) + K_2(\Lambda + B_0)\}$. Combining this result with the expression for $\bar{\epsilon}_c^\infty$ one finds $\bar{\epsilon}_q^\infty = \Delta_g(B_0)$ is a gap similar to the one encountered in the repulsive case, with the important difference that it vanishes at half filling, $\Delta_g(\infty) = 0$. We shall see next that holon-antiholon excitations are gapless but, of course, disappear from the spectrum at half filling. This is precisely where the holon-holon ecitation becomes gapless. We conclude therefore that in the attractive case gapless charge excitations are always present.

The excitation momentum is

$$\Delta P^{quartet}(\Lambda_1^h, \Lambda_2^h) = -\frac{2\pi}{L}(\bar{\nu}(\Lambda_1^h) + \bar{\nu}(\Lambda_2^h)) = \bar{p}_c(\Lambda_1^h) + \bar{p}_c(\Lambda_2^h), \qquad (136)$$

with the counting function $\bar{\nu}$ defined as usual (in this case from eq(127)). The momentum function $\bar{p}_c(\Lambda^h) = -\frac{2\pi}{L}\bar{\nu}_0(\Lambda^h)$ takes the form

$$\begin{aligned}
\bar{p}_c(\Lambda^h) &= -2Re\{\arcsin(\Lambda^h + i\frac{|u|}{4})^2\} - 2\pi \int_{-B_0}^{B_0} d\Lambda' \sigma_0(\Lambda')\Theta(\Lambda^h - \Lambda') \\
&= \bar{p}_c^\infty(\Lambda^h) - i\int_{|\Lambda|>B} d\Lambda' \sigma_1^c(\Lambda'; \Lambda^h) \log\left\{ \frac{\Gamma\left(1 + i\frac{\Lambda^h - \Lambda'}{|u|}\right) \Gamma\left(\frac{1}{2} - i\frac{\Lambda^h - \Lambda'}{|u|}\right)}{\Gamma\left(1 - i\frac{\Lambda^h - \Lambda'}{|u|}\right) \Gamma\left(\frac{1}{2} + i\frac{\Lambda^h - \Lambda'}{|u|}\right)} \right\}.
\end{aligned}$$

In terms of these functions the excitation energy and momentum of the quartet state are

$$\begin{aligned}
\Delta E^{quartet}(\Lambda_1^h, \Lambda_2^h) &= \bar{\epsilon}_c(\Lambda_1^h) + \bar{\epsilon}_c(\Lambda_2^h) + \Delta_g(B_0) \\
\Delta P^{quartet}(\Lambda_1^h, \Lambda_2^h) &= \bar{p}_c(\Lambda_1^h) + \bar{p}_c(\Lambda_2^h)
\end{aligned} \qquad (137)$$

As before, the quartet can be described as the combination of two objects, in this case both are holons. They interact and their scattering is given by

$$\begin{aligned}
\delta^{quartet}(\Lambda_1^h, \Lambda_2^h) &= -L(\bar{\nu}(\Lambda_1^h) - \bar{\nu}_0(\Lambda_1^h)) = \\
&\frac{1}{i}\log\left\{ \frac{1 + i\frac{2}{|u|}(\Lambda_1^h - \Lambda_2^h)}{1 - i\frac{2}{|u|}(\Lambda_1^h - \Lambda_2^h)} \frac{\Gamma\left(1 - i\frac{\Lambda_1^h - \Lambda_2^h}{|u|}\right) \Gamma\left(\frac{1}{2} + i\frac{\Lambda_1^h - \Lambda_2^h}{|u|}\right)}{\Gamma\left(1 + i\frac{\Lambda_1^h - \Lambda_2^h}{|u|}\right) \Gamma\left(\frac{1}{2} - i\frac{\Lambda_1^h - \Lambda_2^h}{|u|}\right)} \right\} \\
&+ \int_{|\Lambda|>B_o} d\Lambda' \sigma_1(\Lambda') \left\{ \frac{1}{i}\log \frac{\Gamma\left(1 - i\frac{\Lambda_1^h - \Lambda'}{|u|}\right) \Gamma\left(\frac{1}{2} + i\frac{\Lambda_1^h - \Lambda'}{|u|}\right)}{\Gamma\left(1 + i\frac{\Lambda_1^h - \Lambda'}{|u|}\right) \Gamma\left(\frac{1}{2} - i\frac{\Lambda_1^h - \Lambda'}{|u|}\right)} + \frac{d\bar{p}_c}{dn} \right\},
\end{aligned}$$

where

$$2\frac{dp_s(\Lambda_{h1})}{dn} = \frac{\Theta(\Lambda_1^h - B_0) + \Theta(\Lambda_1^h + B_0) + \int_{-B_0}^{B_0} d\Lambda' \Psi_0(\Lambda')\Theta(\Lambda_1^h - \Lambda')}{1 + \int_{-B_0}^{B_0} d\Lambda \Psi_0(\Lambda)} \quad (.138)$$

We now turn to compute the quantities at half filling. As $B_0 \to \infty$ we have explicitly,

$$\sigma_1^{c,\infty}(\Lambda) = \frac{4}{|u|} R\left(\frac{4}{|u|}(\Lambda - \Lambda_j^h)\right), \text{ and } \sigma_1^{q,\infty}(\Lambda) = K_1(\Lambda - \Lambda''). \quad (139)$$

Hence

$$\bar{\epsilon}_c^\infty(\Lambda_j^h) = 2t \int_{-\pi}^\pi dk \cos^2 k K_1(\sin k - \Lambda)$$

$$-4t\frac{4}{|u|} \int_{-\infty}^\infty d\Lambda R(\frac{4t}{|u|}(\Lambda - \Lambda_j^h)) \left\{ Re\sqrt{1 - (\Lambda_j^h + i\frac{|u|}{4})^2} - \frac{|u|}{4t} \right\},$$

$$= 2t \int_0^\infty dp \frac{J_1(p)\cos(p\Lambda_{hj})}{p \cosh\left(\frac{|u|}{4}p\right)} \quad , \quad (140)$$

whereas $\epsilon_q^\infty = 0$. The function $\bar{\epsilon}_c^\infty(\Lambda)$ was arrived at by a different route when we discussed the ground state, eq(122).

We also have

$$\bar{p}_c^\infty(\Lambda^h) = -\frac{2\pi}{L}\bar{\nu}_0(\Lambda^h) = -\frac{\pi}{2} + \int_0^\infty dp \frac{J_0(p)\sin(p\Lambda^h)}{p\cosh(\frac{|u|}{4}p)}. \quad (141)$$

and the chemical potential becomes

$$\mu = -|U|/2.$$

We observe that at half filling the holons of the attractive model have the same dispersion as the spinons of the repulsive model. This is quite remarkable in view of the profound differnce in structure of the respective ground states. It is, of course, due to the underlying $SU(2) \times SU(2)$ symmetry that is manifest at half filling, and the $Z_2$ transformation that exchanges them.

The phase shift at half filling follows immediately,

$$\delta_\infty^{quartet}(\Lambda_1^h, \Lambda_2^h) = \frac{1}{i}\log\left\{\frac{1 + i\frac{2}{|u|}(\Lambda_1^h - \Lambda_2^h)}{1 - i\frac{2}{|u|}(\Lambda_1^h - \Lambda_2^h)}\frac{\Gamma\left(1 - i\frac{\Lambda_1^h - \Lambda_2^h}{|u|}\right)\Gamma\left(\frac{1}{2} + i\frac{\Lambda_1^h - \Lambda_2^h}{|u|}\right)}{\Gamma\left(1 + i\frac{\Lambda_1^h - \Lambda_2^h}{|u|}\right)\Gamma\left(\frac{1}{2} - i\frac{\Lambda_1^h - \Lambda_2^h}{|u|}\right)}\right\}. \tag{142}$$

Comparing this result to the spin excitations for the $U > 0$ case, we note again,

$$\delta_\infty^{quartet}(U) = \delta_\pi^{singlet}(|U|), \tag{143}$$

substantiating the identification of holons in the attractive model with spinons in the repulsive model.

*Holon-antiholon.* This excitation is only present away from half-filling, where it is possible to remove one pair with $\Lambda^h \leq B_0$ from the sequence, and add another with $\Lambda^{\bar{h}} \geq B_0$, namely outside the $\{\Lambda_\alpha\}$ set of the ground state. Denoting by $I^{\bar{h}}$, $I^h$ the corresponding quantum numbers, the excitation momentum becomes

$$\Delta P^{\bar{h},h} = \frac{2\pi}{L}(I^{\bar{h}} - I^h), \tag{144}$$

and the integral equations for the $\Lambda$-density

$$\mathcal{L}\sigma(\Lambda) = \phi_0(\Lambda) - \frac{1}{L}\delta(\Lambda - \Lambda^h) + \frac{1}{L}K_2(\Lambda - \Lambda^{\bar{h}}). \tag{145}$$

Again, writing $\sigma(\Lambda) = \sigma_0(\Lambda') + \frac{1}{L}\sigma_1'(\Lambda) - \frac{1}{L}\delta(\Lambda - \Lambda^h)$, the integral equation becomes

$$\mathcal{L}\sigma_1'(\Lambda) = K_2(\Lambda - \Lambda^h) - K_2(\Lambda - \Lambda^{\bar{h}}). \tag{146}$$

with the solution

$$\sigma_1'(\Lambda) = \sigma_1^c(\Lambda - \Lambda^h) - \sigma_1^c(\Lambda - \Lambda^{\bar{h}}), \tag{147}$$

leading to the excitation energy and the momentum

$$\begin{aligned}\Delta E^{h,\bar{h}}(\Lambda^h, \Lambda^{\bar{h}}) &= \bar{\epsilon}_c(\Lambda^h) - \bar{\epsilon}_c(\Lambda^{\bar{h}}) \\ \Delta P^{h,\bar{h}}(\Lambda^h, \Lambda^{\bar{h}}) &= \bar{p}_c(\Lambda^h) - \bar{p}_c(\Lambda^{\bar{h}}).\end{aligned} \tag{148}$$

The holon-antiholon excitation is therefore gapless at $\Lambda^h = \Lambda^{\bar{h}} = B_o$. At this point the holon momentum reaches the charge Fermi momentum (of the attractive model) $\frac{\pi}{2}n = k_F = k_F^c(\text{att})$. We shall find power law behaviour for charge density correlation functions at momentum transfer $q \approx 2k_F$.

Finally, the phase shift is

$$\delta^{h,\bar{h}}(\Lambda^h, \Lambda^{\bar{h}}) = \frac{1}{i}\log\left\{\frac{\Gamma\left(1 - i\frac{\Lambda^h - \Lambda^{\bar{h}}}{|u|}\right)\Gamma\left(\frac{1}{2} + i\frac{\Lambda^h - \Lambda^{\bar{h}}}{|u|}\right)}{\Gamma\left(1 + i\frac{\Lambda^h - \Lambda^{\bar{h}}}{|u|}\right)\Gamma\left(\frac{1}{2} - i\frac{\Lambda^h - \Lambda^{\bar{h}}}{|u|}\right)}\right\}$$
$$+ \int_{|\Lambda|>B_0} d\Lambda' \sigma_1(\Lambda') \left[\frac{1}{i}\log\left\{\frac{\Gamma\left(1 - i\frac{\Lambda^h - \Lambda'}{|u|}\right)\Gamma\left(\frac{1}{2} + i\frac{\Lambda^h - \Lambda'}{|u|}\right)}{\Gamma\left(1 + i\frac{\Lambda^h - \Lambda'}{|u|}\right)\Gamma\left(\frac{1}{2} - i\frac{\Lambda^h - \Lambda'}{|u|}\right)}\right\} + \frac{d\bar{p}_c}{dn}\right] \quad (149)$$

with the first term corresponding to the triplet phase shift in eq(58).

**Spin excitations**

The charge excitations we studied were gapless at half filling and away from it since they essentially involved moving pairs around. The construction of spin excitations, on the other hand, requires breaking of pairs and therefore an energy gap opens up.

*The Triplet.* The $k$ 2-strings describe spin singlets, hence the only way to generate a spin excitation from the ground state is to break one of the pairs creating momenta $k_1$, $k_2$ the real line with the corresponding the quantum numbers $n_1$ and $n_2$. Now $M_1' = N/2 - 1$, hence the state is a spin triplet, consisting of two objects, the spin-1/2 spinons of the attractive model coupled symmetrically to form a spin-1 state.

No hole opens in the $\{I'_\alpha\}$ set since one slot less is available to the reduced number of pairs. The quantum numbers will still be distributed symmetrically around the origin. As a consequence the only contribution to the excitation momentum comes for the real $k$,

$$\Delta P = \frac{2\pi}{L}(n_1 + n_2). \quad (150)$$

Since there are no holes in the $\Lambda$-distribution, $\sigma(\Lambda)$ satisfies the integral equation

$$\mathcal{L}\sigma(\Lambda) = \phi_0(\Lambda) - \frac{1}{L}K_1(\Lambda - \sin k_1) - \frac{1}{L}K_1(\Lambda - \sin k_2) \quad (151)$$

As usual, we introduce an excitation density, $\sigma_1(\Lambda)$, via $\sigma(\Lambda) = \sigma_0(\Lambda) + \frac{1}{L}\sigma_1(\Lambda)$ and we have

$$\sigma_1(\Lambda) = \sigma_1^{k_1}(\Lambda) + \sigma_1^{k_2}(\Lambda), \tag{152}$$

where $\sigma_1^{k_j}$ is the solution of the equation

$$\mathcal{L}\sigma_1^{k_j} = -K_1(\Lambda - \sin k_j), \tag{153}$$

or, equivalently

$$\mathcal{F}\sigma_1^{k_j} = \sigma_1^{k_j,\infty}. \tag{154}$$

The excitation energy is

$$\Delta E^{trip}(k_1, k_2) = \bar{\epsilon}_s(k_1) + \bar{\epsilon}_s(k_2) \quad , \tag{155}$$

with

$$\begin{aligned}
\bar{\epsilon}_s(k_1) &= -2t\cos k_j - 4t\int_{-B_0}^{B_0} d\Lambda \sigma_1^{k_j}(\Lambda)\left\{Re\sqrt{1 - (\Lambda - i\frac{|u|}{4})^2} + \frac{2\mu}{4t}\right\} - \mu \\
&= \bar{\epsilon}_s^\infty(k_1) + \int_{|\Lambda|>B_0} d\Lambda' \sigma_1^{k_j}(\Lambda)\{\bar{\epsilon}_s^\infty(\Lambda) + \frac{|U|}{2} + \mu\},
\end{aligned}$$

and the excitation momentum

$$\Delta P^{trip}(k_1, k_2) = \bar{p}_s(k_1) + \bar{p}_s(k_2) \quad , \tag{156}$$

with

$$\bar{p}_s(k_j) = \bar{p}_s^\infty(k_1) - \int_{|\Lambda|>B_0} d\Lambda' \sigma_1^{k_j}(\Lambda')\left\{\frac{\pi}{2} - 2\arctan\left(e^{\frac{2\pi}{|u|}(\Lambda' - \sin k_j)}\right)\right\}. \tag{157}$$

Finally, the phase shift is

$$\begin{aligned}
\delta^{trip} &= \int_{-B_0}^{B_0} d\Lambda \sigma_1(\Lambda)\left\{\Theta(2(\sin k_1 - \Lambda)) - 2\frac{d\bar{p}_s(k_1)}{dn}\right\} \\
&= \delta_\infty^{trip} - 2\frac{d\bar{p}_s}{dn} + \int_{|\Lambda|>B_0} d\Lambda \sigma_1(\Lambda)\left\{2\arctan\left(e^{\frac{2\pi}{|u|}(\Lambda - \sin k_1)}\right) - \frac{\pi}{2} - \frac{d\bar{p}_s(k_1)}{dn}\right\}
\end{aligned} \tag{158}$$

At half filling we have

$$\sigma_1^k(\Lambda) = -\frac{1}{u}\text{sech}\left(\frac{2\pi}{u}(\Lambda - \sin k)\right), \tag{159}$$

hence

$$\epsilon_s^\infty(k) = \frac{|U|}{2} - 2t\cos k + 4t\int_0^\infty dp \frac{J_1(p)\cos(p\sin k)}{1 + e^{\frac{|u|}{2}|p|}}. \tag{160}$$

$$\bar{p}_s^\infty(k) = k + 2\int_0^\infty dp \frac{J_0(p)\sin(p\sin k)}{1 + e^{\frac{|u|}{2}|p|}}. \tag{161}$$

From (160) and (161) we conclude that the spinons underlying the spin triplet correspond to the charged holons in the repulsive Hubbard model. This is further borne out by the phase shift,

$$\begin{aligned}\delta_\infty^{trip} &= -2\pi\int_{-\infty}^\infty d\Lambda \frac{1}{|u|}\text{sech}\left(\frac{2\pi}{|u|}(\Lambda - \sin k)\right)\Theta(2(\sin k_2 - \Lambda)), \\ &= \frac{1}{i}\log\left\{\frac{\Gamma\left(1 - i\frac{\sin k_1 - \sin k_2}{|u|}\right)\Gamma\left(\frac{1}{2} + i\frac{\sin k_1 - \sin k_2}{|u|}\right)}{\Gamma\left(1 + i\frac{\sin k_1 - \sin k_2}{|u|}\right)\Gamma\left(\frac{1}{2} - i\frac{\sin k_1 - \sin k_2}{|u|}\right)}\right\}. \end{aligned} \tag{162}$$

Away from half filling the symmetry is broken but the identification still survives.

*The Singlet.* We wish to break a pair without changing the spin, hence need to introduce an additional $I_\alpha$. We have $M_1' = N/2 - 1$, $M_1 = 1$. As before, we have two real $k$ and quantum numbers $n_1$, $n_2$. It can be seen from (198) that $I_\alpha = 0$. However, the presence of a new $\Lambda$ implies that the quantum numbers have to be shifted by $\frac{1}{2}$ with respect to those in the triplet, as can be seen from (197). As a result, an extra $\pi$ appears in the phase shift. Also, we determine the position of the added spin momentum to be $\Lambda = \frac{1}{2}(\sin k_1 + \sin k_2)$.

We find that the densities $\sigma(\Lambda)$ and $\sigma_1(\Lambda)$ are the same as in the triplet case. So are the excitation energy and momentum. The only difference with respect the triplet case appears in the phase shift,

$$\delta^{sing} = \delta^{trip} - \Theta(\sin k_1 - \sin k_2) - \pi \tag{163}$$

At half filling, we have

$$\delta_\infty^{sing} = -\pi + \frac{1}{i}\log\left\{\frac{1 - i\frac{2}{|u|}(\sin k_1 - \sin k_2)}{1 + i\frac{2}{|u|}(\sin k_1 - \sin k_2)}\frac{\Gamma\left(1 - i\frac{\sin k_1 - \sin k_2}{|u|}\right)\Gamma\left(\frac{1}{2} + i\frac{\sin k_1 - \sin k_2}{|u|}\right)}{\Gamma\left(1 + i\frac{\sin k_1 - \sin k_2}{|u|}\right)\Gamma\left(\frac{1}{2} - i\frac{\sin k_1 - \sin k_2}{|u|}\right)}\right\}.$$

a result we met as the holon-holon phase shift in the repulsive case, $\delta_\pi^{h,h}$. We must conclude that the spin excitations here are made of the same objects that made the charge excitations in the $u > 0$.

Combining the singlet and triplet reults we find that the spin S-matrix at half filling has the familiar form for spinon scattering,

$$S_\infty^{spin} = \frac{\Gamma\left(1 - i\frac{\sin k_1 - \sin k_2}{|u|}\right)\Gamma\left(\frac{1}{2} + i\frac{\sin k_1 - \sin k_2}{|u|}\right)}{\Gamma\left(1 + i\frac{\sin k_1 - \sin k_2}{|u|}\right)\Gamma\left(\frac{1}{2} - i\frac{\sin k_1 - \sin k_2}{|u|}\right)} \left\{ \frac{(\sin k_1 - \sin k_2)I^{12} + i\frac{|u|}{2}P^{12}}{(\sin k_1 - \sin k_2) + i\frac{|u|}{2}} \right\}. \tag{164}$$

*Charge-spin excitations*

We consider now excitations where the number of electrons changes. Again we add an extra term to the hamiltonian to make apparent the particle-hole symmetry.

To *remove* an electron we must break a pair. Once the electron is removed, we have an unpaired real $k$ left. We have $N-1$ electrons and $M' = N/2 - 1$, hence the number of slots available does not change. As there is one less $k$ 2-string a hole in the $\{\Lambda\}$ sequence will appear. The state is labeled by the holon parameter $k^h$ and the spinon parameter $\Lambda^h$, with the corresponding quantum numbers $n^h$ and $I^h$, and momentum momentum excitation $\Delta P = \frac{2\pi}{L}(n^h - I^h)$ .

The integral equations for $\sigma(\Lambda)$ and $\sigma_1'(\Lambda)$ are straightforward leading to

$$\Delta E^{s,h}(k^h, \Lambda^h) = \bar{\epsilon}_s(k^h) + \bar{\epsilon}_c(\Lambda^h), \quad \Delta P^{s,h}(k^h, \Lambda^h) = \bar{p}_s(k^h) + \bar{p}_c(\Lambda^h) \quad ,(165)$$

Again we find theat an elctron is comopsed of a spinon excitation carrying the spin content and a holon carrying the charge. An electron with definite momentum $p$ removed from the system will be a superposition of the type of state just constructed subject to $\Delta P^{s,h}(k^h, \Lambda^h) = p$, the spread being a measure of its lifetime.

The spinon-holon phase shift is

$$\begin{aligned}
\delta^{s,h}(k^h, \Lambda^h) &= \Theta(2(\sin k^h - \Lambda^h)) \\
&\quad - \int_{-B_0}^{B_0} d\Lambda [\sigma_1^c(\Lambda; \Lambda^h) + \sigma_1^{k^h}(\Lambda)] \left\{ \Theta(2(\sin k^h - \Lambda)) - 2\frac{d\bar{p}_s(k^h)}{dn} \right\} \\
&= \delta_\infty^{s,h} - 2\frac{d\bar{p}_s}{dn} \tag{166}
\end{aligned}$$

$$+ \int_{|\Lambda|>B_0} d\Lambda [\sigma_1^c(\Lambda;\Lambda^h) + \sigma_1^{k^h}(\Lambda)] \left\{ 2\arctan\left(e^{\frac{2\pi}{|u|}(\Lambda_h - \sin k^h)}\right) - \frac{\pi}{2} - \frac{d\bar{p}_s}{dn} \right\}$$

At half filling we have,

1. From (139)
$$\sigma_1^{c,\infty}(\Lambda;\Lambda^h) = \frac{4}{|u|} R\left(\frac{4}{|u|}(\Lambda - \Lambda^h)\right), \qquad (167)$$

2. From (159)
$$\sigma_1^{k^h,\infty}(\Lambda) = -\frac{1}{u}\text{sech}\left(\frac{2\pi}{u}(\Lambda - \sin k^h)\right), \qquad (168)$$

3. From (140)
$$\bar{\epsilon}_c^{\infty}(\Lambda^h) = 2t \int_0^{\infty} dp \frac{J_1(p)\cos(p\Lambda^h)}{p\cosh\left(\frac{|u|}{4}p\right)}, \qquad (169)$$

4. From (160)
$$\bar{\epsilon}_s^{\infty}(k^h) = \frac{|U|}{2} - 2t\cos k^h + 4t \int_0^{\infty} dx \frac{J_1(p)\cos(p\sin k^h)}{1 + e^{\frac{|u|}{2}|p|}}, \qquad (170)$$

5. From (137)
$$\bar{p}_c(\Lambda^h) = \int_0^{\infty} d\Lambda' \frac{J_0(p)\sin(p\Lambda^h)}{p\cosh(\frac{|u|}{4}p)} \qquad (171)$$

6. From (161)
$$\bar{p}_s(k^h) = k^h + 2\int_0^{\infty} dp \frac{J_0(p)\sin(p\sin k^h)}{1 + e^{\frac{|u|}{2}|p|}}. \qquad (172)$$

7. From (166)
$$\delta_{\infty}^{s,h} = \frac{\pi}{2} + 2\arctan\left(e^{\frac{2\pi}{|u|}(\Lambda^h - \sin k^h)}\right) \qquad (173)$$

Once more we see that an energy gap is present at half filling.

When we *add* an electron we introduce a real $k$ describing the antiholon and, from (198), it is clear that the number of available slots for the $\{I'_\alpha\}$-quantum numbers increases by one, while the same amount of $k$ 2-strings are present. As a result, a hole appears in $\{\Lambda\}$. The state is labeles by the spinon parameter $\Lambda^h$ and antiholon parameter $k^{\bar{h}}$.

The integral equations are straightforward and similar to the previous case. However, the value of the integration limit is different, since the number of $k$ 2-strings does not change. We have $\int_{-B}^{B} d\Lambda \sigma(\Lambda) = \int_{-B_0}^{B_0} d\Lambda \sigma_0(\Lambda) = N/2L$, and hence a shift in the integration limit

$$L(B - B_0) = \frac{1 - \int_{-B_0}^{B_0} d\Lambda \sigma'_1(\Lambda)}{\frac{1}{2}\frac{dN_0}{dB_0}}. \tag{174}$$

This change affects the values of the excitation energy and the phase shift, but not excitation momentum,

$$\begin{aligned}
\Delta E^{s,\bar{h}}(k^{\bar{h}}, \Lambda^{\bar{h}}) &= \bar{\epsilon}_s(k^{\bar{h}}) + \bar{\epsilon}_c(\Lambda) + \frac{|U|}{2} + 2\mu \quad, \\
\Delta P^{s,\bar{h}}(k^{\bar{h}}, \Lambda^{\bar{h}}) &= \bar{p}_s(k^{\bar{h}}) + \bar{p}_c(\Lambda_h) \quad, \\
\delta^{s,\bar{h}}(k^{\bar{h}}, \Lambda^{\bar{h}}) &= \Theta(2(\sin k^{\bar{h}} - \Lambda_h)) \\
&\quad + \int_{-B_0}^{B_0} d\Lambda [\sigma_1^c(\Lambda; \Lambda^h) + \sigma_1^{k^{\bar{h}}}(\Lambda)] \left\{ \Theta(2(\sin k^{\bar{h}} - \Lambda)) - 2\frac{dp_c(k^{\bar{h}})}{dn} \right\} + 2\frac{d\bar{p}_s}{dn} \\
&= \delta_\infty^{s,\bar{h}} + \int_{|\Lambda|>B_0} d\Lambda [\sigma_1^c(\Lambda; \Lambda^h) + \sigma_1^{k^{\bar{h}}}(\Lambda)] \left\{ 2\arctan\left(e^{\frac{2\pi}{|u|}(\Lambda_h - \sin k^{\bar{h}})}\right) - \frac{\pi}{2} - \frac{d\bar{p}_s}{dn} \right\}.
\end{aligned} \tag{175}$$

Turning to half filling, the results (167-173) apply here too. In particular, we find that the phase shift is the same

$$\delta_{N=L}^{s,h} = \delta_{N=L}^{s,\bar{h}}$$

reflecting the appearance of the charge $SU(2)$, as the relation between the energies

$$\Delta E^{s,h} = \Delta E^{s,\bar{h}} \tag{176}$$

reflects the particle hole symmetry.

We have completed a detailed discussion of the elementary excitations and their varoius characteristics. These are the low lying solutions of the Bethe-Ansatz equations. To compute the free energy, however, we need the complete set of solutions as given by the string hypothesis.

### The Thermodynamics of the Hubbard Model

The thermodynamics is derived [5] by the method [22] we already employed in Lecture 3. We shall merely outline the main steps and write down the answer.

Consider the thermodynamic potential, $\Omega = E - AN - T\mathcal{S}$, with $\mathcal{S}$ the entropy and $A$ the (external) chemical potential. One calculates it summing over all energy eigenstates, or equivalently integrating over all allowed solution densities. Denote by $\rho(k), \sigma_n(\Lambda)$ and $\sigma'_n(\Lambda)$ the distribution functions of $k, \Lambda^{(n)}, \Lambda'^{(n)}$ respectively, where $\Lambda^{(n)}$ is the real part of the $\Lambda$ $n$-string and $\Lambda'^{(n)}$ is the real part of the $k - \Lambda$ $n$-string, and by $\rho^h(k), \sigma_n^h(\Lambda)$ and $\sigma_n'^h(\Lambda)$ the corresponding hole-distributions, see Appendix. Further define:

$\zeta(k) = \rho^h(k)/\rho(k),\ \eta_n(\Lambda) = \sigma_n^h(\Lambda)/\sigma_n(\Lambda),\ \eta'_n(\Lambda) = \sigma_n'^h(\Lambda)/\sigma'_n(\Lambda).$

The same steps that led to the thermodynamic equations in Lecture 3 lead now to the following expression for the thermodynamoc potential,

$$\Omega/L = E_o - A - T\left[\int_{-\pi}^{\pi} \rho_o(k)\ln(1 + \zeta(k))dk + \int_{-\infty}^{\infty} \sigma_o(\Lambda)\ln(1 - \eta_1(\Lambda))d\Lambda\right]$$

with $E_o, \sigma_o(\Lambda), \rho_o(k)$ being the ground state energy and densities at half filling, and the functions $\zeta(k)$ and $\eta_1(\Lambda)$ are determined from the following set of coupled integral equations,

$$\ln \eta_n = G[\ln(1 + \eta_{n+1}) + \ln(1 + \eta_{n-1})]\quad n = 2, 3, ...$$
$$\ln \eta'_n = G[\ln(1 + \eta'_{n+1}) + \ln(1 + \eta'_{n-1})]\ n\ n = 2, 3, ...$$
$$\ln \eta_1 = G[\ln(1 + \eta_2) - \int_{-\pi}^{\pi} \delta(\Lambda - \sin k)\ln(1 + \zeta^{-1})\cos k\ dk]$$
$$\ln \eta'_1 = G[\ln(1 + \eta'_2) - \int_{-\pi}^{\pi} \delta(\Lambda - \sin k)\ln(1 + \zeta)\cos k\ dk]$$
$$\ln \zeta = -\frac{2\cos k}{T} + \frac{1}{u}\int_{-\infty}^{\infty} d\Lambda\ \text{sech}\frac{2\pi}{u}(\Lambda - \sin k)\left[\frac{4}{T} - (\text{sgn } U)\ Re\sqrt{1 - (\Lambda - \frac{u}{4}i)^2} + \ln\frac{1+\eta'_1}{1+\eta_1}\right]$$

where $G$ is the integral operator, $Gf(\Lambda) = \frac{1}{u}\int \frac{1}{\cosh\frac{2\pi}{u}(\Lambda-\Lambda')}f(\Lambda')$. The asymptotic conditions are,

$$\ln \eta_n \;\to\; n\frac{2h}{T}, \quad \text{as} \;\; n \to \infty \tag{177}$$

$$\ln \eta'_n \;\to\; n\frac{u-2A}{T}, \quad \text{as} \;\; n \to \infty \tag{178}$$

We leave it as an exercise to show that in the limit $T \to 0$ and no magnetic field the equations collapse, depending on the the sign of $U$, to their respective ground state equations: eqs(13,14) in the case of repulsion, eq(111) in the case of attraction. If the zero temperature limit is taken in the presence of a magnetic field one obtains the corresponding magnetization equations.

In the repulsive case,

$$\rho(k) \;=\; \frac{1}{2\pi} + \cos k \int_{-B(h)}^{B(h)} d\Lambda \sigma(\Lambda) K_1(\sin k - \Lambda)$$

$$\sigma(\Lambda) \;=\; \int_{-Q}^{Q} dk \rho_o(k) K_1(\sin k - \Lambda) - \int_{-B(h)}^{B(h)} d\Lambda' \sigma(\Lambda') K_2(\Lambda - \Lambda'),$$

In the attractive case we have

$$\sigma(\Lambda) \;=\; \frac{1}{\pi} Re \left\{ \frac{1}{\sqrt{1-(\Lambda+i\frac{|u|}{4})^2}} \right\} - \int_{-Q(h)}^{Q(h)} K_1(\Lambda - \sin k')\rho(k')dk' - \int_{-B}^{B} d\Lambda' \sigma(\Lambda') K_2(\Lambda - \Lambda')$$

$$\rho(k) \;=\; \frac{1}{2\pi} + \cos k \int_{-B}^{B} d\Lambda \sigma(\Lambda) K_1(\sin k - \Lambda)$$

The integration limits are now determined by the imposed magnetic field $h$. In the repulsive case it mainly affects $B = B(h) < \infty$ ( $Q$ is mainly determined by the density though there is also a weak dependence on $h$.) The situation is reversed in the attractive model where $B$ is mainly determined by the density and $Q$ mainly by the magnetic field.

Unlike the magnetization equation discussed in lecture 3 the limits are imposed symmetrically and therefore analytical results are available only in some limits. For extensive numerical work see [17] and [4] in the repulsive case, and [18] [19] in the attractive case.

The thermodyanmic equations can now be analysed in a manner discussed in Lecture 3 to determine the behavior of the model in the infra red. The physics is determined by the gapless excitations: charge excitaions in the attractive model, spin excitations in the repulsive model at half filling and both charge and spin away from it, and will flow to a fixed point accordingly.

Alternatively, the nature of the fixed point can be identified using Bethe-Ansatz finite size calculations combined with methods of conformal field theory [20][21]. But that would be the subject of Lecture 5.

**Appendix**

Here we discuss the different species of $\{k_j\}$ and $\{\Lambda_\alpha\}$ that appear in the attractive and repulsive cases. We can classify the eigenvalues in the following categories:

1. *Real $k_j$*. They have associated quantum numbers $n_j$.

2. *k 2-strings*. We may have solutions with pairs of complex $k$, $k_\alpha = \kappa_n^\pm \pm i\chi_n$.

   To be solutions of the Bethe-Ansatz equations, eqs(1,2), in the thermodynamic limit we must have [6],

   $$\sin\left(\kappa_n^{(\pm)} \pm i\chi_n^{(\pm)}\right) = \Lambda_n \mp i\frac{U}{4} + \mathcal{O}\left(e^{-\eta_n^{(\pm)}N}\right) \qquad (179)$$

   This is valid only if we can neglect $\mathcal{O}\left(e^{-\eta_n^{(\pm)}N}\right)$ terms, where

   $$\eta_n^{(\pm)} = \chi_n^{(\pm)} \pm \frac{1}{N}Im\left\{\sum_{m\neq n} 2\arctan\left(\frac{4}{U}\left(\sin\left(\kappa_n^{(\pm)} \pm i\chi_n^{(\pm)}\right) - \Lambda_m\right)\right)\right\} \qquad (180)$$

   Therefore, we must have $\chi_n^{(\pm)} \geq 0$. Let's prove this point. That is, the values described by the string hypothesis (180) are indeed solutions of the eigenvalue problem. We will start with the spin equation (2). Let's write

   $$\sin k_\pm = \sin\left(\kappa \pm i\chi_\alpha\right) = \lambda + \epsilon \mp i\frac{c}{2}, \quad \epsilon \ll \lambda \qquad (181)$$

   where $\lambda$ belongs to the $\{\Lambda_\alpha\}$ set and, for simplicity, we let all the $\Lambda$'s to be real. Then (2) becomes

$$-\left(\prod_{\delta=1}^{M-1}\frac{\Lambda_\delta-\Lambda_\gamma+i\frac{u}{2}}{\Lambda_\delta-\Lambda_\gamma-i\frac{u}{2}}\right)\frac{\lambda-\Lambda_\gamma+i\frac{u}{2}}{\lambda-\Lambda_\gamma-i\frac{u}{2}}$$

$$=\left(\frac{1+i\frac{u}{2\epsilon}}{1-i\frac{u}{2\epsilon}}\right)\left(\prod_{j=1}^{N-2}\frac{\Lambda_\gamma-\sin k_j-i\frac{u}{4}}{\Lambda_\gamma-\sin k_j+i\frac{u}{4}}\right)\frac{\Lambda_\gamma-\lambda-i\frac{u}{2}}{\Lambda_\gamma-\lambda+i\frac{u}{2}}$$

$$\overset{\epsilon\to 0}{\simeq}\ e^{i\pi}\prod_{j=1}^{N-2}\frac{\Lambda_\gamma-\sin k_j-i\frac{u}{4}}{\Lambda_\gamma-\sin k_j+i\frac{u}{4}} \tag{182}$$

which is again spin equation, but for $N-2$ particles see eq(79).

We now turn to the charge equation, (1). After substitution, we get

$$e^{i(\kappa_\alpha+i\chi_\alpha)L}=\left(\frac{\epsilon}{\epsilon-i\frac{u}{2}}\right)\prod_{\gamma\ne\alpha}\frac{\Lambda_\gamma-\sin(\kappa_\alpha+i\chi_\alpha)-i\frac{u}{4}}{\Lambda_\gamma-\sin(\kappa_\alpha+i\chi_\alpha)+i\frac{u}{4}} \tag{183}$$

$$e^{i(\kappa_\alpha-i\chi_\alpha)L}=\left(\frac{\epsilon+i\frac{u}{2}}{\epsilon}\right)\prod_{\gamma\ne\alpha}\frac{\Lambda_\gamma-\sin(\kappa_\alpha-i\chi_\alpha)-i\frac{u}{4}}{\Lambda_\gamma-\sin(\kappa_\alpha-i\chi_\alpha)+i\frac{u}{4}} \tag{184}$$

Dividing (184) by (183), and taking logarithms we get

$$\chi_\alpha L\simeq-\log\epsilon+\log\frac{u}{4}+\sum_{\gamma\ne\alpha}Im\left(-2\arctan\left(\sin(\kappa_\alpha+i\chi_\alpha)-\Lambda_\gamma\right)\right) \tag{185}$$

Notice that $\log\frac{u}{4}\ll L$. Therefore, in order for (185) to be satisfied, we must have

$$\epsilon\sim\mathcal{O}\left(e^{-\eta L}\right) \tag{186}$$

with $\eta_\alpha$ given by (180).

When we neglect the exponential term, (179) has two possible solutions

(a)

$$\kappa_n^{(\pm)}\pm i\chi_n^{(\pm)}=\arcsin\left(\Lambda_n\mp i\frac{U}{4}\right),\quad -\frac{\pi}{2}\le\kappa_n^{(\pm)}\le\frac{\pi}{2}\implies\cos\kappa_n^{(\pm)}\ge 0 \tag{187}$$

This gives

$$\sin\left(\kappa_n^{(\pm)} \pm i\chi_n^{(\pm)}\right) = \sin\kappa_n^{(\pm)} \cosh\chi_n^{(\pm)} \pm i\cos\kappa_n^{(\pm)} \sinh\chi_n^{(\pm)} = \Lambda_n \mp i\frac{U}{4} \tag{188}$$

Since $\cos\kappa_n^{(\pm)} \geq 0$,

$$\chi_n^{(\pm)} > 0 \iff U < 0 \tag{189}$$

(b)

$$\kappa_n'^{(\pm)} \pm i\chi_n^{(\pm)} = \pi - \arcsin\left(\Lambda_n \mp i\frac{U}{4}\right), \quad \frac{\pi}{2} \leq \kappa_n'^{(\pm)} \leq \frac{3\pi}{2} \implies \cos\kappa_n'^{(\pm)} \leq 0 \tag{190}$$

This gives

$$\sin\left(\kappa_n'^{(\pm)} \pm i\chi_n^{(\pm)}\right) = \sin\kappa_n'^{(\pm)} \cosh\chi_n^{(\pm)} \pm i\cos\kappa_n'^{(\pm)} \sinh\chi_n^{(\pm)} = \Lambda_n \mp i\frac{U}{4} \tag{191}$$

Since $\cos\kappa_n'^{(\pm)} \leq 0$,

$$\chi_n^{(\pm)} > 0 \iff U > 0 \tag{192}$$

We conclude that the forms of the k 2-string that is a solution of the problem depends on the sign of the interaction.

Therefore, the k 2-strings are of the form

i) $u > 0$.

$$\begin{aligned} k_\alpha^- &= \pi - \arcsin(\Lambda' - i\frac{u}{4}) \\ k_\alpha^+ &= \pi - \arcsin(\Lambda' + i\frac{u}{4}) \end{aligned}, \tag{193}$$

ii) $u < 0$.

$$\begin{aligned} k_\alpha^- &= \arcsin(\Lambda' - i\frac{|u|}{4}) \\ k_\alpha^+ &= \arcsin(\Lambda' + i\frac{|u|}{4}) \end{aligned}. \tag{194}$$

Notice that each pair is described by only one parameter, $\Lambda'_\alpha$. The quantum numbers associated to the pairs will be denoted $I'_\alpha$.

3. *Quartet, $k-\Lambda$ 2-string.* One of the possible excitations in the attractive case involves a group of four complex $k_\alpha$ such that

$$\begin{aligned} k^1_\alpha &= \arcsin(\Lambda''_\alpha + i\frac{|u|}{2}) \quad, \\ k^2_\alpha &= \pi - \arcsin(\Lambda''_\alpha) \quad, \\ k^3_\alpha &= \arcsin(\Lambda''_\alpha) \quad, \\ k^4_\alpha &= \arcsin(\Lambda''_\alpha - i\frac{|u|}{2}) \end{aligned} \qquad (195)$$

parametrized by a single variable $\Lambda''_\alpha$ and the quantum number associated is $I''_\alpha$.

4. *Real $\Lambda_\alpha$.* The corresponding quantum numbers are denoted $I_\alpha$, as in the ground state.

5. *$\Lambda$ 2-string.* It will only appear in the elementary excitations for $u > 0$. We have

$$\Lambda^\pm_\alpha = \lambda_\alpha \pm i\frac{u}{4} \ . \qquad (196)$$

The pair is parametrized by the real $\lambda_\alpha$ and has quantum number $J_\alpha$ associated to it.

In the thermodynamic limit, where the *string hypothesis* is valid, the objects enumerated are allowed solutions of the Bethe-Ansatz equations, and the sets $\{\Lambda'_\alpha\}$, $\{\Lambda''_\alpha\}$, and $\{\lambda_\alpha\}$ are subsets of $\{\Lambda_\alpha\}$.

We wish to count the number of slots available for each configuration. Denoting,

$$\begin{aligned} M &= \text{number of real } \Lambda_\alpha, \\ \bar{M} &= \text{number of } \lambda_\alpha, \ \Lambda \text{ 2-strings} \\ M' &= \text{number of } \Lambda'_\alpha, \ k \text{ 2-strings} \\ M'' &= \text{number of } \Lambda''_\alpha, \text{quartets} \\ M''' &\equiv M' + 2M'', \\ N - 2M''' &= \text{number of real } k_j, \\ N_\downarrow &= M + 2\bar{M} + M''', \\ N_\uparrow &= N - N_\downarrow. \end{aligned}$$

we have (assuming $L$ of the form, $L = 4\nu + 2$, $\nu$ integer)[5][9],

$$n_j = \begin{cases} \text{integer} & (M + \bar{M} + M' + M'' \quad \text{even}), \quad |n_j| \leq L/2 \\ \text{h.o.i} & (M + \bar{M} + M' + M'' \quad \text{odd}), \quad |n_j| \leq (L-1)/2 \end{cases} \quad (197)$$

$$|I_\alpha| \leq \frac{1}{2}(N - 2M''' - (M + 2\bar{M}) - 1) \quad , \quad I_\alpha = \begin{cases} \text{integer} & (N - M \quad \text{odd}) \\ \text{h.o.i} & (N - M \quad \text{even}) \end{cases}$$

$$|J_\alpha| \leq \frac{1}{2}(N - 2M''' - 2M - \bar{M} - 1) \quad , \quad J_\alpha = \begin{cases} \text{integer} & (N - \bar{M} \quad \text{odd}) \\ \text{h.o.i} & (N - \bar{M} \quad \text{even}) \end{cases}$$

$$|I'_\alpha| \leq \frac{1}{2}(L - N + 2M''' - M' + 2M'' - 1) \quad , \quad I_\alpha = \begin{cases} \text{integer} & (L - (N - M') \quad \text{odd}) \\ \text{h.o.i} & (L - (N - M') \quad \text{even}) \end{cases}$$

$$|I''_\alpha| \leq \frac{1}{2}(L - N + 2M''' - 2M' + M - 1) \quad , \quad I''_\alpha = \begin{cases} \text{integer} & (L - (N - M'') \quad \text{odd}) \\ \text{h.o.i} & (L - (N - M'') \quad \text{even}) \end{cases} \quad (198)$$

ACKNOWLEGEMENTS

I wish to thank Yu Lu for his invitation to deliver these lectures, and his kind patience while they were written up. I am grateful to G. Kotliar and A. Ruckenstein for their comments, to M. Rozenberg for teaching me how to draw, to C. Destri for ongoing discussions, and to A. Jerez for reading the manuscript and for his many many suggestions.

# References

[1]     E.L. Lieb and F.Y. Wu, Phys. Rev. Lett **20**, 1445 (1968).

[2]     A. A. Ovchinikov, JETP **30**, 1160 (1970).

[3]     C. F. Coll, phys. Rev. **9**, 2150 (1974).

[4]     H. Shiba, Phys. Rev. **B 6**, 930 (1972).

[5]     M. Takahashi, Prog. Theor. Phys. **47**, 69 (1972).

[6]     F. Woynarovich, J. Phys. **C 15**, 85 (1982), ibid., 97 (1982).

[7] A. Kluemper, A. Schadschneider and J. Zittarz Z. Phys. bf B 78, 99 (1990).

[8] A. Jerez and N. Andrei, Rutgers Preprint (in preparation).

[9] F.H.L. Essler and V.E. Korepin, preprint ITP-SB-93-45.

[10] C. N. Yang, Phys. Rev. Lett.**63**, 2144 (1989). I. Affleck, talk given at Nato Advanced Study Institute on *Physics, Geometry and Topology*, Banff, (1989).

[11] F.H.L. Essler, V.E. Korepin and K. Schoutens, Nuc. Phys. **B384**, 431 (1992).

[12] J. Carmelo and D. Baeriswyl, Phys. Rev. **B 37**, 7541 (1987).

[13] V. J. Emery, in *Highly Conducting One-Dimensional Solids*, J. T. Devreese et al. eds, Plenum, N. Y. 1979. J. Solyom, Adv. in Phys. **28**, 201 (1979).

[14] N. Andrei, J.H. Lowenstein, Phys. Rev. Lett. bf 43, 1693 (1979).

[15] N. Andrei, J. H. Lowenstein, Phys. Lett. **B 91**, 401 (1980).

[16] R. B. Griffiths, Phys. Rev. **A 3** (1964), 768.

[17] M. Takahashi, Prog. Theo. Phys. **42** (1969), 1099, **43** (1970), 1619.

[18] F. Woynarovich, K. Penc, Z. Phys. B **85** (1991), 269.

[19] K-J-B. Lee and P. Schlottmann, Phys. Rev. **B 40**, 9104 (1989).

[20] F. Woynarovich, J. Phys. **A 22**, 2615 (1989) F. Woynarovich and H.-P. Eckle, J. Phys. **A 20**, L443 (1987).

[21] H. Frahm and V. V. Korepin, Phys. Rev.**B 42**, 10553 (1990). N. Kawakami and Sung-Kil Yang J. Phys: Condens. Matter **3**, 5983 (1991).

[22] C. N. Yang and C. P. Yang, J. Math. Phys. **10**, 1115 (1969)